\documentclass[preprint,11pt]{elsarticle}
\makeatletter

\let\ps@pprintTitle\ps@plain

\makeatother
 \journal{ }
\usepackage[dvipsnames, svgnames, x11names]{xcolor}
\usepackage{graphicx}
\usepackage{pifont,latexsym,ifthen,amsthm,rotating,calc,textcase,booktabs}
\usepackage{amsfonts,amssymb,amsbsy,amsmath,extarrows,bm,mathtools}
\newtheorem{theorem}{Theorem}[section]
\newtheorem{lemma}[theorem]{Lemma}
\newtheorem{corollary}[theorem]{Corollary}

\newtheorem{proposition}[theorem]{Proposition}

\newtheorem{prob}{RH problem}[section]
\newtheorem{prob2}{$\bar{\partial}$-RH problem}[section]
\newtheorem{prob3}{$\bar{\partial}$-problem}[section]

\numberwithin{equation}{section}
\usepackage{natbib}
\usepackage{tikz}
\usetikzlibrary{decorations.markings,arrows.meta}
\usepackage{stmaryrd}
\usepackage[T1]{fontenc}
\usepackage[english]{babel}
\usepackage{listings}
\usepackage{xcolor,mathrsfs,url}
\usepackage{ifthen}
\usepackage{fancyhdr}
\usepackage{verbatim}
\usepackage{todonotes}
\usepackage{float}
\usepackage{subfigure}
\usepackage[subfigure]{tocloft}
\usepackage{hyperref}
\usepackage{appendix}
\usepackage[percent]{overpic}
\usepackage {caption}

\hypersetup{hypertex=true,
            colorlinks=true,
            linkcolor=blue,
            anchorcolor=blue,
            citecolor=red}

\DeclareMathOperator*{\im}{Im}

\DeclareFontFamily{U}{mathx}{}
\DeclareFontShape{U}{mathx}{m}{n}{<-> mathx10}{}
\DeclareSymbolFont{mathx}{U}{mathx}{m}{n}
\DeclareMathAccent{\widehat}{0}{mathx}{"70}
\DeclareMathAccent{\widecheck}{0}{mathx}{"71}

\usetikzlibrary{decorations.markings}
\definecolor{sectorblue}{RGB}{225,238,248}

\begin{document}

\begin{frontmatter}
\title{\LARGE Painlev\'e   transcendents  in  the  defocusing Hirota equation  with  non-zero boundary conditions}

\author[inst1]{Shuning Lin}

\author[inst1]{Engui Fan$^{\star,}$  }

\address[inst1]{ School of Mathematical Sciences and Key Laboratory of Mathematics for Nonlinear Science, Fudan University, Shanghai, 200433, China.\\$\star$ Corresponding author and e-mail address: faneg@fudan.edu.cn}


\begin{abstract}

We consider the Cauchy problem for the defocusing Hirota equation with non-zero boundary conditions
\begin{align*}
&iq_t+ q_{xx}-2\left(|q|^2-1\right)q
+i\left(q_{xxx}-6|q|^2q_x\right)=0,\\
&q(x,0)=q_0(x)\to\pm 1, \quad x\to\pm\infty,
\end{align*}
which can be characterized using a Riemann-Hilbert problem through the inverse scattering transform. With the $\bar{\partial}$-steepest
descent method and double scaling limit technique, we derive the long-time asymptotics of the solution in two transition regions
\begin{align*}
    \mathcal{P}_{a}
    :=\left\{(x,t)\in\mathbb{R}\times\mathbb{R}^{+}:
    \left| {x}/{t}-a\right|t^{2/3}<C\right\},
    \quad a\in\{-4,-8\},
\end{align*}
where $C>0$ is a constant. It is shown that the asymptotic expansions  in two transition
 regions can be  expressed in terms of the Hastings--McLeod solution
of the Painlev\'e II equation in the generic case and Ablowitz--Segur solutions in the non-generic case.

\end{abstract}

\begin{keyword}
Defocusing Hirota equation, \ Cauchy problem,\  Riemann-Hilbert problem, \ $\bar{\partial}$-steepest descent method, \ Painlev\'{e}  transcendents, \ Long-time asymptotics

\textit{Mathematics Subject Classification:} 35P25; 35Q51; 35Q15; 35B40; 35C20.
\end{keyword}

\end{frontmatter}

\tableofcontents

\section{Introduction }

We investigate the Painlev\'{e} asymptotics of the defocusing Hirota equation with nonzero boundary conditions (NZBCs)
\begin{align}\label{Hirota}
&iq_t+ q_{xx}-2\left(|q|^2-1\right)q
+i\left(q_{xxx}-6|q|^2q_x\right)=0,\\
&q(x,0)=q_0(x)\to\pm 1, \quad x\to\pm\infty,\label{Hirota-bc}
\end{align}
where $q=q(x,t)$ is a complex-valued function,    and  $q_0(x)-\tanh{(x)}\in H^{4,4}(\mathbb{R})$. The Hirota equation  is an important integrable higher-order extension of the nonlinear Schr\"odinger (NLS) equation. It incorporates third-order dispersion and nonlinear corrections of self-steepening type into the standard NLS equation, providing an important mathematical model for studying the propagation of ultrashort optical pulses in the presence of higher-order effects \cite{KivsharAgrawal2003,Kodama1985,KodamaHasegawa1987,YanDai2013}.
 The usual form of the Hirota equation is \cite{Hirota1973}
\begin{align}\label{usual-Hirota}
iq_t+\alpha \left[q_{xx}-2|q|^2q\right]
+i\beta \left(q_{xxx}-6|q|^2q_x\right)=0,	
\end{align}
where the real constants $\alpha$ and $\beta$ are the coefficients of second- and third-order dispersion, respectively. When $\alpha=1$ and $\beta=0$, the defocusing Hirota equation (\ref{usual-Hirota}) reduces to the defocusing NLS equation, whereas when $\alpha=0$ and $\beta=1$, it reduces to the defocusing complex modified Korteweg-de Vries   equation. For $\alpha\beta\neq0$,
using  the  scaling transformation $q\mapsto\frac{\alpha}{\beta}q, x\mapsto\frac{\beta}{\alpha}x, t\mapsto\frac{\beta^{2}}{\alpha^{3}}t$,  we can change  the equation
 \eqref{usual-Hirota}   into a simple  form
\begin{align}\label{Hirota-3}
iq_t+q_{xx}-2|q|^{2}q+i\left(q_{xxx}-6|q|^{2}q_x\right)=0,
\end{align}
which further   reduces to the equation \eqref{Hirota} by  the change of variable $q\mapsto qe^{-2it}$.

Considerable progress has been made in the study of the defocusing Hirota equation \eqref{Hirota}. Regarding exact solutions, Zhang and Yuan constructed dark soliton solutions using a binary Darboux transformation \cite{ZhangYuan2017}.
Zhang et al. further investigated double-valley dark solitons and their inelastic interactions \cite{ZhangQinLingZhao2021}.
 Soliton solutions of the defocusing Hirota equation with nonzero boundary conditions were obtained in \cite{ZhangChenYan2020}
  and \cite{HuangDiYao2023} via the inverse scattering transform and the $\bar{\partial}$-dressing method, respectively.
  Wang and He constructed dark soliton solutions on elliptic-function backgrounds \cite{WangHe2025}.
   Cao and Zhao obtained algebro-geometric solutions using the Riemann-Hilbert (RH) method \cite{CaoZhao2024},
   while Li et al. developed Whitham modulation theory for periodic solutions and investigated the corresponding dam-breaking problem \cite{LiBaiZhao2024}.  Huang et al.
    considered  the initial-value problem  of the Hirota equation \eqref{Hirota}
    with zero boundary conditions   and derived the long-time  asymptotics  using the Deift-Zhou nonlinear steepest descent method \cite{HuangXuFan2015}. Higher-order asymptotic expansions for the same zero-boundary problem were obtained in \cite{HuangXuFan2015Higher}.

\begin{figure}[htbp]
    \centering
    \begin{tikzpicture}[node distance=1.5cm]

    \fill[white]
        (-5,0)--(4.5,0)--(4.5,3)--(-5,3)--cycle;

    \fill[white]
        (0,0)--(-5,{5/4})--(-5,{20/7})--cycle;

    \fill[white]
        (0,0)--(-5,{5/8})--(-5,{5/4})--cycle;

    \draw[blue!15,fill=blue!15]
        (0,0)--(-5,1.125)--(-5,1.375)--cycle;

    \draw[blue!15,fill=blue!15]
        (0,0)--(-5,0.500)--(-5,0.750)--cycle;

    \draw[-latex]
        (-5.5,0)--(5,0)
        node[right] {$x$};

    \draw[-latex]
        (0,0)--(0,3)
        node[above] {$t$};

    \node[below] at (0,0) {$0$};

    \draw[ ]
        (0,0)--(-5,{20/7})
        node[left,black] {\footnotesize$\xi=-\frac{7}{4}$};

    \draw[]
        (0,0)--(-5,{5/4});

    \node[anchor=east] at (-5.12,1.28) {\footnotesize$\xi=-4$};

    \draw[ ]
        (0,0)--(-5,{5/8});

    \node[anchor=east] at (-5.12,0.64) {\footnotesize$\xi=-8$};

    \node at (-3.6,1.50) {\footnotesize $\mathcal{R}_{\mathrm{III}}$};

    \node at (-4.0,0.75) {\footnotesize$\mathcal{R}_{\mathrm{II}}$};

    \node at (-4.0,0.19) {\footnotesize$\mathcal{R}_{\mathrm{I}}$};
    \node at (1.8,1.5) {\footnotesize$\mathcal{R}_{\mathrm{I}}$};



    \end{tikzpicture}

    \caption{The $(x,t)$-plane is divided into five kinds of asymptotic regions:
    $\mathcal{R}_{\mathrm{I}}$ (solitonless regions): $\xi<-8$ or $\xi>-7/4$;
    $\mathcal{R}_{\mathrm{II}}$ (Zakharov--Manakov region): $-8<\xi<-4$;
    $\mathcal{R}_{\mathrm{III}}$ (solitonic region): $-4<\xi\leq-7/4$;
    and two transition regions: $\mathcal{P}_{-4}$ ($\xi\approx-4$) and $\mathcal{P}_{-8}$ ($\xi\approx-8$).}
    \label{div-region}
\end{figure}
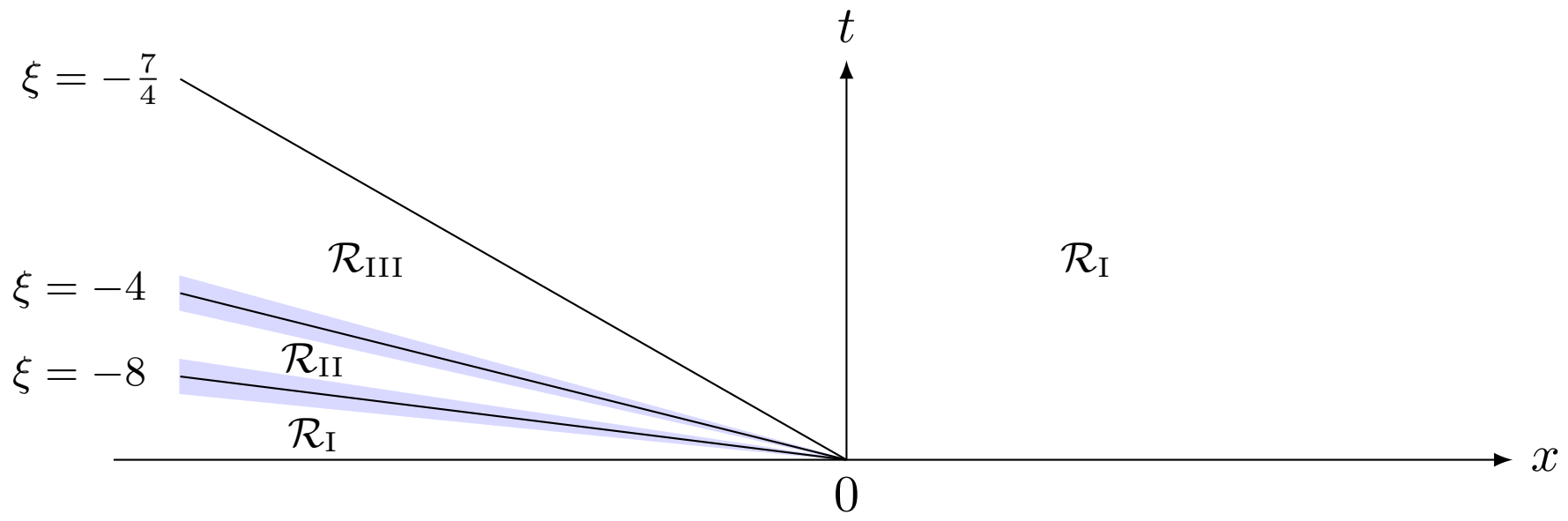
    Recently,  Peng and Chen considered   the Cauchy problem \eqref{Hirota}-\eqref{Hirota-bc} and   divided the $(x,t)$-plane into three  distinct asymptotic regions \cite{PengChen2026}.
More precisely, with  $\xi=x/t$ and $t>0$,  these regions are described as follows (see Fig. \ref{div-region}):
\begin{itemize}
\item Solitonless regions $\mathcal{R}_{\mathrm{I}}:=\left\{(x,t):\xi<-8\right\}\cup\left\{(x,t):\xi>-\frac{7}{4}\right\}$;
\item Zakharov--Manakov region $\mathcal{R}_{\mathrm{II}}:=\left\{(x,t):-8<\xi<-4\right\}$;
\item Solitonic region $\mathcal{R}_{\mathrm{III}}:=\left\{(x,t):-4<\xi \leq -\frac{7}{4}\right\}$.
\end{itemize}
In particular, they derived the long-time asymptotics for the solitonless region $\mathcal{R}_{\mathrm{I}}$.
Thus, a   remaining important  problem is how  to characterize the long-time asymptotics  in the following  two transition regions  near the critical
velocities   $\xi=-4$ and $\xi=-8$:
\begin{align}
    \mathcal{P}_{-4}
    &:= \left\{(x,t)\in\mathbb{R}\times\mathbb{R}^+:
    \left|\xi+4\right|t^{2/3}<C\right\},\\
   \mathcal{P}_{-8}
    &:= \left\{(x,t)\in\mathbb{R}\times\mathbb{R}^+:
    \left|\xi+8\right|t^{2/3}<C\right\},
\end{align}
where    $C>0$ is a fixed constant. The aim of  this paper is to  study the  Painlev\'e asymptotics in two transition regions $\mathcal{P}_{-4}$ and $\mathcal{P}_{-8}$    by using the $\bar{\partial}$-steepest descent method and the double scaling limit technique.

The Painlev\'{e} transcendents play an important role in the long-time asymptotic analysis of many integrable nonlinear evolution equations, particularly in characterizing the behavior of solutions in transition regions associated with coalescing stationary phase points. In early work, Segur and Ablowitz \cite{SegurAblowitz1981} described the asymptotic behavior of the Korteweg-de Vries   equation in transition regions using Painlev\'e functions. Deift and Zhou \cite{DeiftZhou1993} developed the nonlinear steepest descent method in their study of the modified Korteweg-de Vries (mKdV) equation and rigorously analyzed its long-time asymptotics, including those in the transition region. Subsequently, Boutet de Monvel et al. \cite{BoutetdeMonvelItsShepelsky2010} applied this method to investigate Painlev\'e-type asymptotics for the Camassa-Holm equation. Charlier and Lenells \cite{CharlierLenells2020} obtained higher-order Airy and Painlev\'e asymptotics for the mKdV equation, and Huang and Zhang \cite{HuangZhang2022} extended this analysis to the mKdV hierarchy. Under   nonzero boundary conditions, Painlev\'e asymptotic formulas in transition regions were established by Wang and Fan for the defocusing NLS equation \cite{WangFan2023} and by Wang, Xu, and Fan for the defocusing mKdV equation \cite{WangXuFan2025}.  In a transition region for   the defocusing Hirota equation \eqref{Hirota}  with zero boundary conditions,  Xun and Fan \cite{XunJuFan2022}   derived long-time asymptotic expansions in terms of solutions of the Painlev\'e II equation.

\subsection{Main results}

We investigate the long-time asymptotics  of the  Hirota equation \eqref{Hirota}
  in two transition regions $\mathcal{P}_{-4}$ and $\mathcal{P}_{-8}$
 and  the main results are stated as follows.

\begin{theorem}\label{Theo-1}
For initial data $q_0(x)-\tanh{(x)}\in H^{4,4}(\mathbb{R})$, let $r(z)$ and $\{(\eta_j,c_j)\}_{j=1}^N$ be the associated reflection coefficient and discrete scattering data, respectively. Then the long-time asymptotics of the solution to the Cauchy problem \eqref{Hirota}-\eqref{Hirota-bc} in two transition regions are given as follows.
\begin{itemize}
  \item Case I.  For $(x,t)\in\mathcal{P}_{-4}$, we have
\begin{align}\label{q-case1}
q(x,t) = q^{sol}_\Lambda(x,t)-iT^2(\infty)\left( {9t}/{4}\right)^{-\frac{1}{3}} f_{12}+\mathcal{O}\left(t^{-\frac{1}{3}-\tau}\right),
\ \ 0<\tau< {1}/{15},
\end{align}
where  $q^{sol}_\Lambda(x,t)$ is   a   soliton  solution    given by \eqref{soliton-reduction},   and
\begin{align*}
   &T(\infty)=\left(\prod_{j\in\Delta_{-4}}\overline{\eta}_j\right)
    \exp\left\{\frac{i}{2}\int_{\mathbb{R}}\frac{\nu(\zeta)}{\zeta}\,d\zeta\right\},   \\
     &\begin{aligned}
     &   f_{12}=\frac{i}{2}\Big\{
        \alpha_1\bigl[(1+i\overline{\beta}_0)\alpha_1-i\alpha_0\overline{\beta}_1\bigr]u(s)e^{i\varphi_0} +\beta_1\bigl[(1+i\overline{\beta}_0)\beta_1-i\alpha_0\overline{\alpha}_1\bigr]u(s)e^{-i\varphi_0}\\
        &+\bigl[i\alpha_0\bigl(|\alpha_1|^2+|\beta_1|^2\bigr)-2(1+i\overline{\beta}_0)\alpha_1\beta_1\bigr]
        \int_s^\infty u(\zeta)^2\,\mathrm{d}\zeta\Big\},\quad \varphi_0=\arg\widetilde{r}(1),
    \end{aligned}
\end{align*}
with parameters $\alpha_j, \beta_j, j=0,1$ being determined  by (\ref{mout}), and the   function $u(s)$ being
a  solution of the Painlev\'{e} \uppercase\expandafter{\romannumeral2}  equation
\begin{align}\label{equ-pain}
u''(s)=2u^3(s)+su(s), \ \ s={4}/{9}\left({9t}/{4}\right)^{2/3}\left({x}/{t}+4\right),
\end{align}
characterized by the asymptotic condition
\begin{align}\label{equ-pain-asy}
 u(s) \sim -|\widetilde{r}(1)| \mathrm{Ai}(s), \quad s \to +\infty,
\end{align}
where $\operatorname{Ai}(s)$ is the classical Airy function. The solution $u(s)$ of \eqref{equ-pain}-\eqref{equ-pain-asy} is known as the Hastings--McLeod solution \cite{p1} in the generic case, i.e., $|r(1)|=1$, and as an Ablowitz--Segur solution \cite{SegurAblowitz1981} in the non-generic case, i.e., $|r(1)|<1$.

\item Case II.   For $(x,t)\in\mathcal{P}_{-8}$, we have
\begin{align}
    q(x,t)=
    -T^2(\infty)
    +\frac{T^2(\infty)}{2}
    \left(\frac{15t}{4}\right)^{-\frac{1}{3}}
    \left[
        u(s)e^{i\varphi_0}
        -i\int_s^\infty u(\zeta)^2\,d\zeta
    \right]
    +\mathcal{O}\left(t^{-\frac{1}{3}-\tau}\right),\nonumber
\end{align}
where $0<\tau<1/15$ and
\begin{align*}
&T(\infty)=
    \exp\left\{
        \frac{i}{2}\int_J\frac{\nu(\zeta)}{\zeta}\,d\zeta
    \right\},
    \quad
    J=\mathbb{R}\setminus[z_2,z_1],	\\
&s=\frac{4}{15}
    \left(\frac{15t}{4}\right)^{2/3}
    \left(\frac{x}{t}+8\right),
    \quad
    \varphi_0=\arg\widetilde{r}(-1),
\end{align*}
and $u(s)$ is the   solution of \eqref{equ-pain} characterized by the asymptotic condition
\begin{align}\label{equ-pain-asy2}
    u(s)\sim-|\widetilde{r}(-1)|\operatorname{Ai}(s).
    \quad s\to+\infty,
\end{align}
\end{itemize}
\end{theorem}

 Compared with the NLS equation and mKdV equation associated with the   Hirota equation (\ref{Hirota}),
 the highlights of  our  work are summarized in the following points.
\begin{itemize}
    \item     Compared with NLS equation and mKdV equation \cite{WangFan2023,WangXuFan2025},
     the  phase function of the   Hirota equation (\ref{Hirota})
     admits   more intricate   signature table,
     which  require  more detailed   adjustments to the RH  deformations in the asymptotic analysis.

    \item  By contrast to   the defocusing NLS and mKdV equations \cite{WangFan2023,WangXuFan2025},
    where  no   solitons   appear  in   asymptotic expansion  in  transition regions since
     solitons  can be removed.  While    solitons will  persist   in the transition region $\mathcal{P}_{-4}$ for the  Hirota equation (\ref{Hirota}),  as shown   in \eqref{q-case1}.

	\item In contrast to   the zero-boundary case    \cite{XunJuFan2022},   we
	have to deal with   both the generic and non-generic cases for our present   nonzero-boundary case.   In the generic case, the condition $|r(\pm1)|=1$ gives rise to the  singularities of   the jump factorization at the coalescence points $z=\pm1$. We overcome this difficulty through a regular reformulation of the scattering data that removes these singularities while preserving the estimates required for the $\bar{\partial}$-extensions.
\end{itemize}

\subsection{Outline of this paper}

The paper is organized as follows. In Section  \ref{sec-IST}, we recall the basic results of scattering
transform for the Cauchy problem \eqref{Hirota}-\eqref{Hirota-bc}, including the properties of the Jost functions and scattering data, and formulate the associated RH problem. We further
 distinguish the solitonless, Zakharov--Manakov, solitonic, and transition regions according to
  the distribution of stationary phase points. Section \ref{sec-P-4} is devoted to the long-time asymptotics of the defocusing Hirota equation in the transition region $\mathcal{P}_{-4}$. In Section \ref{sec3.1}, we apply a diagonal conjugation and remove the singularity at $z=0$ while retaining the poles, obtaining a meromorphic RH problem for $M^{(2)}(z)$. In Section  \ref{sec-hybrid}, we construct continuous extensions of the jump matrices to transform  this RH problem into a hybrid $\bar{\partial}$-RH problem for $M^{(3)}(z)$. The latter is then decomposed into a pure RH problem for $M^{rhp}(z)$ and a pure $\bar{\partial}$-problem for $M^{(4)}(z)$, which are analyzed separately. In Sec. \ref{sec-pureRH}, we construct the outer model $M^{out}(z)$ associated with the discrete spectrum and a Painlev\'e II local parametrix $M^{loc}(z)$ near $z=1$, where the two stationary points coalesce as $t\to\infty$. We then estimate the error matrix $E(z)$ through a small-norm RH problem. In Sec. \ref{sec-Dbar}, we establish solvability of the pure $\bar{\partial}$-problem for $M^{(4)}(z)$ and estimate its contribution to the reconstruction of $q(x,t)$. Finally, in Section \ref{sec-proof}, we combine these estimates to complete the proof of Case I of Theorem \ref{Theo-1}. In Section \ref{sec-P-8}, we investigate the asymptotics of the solution in the transition region $\mathcal{P}_{-8}$ using an approach similar to that in Section \ref{sec-P-4}. The result is stated in Case II of Theorem \ref{Theo-1}.

\subsection{Some notation}

We introduce some notation that will be used in this paper:
\begin{itemize}
    \item  $L^{p,s}(\mathbb{R})$ defined with the norm $\|q\|_{L^{p,s}(\mathbb{R})} := \|\langle x \rangle^s q\|_{L^{p}(\mathbb{R}) }$,  where $\langle x \rangle=\sqrt{1+x^2}$.
    \item $W^{k,p}(\mathbb{R})$ defined with the norm $\|q\|_{W^{k,p}(\mathbb{R})}:=\sum_{j=0}^k \|\partial^jq\|_{L^p(\mathbb{R})}$, where $\partial^jq$ is the $j^{th}$ weak derivative of $q$.
   \item$H^k(\mathbb{R})$ defined with the norm $\|q\|_{H^k(\mathbb{R})}:=\|\langle x \rangle^k \widehat q\|_{L^2(\mathbb{R})}$, where $\widehat q$ is the Fourier transform of $q$.
  \item $H^{k,s} (\mathbb{R}) : = H^k(\mathbb{R}) \cap L^{2,s}(\mathbb{R})$ defined with the norm
    $\|q\|_{H^{k, s}(\mathbb{R})}  :=     \|  q \|_{H^k(\mathbb{R})}+\|\langle x \rangle^s q\|_{L^{2}(\mathbb{R}) }.$
    \item As usual, the three Pauli matrices are defined by
    \begin{equation}
    \sigma_1 = \left(\begin{array}{cc}
  0 & 1\\
  1 & 0\\
 \end{array}\right), \quad \sigma_2 = \left(\begin{array}{cc}
  0 & -i\\
  i & 0\\
 \end{array}\right), \quad \sigma_3 = \left(\begin{array}{cc}
  1 & 0\\
  0 & -1\\
 \end{array}\right). \nonumber
    \end{equation}
    \item $\hat{\sigma}_3 $ acts on a matrix $A$ by   $e^{\hat{\sigma}_3 }A = e^{\sigma_3}  A e^{-\sigma_3} $.
    \item For a matrix $A$, we denote its $(i,j)$-entry by $A_{ij}$ and its $j$-th column by $A_j$, unless otherwise specified.
    \item We use $a \lesssim b$ ($a \gtrsim b$) to denote $a \leq Cb$ ($a \geq Cb$) for constants $C> 0$.
\end{itemize}

\section{Inverse Scattering Transform}\label{sec-IST}

In this section, we summarize the main results on the inverse scattering transform for the Cauchy problem \eqref{Hirota}-\eqref{Hirota-bc}. Further details can be found in \cite{ZhangChenYan2020, PengChen2026}.

\subsection{Jost functions}

The Lax pair of the defocusing Hirota equation \eqref{Hirota} is given by   \cite{ZhangChenYan2020, PengChen2026}
\begin{align}\label{Hirota-Lax}
\Phi_x&=X\Phi=(ik\sigma_3+Q)\Phi,
\quad
\Phi_t=T\Phi
=\left(T_{\mathrm{NLS}}+T_{\mathrm{CMKdV}}\right)\Phi,
\end{align}
where $\Phi = \Phi(x,t,k)$ is a $2 \times 2$ matrix-valued function and $k \in \mathbb{C}$ is the spectral parameter. Here, the matrices $Q$, $T_{\mathrm{NLS}}$, and $T_{\mathrm{CMKdV}}$ are defined by
\begin{align}
&Q=Q(x,t)=\sigma_1\operatorname{diag}\left(\bar{q}(x,t),q(x,t)\right),\nonumber\\
&T_{\mathrm{NLS}}
=-2kX+i\sigma_3\left(Q_x-Q^2+I\right),
\nonumber\\
&T_{\mathrm{CMKdV}}
=-2k\left(T_{\mathrm{NLS}}-i\sigma_3\right)
+[Q_x,Q]+2Q^3-Q_{xx}.\nonumber
\end{align}

In view of the boundary conditions \eqref{Hirota-bc}, the Lax pair (\ref{Hirota-Lax}) reduces to the following asymptotic spectral problems as $x\to\pm\infty$:
\begin{align}
\Psi_{\pm,x}=X_\pm\Psi_\pm,
\quad
\Psi_{\pm,t}=T_\pm\Psi_\pm, \label{back}
\end{align}
where
\begin{align}
X_\pm=ik\sigma_3+Q_\pm,
\quad
T_\pm=(4k^2-2k+2)X_\pm,
\quad
Q_{\pm} = \pm \sigma_1.
\end{align}
The eigenvalues of $X_\pm$ are $\pm i\lambda$, where $\lambda$ satisfies
\begin{equation}
    \lambda^2=k^2-1.
\end{equation}
Since $\lambda$ is a two-valued function of $k$, we introduce the uniformizing variable
\begin{equation}
    z=k+\lambda.
\end{equation}
Then $k$ and $\lambda$ become the following single-valued functions of $z$:
\begin{equation}
    \lambda(z)=\frac{1}{2}(z-z^{-1}),\quad k(z)=\frac{1}{2}(z+z^{-1}).
\end{equation}
Consequently,  it follows from ( \ref{back}) that the background solutions are given by
\begin{align}
\Psi_{\pm}(x,t,z)
= E_{\pm}(z)e^{it\theta(x,t,z)\sigma_3},
\quad z \in \mathbb{C}\setminus\{0\},
\end{align}
where
\begin{align}
E_\pm&=
\begin{pmatrix}
1 & \pm i z^{-1}\\
\mp i z^{-1} & 1
\end{pmatrix},
\quad
\theta(x,t,z)
=\lambda(z)\left(\frac{x}{t}+4k^{2}(z)-2 k(z)+2\right).
\end{align}

For $z \in \mathbb{R}\setminus\{0,\pm 1\}$,   the solutions $\Phi_\pm$   of the Lax pair \eqref{Hirota-Lax} is characterized by the asymptotic behavior
\begin{align}
\Phi_\pm(x,t,z)\sim\Psi_\pm(x,t,z),
\quad x\to\pm\infty.
\end{align}
For $z=\pm1$, their values are defined by continuous extension. We introduce the gauge transformation
\begin{align}
\mu_\pm(x,t,z)=\Phi_\pm(x,t,z)\mathrm{e}^{-i t\theta(x,t,z) \sigma_3}. \label{gauge}
\end{align}
 It follows that
\begin{align}
\mu_\pm(z)&\sim E_\pm(z),
\quad x\to\pm\infty,
\end{align}
and
\begin{align}
\det(\Phi_\pm)=\det(\mu_\pm)=1-z^{-2}.
\end{align}
Under the  transformation (\ref{gauge}), the Lax pair takes the form
\begin{align}
&\left(E_\pm^{-1}\mu_\pm\right)_x
+i\lambda\left[E_\pm^{-1}\mu_\pm,\sigma_3\right]
=E_\pm^{-1}\Delta Q_\pm\mu_\pm,
\quad z\neq 0,\pm1,
\\
&\left(E_\pm^{-1}\mu_\pm\right)_t
+i\lambda (4k^2-2k+2)
\left[E_\pm^{-1}\mu_\pm,\sigma_3\right]
=E_\pm^{-1}\Delta T_\pm\mu_\pm,
\quad z\neq 0,\pm1,
\end{align}
where
\begin{align}
\Delta Q_\pm&=Q-Q_\pm,
\quad
\Delta T_\pm=T-T_\pm.
\end{align}
The eigenfunctions $\mu_\pm$ satisfy the following Volterra-type integral equations:
\begin{align}
\mu_\pm(z)
&=
\begin{cases}
\displaystyle
E_\pm+E_\pm\int_{\pm\infty}^{x}
\mathrm{e}^{i\lambda(x-y)\widehat{\sigma}_3}
\left[E_\pm^{-1}\Delta Q_\pm\mu_\pm\right]\mathrm{d}y,
& z\neq 0, \pm1,
\\[8pt]
\displaystyle
E_\pm+\int_{\pm\infty}^{x}
\left[I+(x-y)(Q_\pm + i z \sigma_3)\right]
\Delta Q_\pm\mu_\pm\,\mathrm{d}y,
& z=\pm1.
\end{cases}
\end{align}

Denote $\mathbb{C}^\pm =\{z \in \mathbb{C}:\pm \im z >0\}$ and   $\mu_{\pm}(z)=\left( \mu_{\pm,1}(z),  \mu_{\pm,2}(z)\right)$,  we summarize  the properties of $\mu_{\pm} (z)$   in the following lemma.
\begin{lemma}\label{lemma-mu}
Given $n\in\mathbb{N}_0$, suppose that $q_0-\tanh{(x)}\in L^{1,n+1}(\mathbb{R})$ and $q'_0(x) \in W^{1,1}(\mathbb{R})$. Then we have

\begin{itemize}
\item {Analyticity:} The functions $\mu_{+,1}(z)$ and $\mu_{-,2}(z)$ can be analytically extended to $\mathbb{C}^+$ and continuously extended to $\mathbb{C}^+\cup (\mathbb{R}\setminus\{0\})$, whereas $\mu_{-,1}(z)$ and $\mu_{+,2}(z)$ can be analytically extended to $\mathbb{C}^-$ and continuously extended to $\mathbb{C}^-\cup (\mathbb{R}\setminus\{0\})$.

\item {Symmetry:} For $z\in\mathbb{R}\setminus\{0,\pm1\}$, $\mu_{\pm}(z)$ satisfies the symmetries
\begin{align}
\mu_{\pm}(z)
=
\sigma_{1}\overline{\mu_{\pm}(\overline{z})}\sigma_{1}
=\mp z^{-1} \mu_\pm\left(z^{-1}\right) \sigma_2.
\end{align}
These symmetries extend columnwise to the corresponding domains of analyticity.

\item {Asymptotics:} $\mu_{\pm}(z)$ possess the following asymptotic behaviour:
\begin{align}
\mu_{\pm}(z)&=I+\mathcal{O}\left(z^{-1}\right),\quad z\to\infty,
\\
\mu_{\pm}(z)&=\mp \frac{\sigma_2}{z}+\mathcal{O}(1),\quad z\to0.
\end{align}
\end{itemize}

\end{lemma}

\subsection{The scattering data}

Since $\Phi_{+}(z)$ and $\Phi_{-}(z)$ are two fundamental matrix solutions of the Lax pair (\ref{Hirota-Lax}), they are connected by
\begin{align}
    \Phi_{+}(z)=\Phi_{-}(z)S(z),
    \quad z\in\mathbb{R}\setminus\{0,\pm1\},
\end{align}
where the spectral matrix $S(z)$ takes the form
\begin{align}
    S(z)=
    \begin{pmatrix}
        a(z) & \overline{b(z)}\\
        b(z) & \overline{a(z)}
    \end{pmatrix}.
\end{align}
Here, $a(z)$ and $b(z)$ denote the scattering coefficients. The associated reflection coefficient is introduced as
\begin{align}
    r(z)=\frac{b(z)}{a(z)}.
\end{align}
The main properties of the scattering coefficients and the reflection coefficient are summarized in the following lemma.

\begin{lemma}\label{lemma-sd}
Let $q_0(x)-\tanh(x)\in L^{1,2}(\mathbb{R})$ and
$q_0'(x)\in W^{1,1}(\mathbb{R})$. Then
\begin{itemize}

\item The scattering coefficients can be expressed in terms of the
Jost functions as
\begin{align}
a(z)=\frac{\det\bigl(\Phi_{+,1}(z),\Phi_{-,2}(z)\bigr)}
{1-z^{-2}},\quad
b(z)=\frac{\det\bigl(\Phi_{-,1}(z),\Phi_{+,1}(z)\bigr)}
{1-z^{-2}}.
\end{align}

\item The function $a(z)$ can be analytically extended to
$\mathbb{C}^+$, whereas $b(z)$ and $r(z)$ are defined for $z\in\mathbb{R}\setminus\{0,\pm1\}$.

\item For each $z\in\mathbb{R}\setminus\{0,\pm1\}$, we have
\begin{align}
\det S(z)=|a(z)|^2-|b(z)|^2=1,
\quad
|r(z)|^2=1-|a(z)|^{-2}<1.
\end{align}

\item The functions $a(z)$, $b(z)$, and $r(z)$ satisfy the symmetries
\begin{align}
a(z)=-\overline{a\left(\overline z^{-1}\right)},\quad
b(z)=\overline{b\left(\overline z^{-1}\right)},\quad
r(z)=-\overline{r\left(\overline z^{-1}\right)}.
\end{align}
\item The scattering data have the following asymptotic behaviour:
\begin{align}
&a(z)=
1+\mathcal{O}\left(z^{-1}\right),
\quad z\to\infty,\quad z\in\overline{\mathbb{C}^{+}},\\
&a(z)=
-1+\mathcal{O}(z),
\quad z\to0,\quad z\in\overline{\mathbb{C}^{+}},\\
&b(z)=
\mathcal{O}\left(z^{-2}\right),
\quad z\to\infty,\quad z\in\mathbb{R},\\
&b(z)=
\mathcal{O}\left(z^2\right),
\quad z\to0,\quad z\in\mathbb{R}.
\end{align}

\item In the generic case, $a(z)$ and $b(z)$ have the same order singularities at $z=\pm1$,
\begin{equation}
a(z)=\frac{s_\pm}{z\mp1}+\mathcal{O}(1),\quad
b(z)=\mp\frac{i s_\pm}{z\mp1}+\mathcal{O}(1), \quad
\lim_{z\to\pm1}r(z)=\mp i,
\end{equation}
where
\begin{equation}
s_\pm=\pm\frac{1}{2}
\det\left[\Phi_{+,1}(\pm1),\Phi_{-,2}(\pm1)\right]
\neq0.
\end{equation}

\item In the non-generic case, $a(z)$ and $b(z)$  are continuous at $z=\pm1$, and $|r(\pm1)|<1$.

\end{itemize}
\end{lemma}

We extend $r(z)$ continuously to $\mathbb{R}$ by setting $r(0)=0$ and defining $r(\pm1)$ by the corresponding limits.

\begin{lemma}
	Suppose that $q_0(x)-\tanh{(x)}\in L^{1,2}(\mathbb{R})$ and $q'_0(x) \in W^{1,1}(\mathbb{R})$. Then $r(z)  \in H^{1}(\mathbb{R})$.
\end{lemma}

The zeros of $a(z)$ in $\mathbb C^+$ are simple and finite in number, and lie on the unit circle. Let $\eta_1,\eta_2,\ldots,\eta_N$ denote these zeros; namely,
\begin{equation*}
a(\eta_n)=0,\quad \eta_n\in\mathbb C^+\cap\{z\in\mathbb C:|z|=1\},\quad n=1,\ldots,N.
\end{equation*}
By the symmetries of the scattering data, the corresponding discrete spectral points in $\mathbb C^-$ are given by $\overline{\eta}_n$. We therefore define
\begin{align}
\mathcal{Z}^+ = \{ \eta_n \}_{n=1}^N \subset \mathbb{C}^+ , \quad \mathcal{Z}^- = \{ \overline{\eta}_n \}_{n=1}^N \subset \mathbb{C}^- ,
\end{align}
and collect the discrete spectrum as
\begin{align}
\mathcal{Z}=\mathcal{Z}^+ \cup\mathcal Z^- =\left\{\eta_n, \overline{\eta}_n\right\}_{n=1}^{N}.
\end{align}
At each zero $\eta_n$ of $a(z)$, the columns $\Phi_{+,1}(\eta_n)$ and $\Phi_{-,2}(\eta_n)$ are linearly dependent. Hence, there exists a nonzero constant $c(\eta_n)$, independent of $x$ and $t$, such that
\begin{align}
    \Phi_{+,1}(\eta_n)=c(\eta_n)\Phi_{-,2}(\eta_n),
    \quad
    \Phi_{+,2}(\overline{\eta}_n)
    =\overline{c(\eta_n)}\,\Phi_{-,1}(\overline{\eta}_n).
\end{align}
Moreover, $a(z)$ admits the following trace formula:
\begin{align}\label{equ-trace}
a(z) = \prod_{j=1}^{N} \frac{z - \eta_j}{z - \overline{\eta}_j} \exp\left(-i \int_{\mathbb{R}} \frac{\nu(\zeta)}{\zeta - z} \, d\zeta\right), \quad z\in\mathbb{C}^+
\end{align}
where
\begin{align}\label{eq:nu-def}
\nu(\zeta) = -\frac{1}{2\pi} \log\left(1 - |r(\zeta)|^2\right).
\end{align}

\subsection{A basic RH problem }

Based on the Jost solutions $\mu_{\pm,j}(z),\, j=1,2$, and the scattering relation, we define a sectionally meromorphic matrix as follows:
\begin{align}
M(z)=M(z;x,t):=
\begin{cases}
\left(
\dfrac{\mu_{+,1}(z)}{a(z)},
\mu_{-,2}(z)
\right),
& z\in\mathbb{C}^{+},
\\
\left(
\mu_{-,1}(z),
\dfrac{\mu_{+,2}(z)}{\overline{a(\overline z)}}
\right),
& z\in\mathbb{C}^{-},
\end{cases}
\end{align}
which solves the following RH problem.
\begin{prob}\label{RH1}
Find a matrix-valued function $M(z)=M(z;x,t)$ which satisfies
\begin{itemize}
\item  Analyticity: $M(z)$  is analytic in $\mathbb{C}\setminus (\mathbb{R}\cup\mathcal{Z})$ and has simple poles at the points in $\mathcal{Z}$.
\item  Jump condition: $M(z)$ satisfies the jump condition
	$$M_+(z)=M_-(z)V(z), \quad z\in\mathbb{R}\setminus\{0\},$$
      	where
      	\begin{equation}\label{V1}
      		V(z)=\left(\begin{array}{cc}
      			1-|r(z)|^2 & -\overline{r(z)}e^{2it \theta(z)}\\
      		r(z)	e^{-2it \theta(z)} & 1
      		\end{array}\right),
      	\end{equation}
      with the phase function $\theta(z) =\lambda(z)\left(\frac{x}{t}+4k^2(z)-2k(z)+2\right).$

 \item  Asymptotic behaviors:
\begin{align}
M(z)&=I+\mathcal{O}\left(z^{-1}\right),
\quad z\to\infty,
\\
M(z)&=
\frac{\sigma_{2}}{z}+\mathcal{O}(1),
\quad z\to0.
\end{align}
\item  Residue conditions:
\begin{align}
&\underset{z=\eta_n}{\operatorname{Res}}M(z)=
\lim_{z\to \eta_n}
M(z)
\begin{pmatrix}
0 & 0 \\
c_n\mathrm{e}^{-2it\theta(\eta_n)} & 0
\end{pmatrix},
\\
&\underset{z=\overline{\eta}_n}{\operatorname{Res}}M(z)=
\lim_{z\to\overline{\eta}_n}
M(z)
\begin{pmatrix}
0 & \overline{c}_n\mathrm{e}^{2it\theta(\overline{\eta}_n)} \\
0 & 0
\end{pmatrix},
\end{align}
where $c_n=\frac{b(\eta_n)}{a'(\eta_n)}$.

\end{itemize}
\end{prob}

The potential $q(x,t)$ can be recovered from $M(z)$ through the reconstruction formula
\begin{align}\label{recon-sol}
q(x,t)=-\lim_{z\to\infty}i\bigl(zM(z)\bigr)_{12}.
\end{align}
The symmetries established in Lemmas \ref{lemma-mu} and \ref{lemma-sd}, together with the uniqueness of the solution to RH problem \ref{RH1}, imply that
\begin{align}\label{symmetry-M}
M(z)=
\sigma_{1}\overline{M(\overline{z})}\sigma_{1}=
z^{-1}M\left(z^{-1}\right)\sigma_{2}.
\end{align}

\subsection{Classification of asymptotic regions}

\begin{figure}[H]
  \centering

  \subfigure[$\xi<-8$]{
    \label{fig3-a}
    \includegraphics[width=0.225\linewidth]{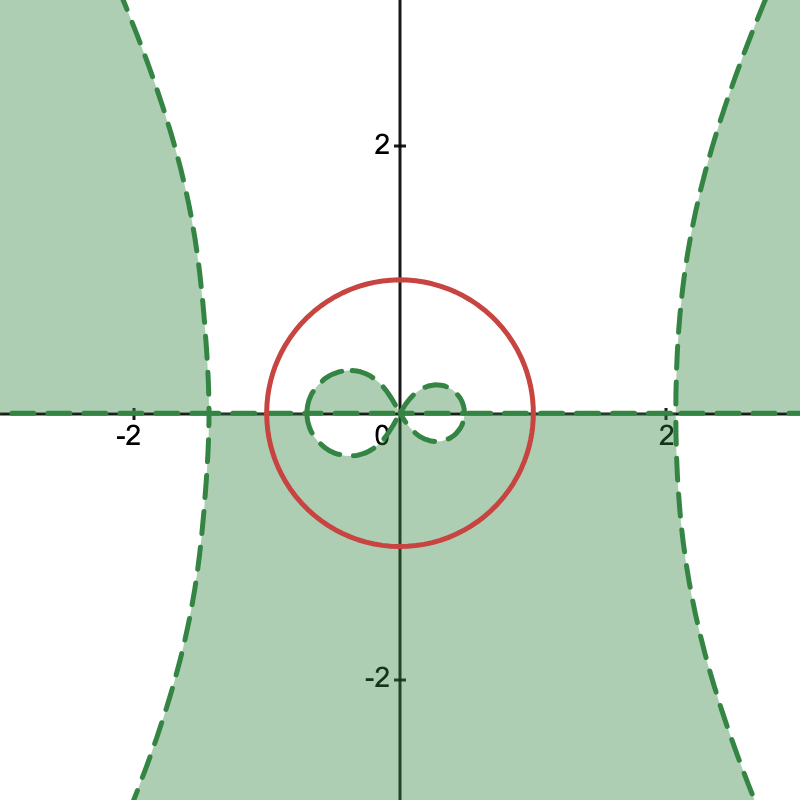}
  }\hfill
  \subfigure[$\xi=-8$]{
    \label{fig3-b}
    \includegraphics[width=0.225\linewidth]{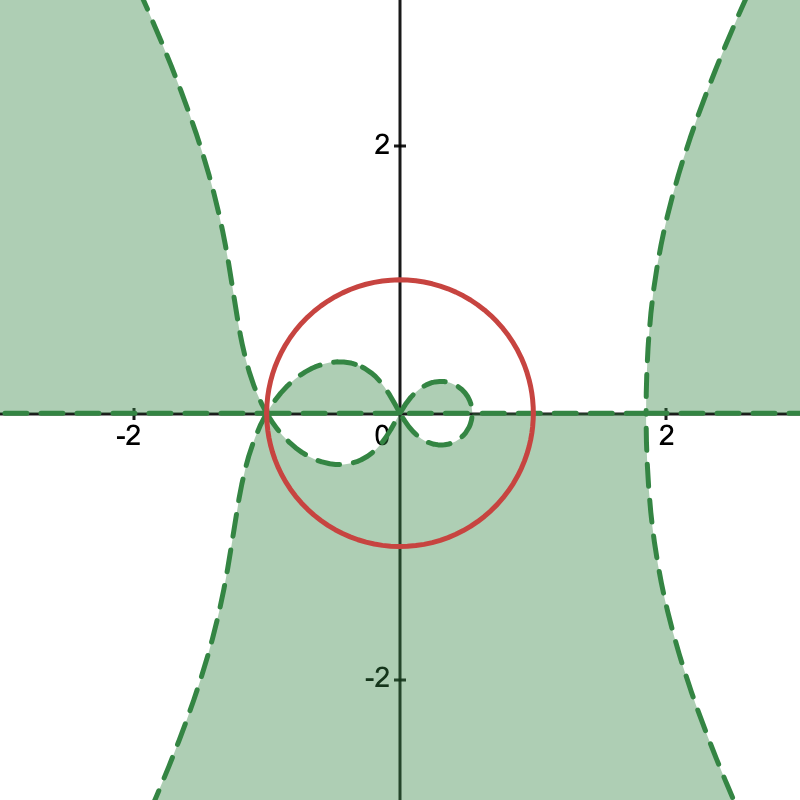}
  }\hfill
  \subfigure[$-8<\xi<-4$]{
    \label{fig3-c}
    \includegraphics[width=0.225\linewidth]{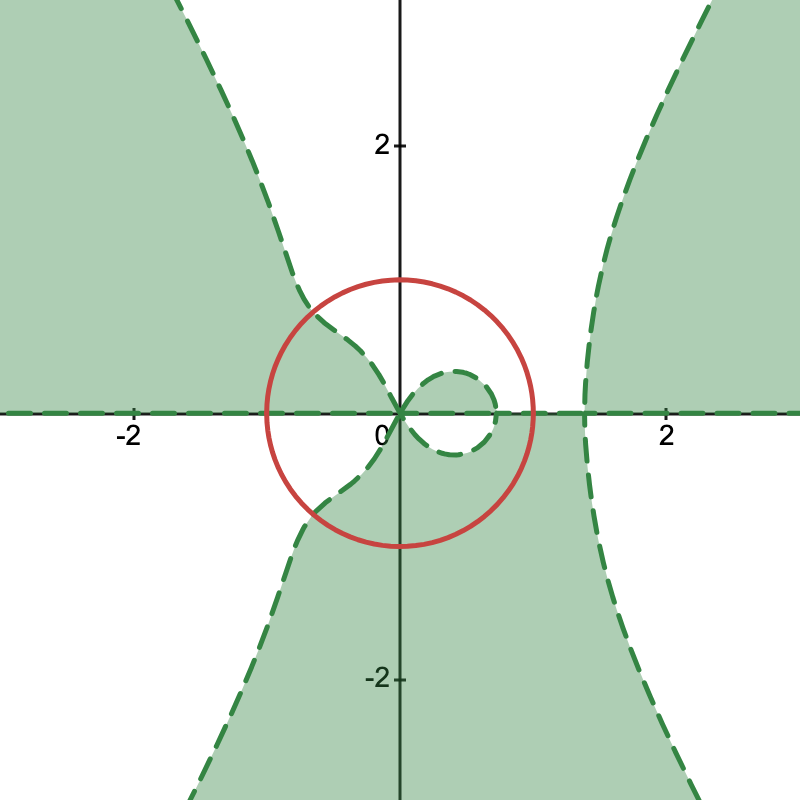}
  }\hfill
  \subfigure[$\xi=-4$]{
    \label{fig3-d}
    \includegraphics[width=0.225\linewidth]{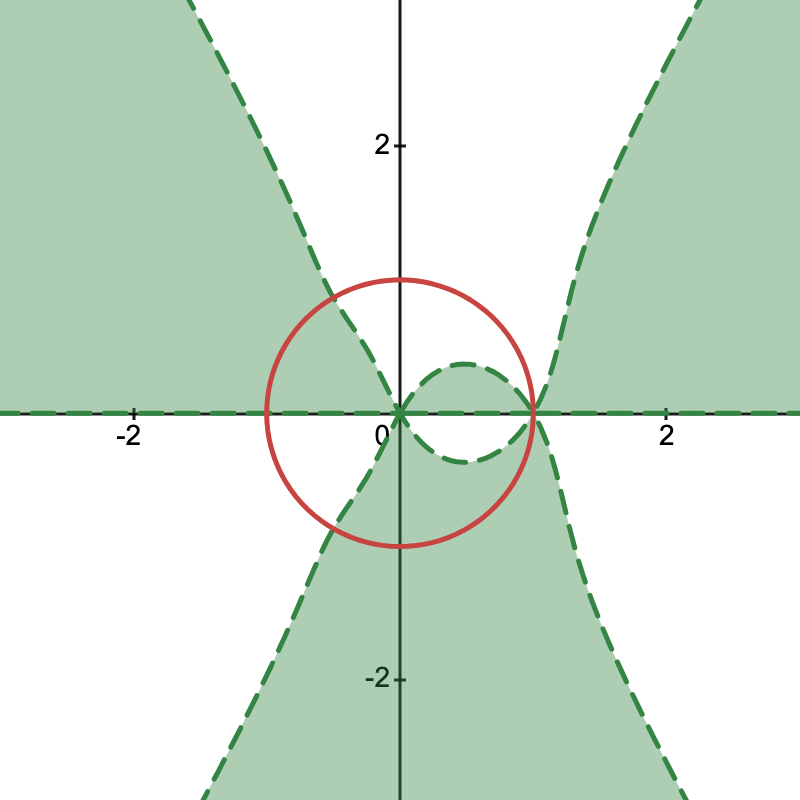}
  }

  \vspace{0.3cm}

  \subfigure[$-4<\xi<-\frac{7}{4}$]{
    \label{fig3-e}
    \includegraphics[width=0.225\linewidth]{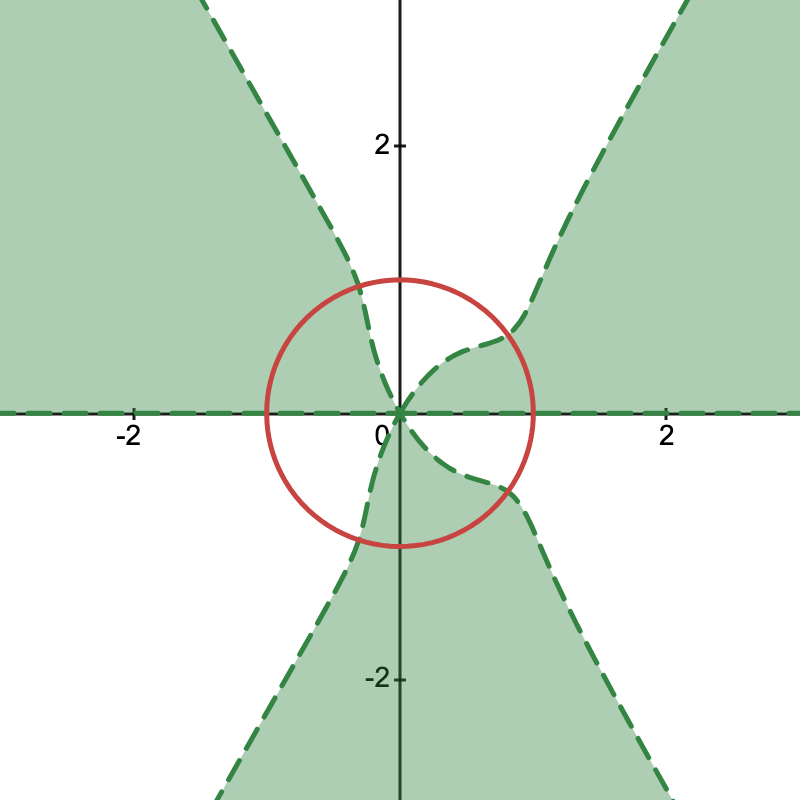}
  }
  \hspace{0.025\linewidth}
  \subfigure[$\xi=-\frac{7}{4}$]{
    \label{fig3-f}
    \includegraphics[width=0.225\linewidth]{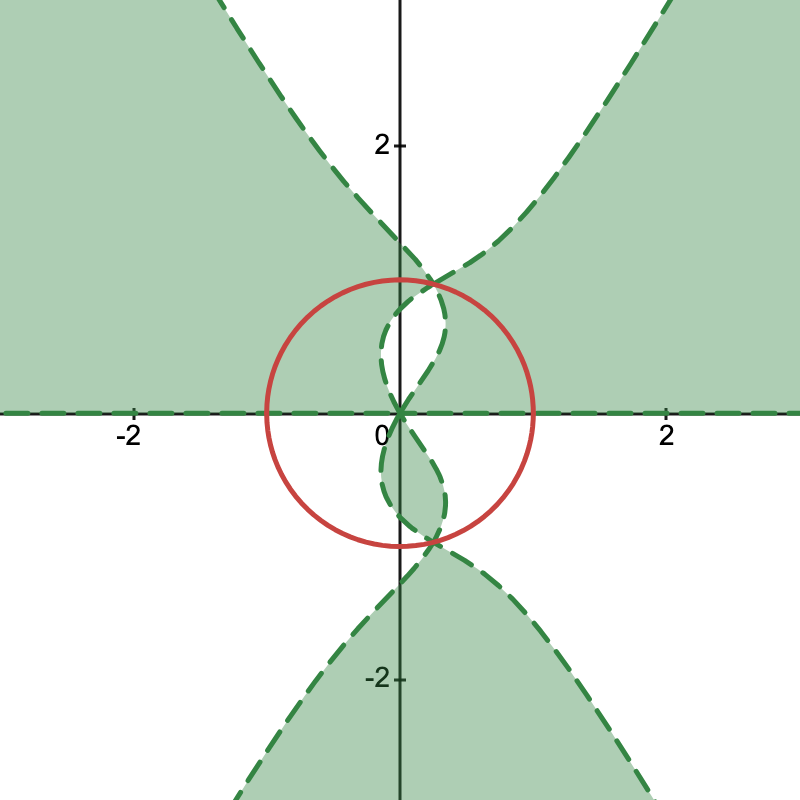}
  }
  \hspace{0.025\linewidth}
  \subfigure[$\xi>-\frac{7}{4}$]{
    \label{fig3-g}
    \includegraphics[width=0.225\linewidth]{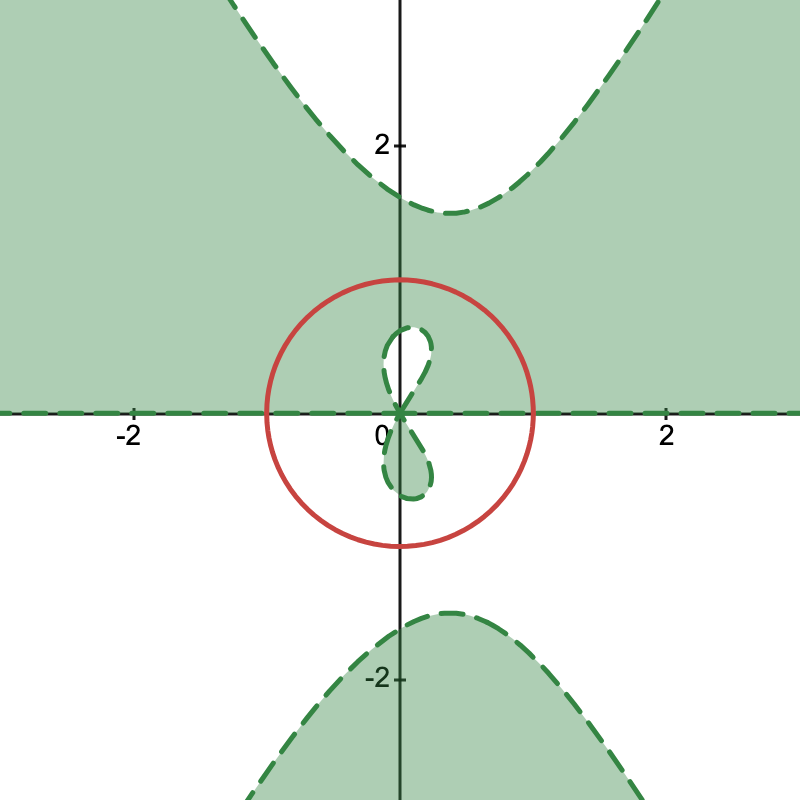}
  }

  \caption{\footnotesize
  The distribution of saddle points and the signature table of $\operatorname{Re}(2i\theta(z))$, where $\operatorname{Re}(2i\theta(z))<0$ in the green regions and $\operatorname{Re}(2i\theta(z))>0$ in the white regions. Figure (a):  There are four saddle points on $\mathbb{R}$ for the solitonless  region; Figure (c):  There are two saddle points on $\mathbb{R}$ for the Zakharov--Manakov region; Figure (e): There are four phase points on the unit circle for the solitonic region; Figure (g): There are no real saddle points for the solitonless region; Figures (b), (d) and (f) are three critical cases.
  }
  \label{proptheta}
\end{figure}

The jump matrix $V(z)$ in \eqref{V1} has the following two factorizations:
       \begin{equation}\label{V_factorization}
         V(z)=        \begin{cases}
           \left(\begin{array}{cc} 1&-\overline{r(z)}e^{2it\theta(z) }\\ 0&1\end{array}  \right)
           \left(\begin{array}{cc} 1&0\\   r(z)e^{-2it\theta(z) }&1\end{array}  \right),\\
           \left(\begin{array}{cc} 1 & 0 \\ \frac{r(z)}{1-|r(z)|^2}e^{-2it\theta(z) } & 1\end{array}  \right)
           \left(1-|r(z)|^2 \right)^{\sigma_3} \left(\begin{array}{cc} 1 &
             \frac{-\overline{r(z)}}{1-|r(z)|^2}e^{2it\theta(z) } \\ 0 & 1\end{array}  \right).
         \end{cases}
       \end{equation}
The exponential factors $\mathrm{e}^{\pm 2it\theta(z)}$  play a central role in the long-time analysis of RH problem \ref{RH1}. The phase function is given by
\begin{align}
\theta(z)=\frac12\left(z-\frac1z\right)
\left[\frac{x}{t}+\left(z+\frac1z\right)^2-\left(z+\frac1z\right)+2\right].
\end{align}
Setting $\xi=x/t$ and $z=u+iv$, a direct calculation yields
\begin{align}
    \operatorname{Re}\bigl(2i\theta(z)\bigr)
    =
    v\left(
        v^2-3u^2+2u-\xi-3
        -\frac{4u^2}{(u^2+v^2)^3}
        +\frac{2u+1}{(u^2+v^2)^2}
        -\frac{\xi+3}{u^2+v^2}
    \right),
\end{align}
the signature table of which is shown in Fig. \ref{proptheta}.

The asymptotic regions are determined by the stationary points on the real jump contour and the exponential factors associated with the discrete spectrum on the unit circle. We first examine the real stationary points of the phase function.

The stationary equation is
\begin{align}
    \theta'(z)=\frac{1}{2z}\left[3\omega^3-2\omega^2-(6-\xi)\omega+4\right]=0,
    \quad
    \omega=z+z^{-1}.
\end{align}
Accordingly, we introduce the cubic equation
\begin{align}\label{equ-xi}
    P(\omega;\xi):=3\omega^3-2\omega^2-(6-\xi)\omega+4=0.
\end{align}
For $z \in \mathbb{R}\setminus\{0\}$, the variable $\omega$ belongs to $(-\infty,-2]\cup[2,\infty)$, and the stationary equation can be written as
\begin{align*}
    \xi = -3\omega^2 + 2\omega + 6 - \frac{4}{\omega}.
\end{align*}
The right-hand side increases strictly from $-\infty$ to $-8$ on $(-\infty,-2]$ and decreases strictly from $-4$ to $-\infty$ on $[2,\infty)$. The relation $z^2-\omega z+1=0$ gives two distinct real stationary points for $|\omega|>2$, which coalesce at $z=-1$ when $\omega=-2$ and at $z=1$ when $\omega=2$. Thus, $\xi=-8$ and $\xi=-4$ are the critical values at which pairs of real stationary points coalesce.

For $\xi\leq -4$, let $\omega_1>\omega_2>\omega_3$ denote the three distinct real roots of $P(\omega;\xi)=0$. When $\xi<-8$, we have $\omega_1>2$ and $\omega_3<-2$, giving four real stationary points:
\begin{align}\label{eq:positive-stationary-points}
    z_1=
    \frac{\omega_1+\sqrt{\omega_1^2-4}}{2},\quad
    z_2=
    \frac{\omega_1-\sqrt{\omega_1^2-4}}{2},\\
    z_3=
    \frac{\omega_3+\sqrt{\omega_3^2-4}}{2},\quad
    z_4=
    \frac{\omega_3-\sqrt{\omega_3^2-4}}{2}.
\end{align}
They satisfy
\begin{align*}
    z_4<-1<z_3<0<z_2<1<z_1,
    \quad z_1z_2=z_3z_4=1.
\end{align*}

Since the discrete spectrum lies on the unit circle, we next examine the intersections of the critical line $\mathcal{L}$ with the unit circle $\mathcal{C}$, where
\begin{align*}
    \mathcal{L}
    &:= \{z\in\mathbb{C}\setminus\{0\}:
   \operatorname{Re}(2i\theta(z))=0\},
  \quad
   \mathcal{C}:=\{z\in\mathbb{C}:|z|=1\}.
\end{align*}
Combining the distribution of real stationary points with the intersection pattern of $\mathcal{L}$ and $\mathcal{C}$, we distinguish the following asymptotic regions.

\begin{itemize}

\item \textbf{Solitonless region:} For $\xi<-8$ or $\xi>-\frac{7}{4}$, there is no nontrivial intersection between $\mathcal{L}$ and $\mathcal{C}$. These two cases correspond to two distinct solitonless regions, as depicted in Fig. \ref{proptheta} (a) and Fig. \ref{proptheta} (g).

\item \textbf{Zakharov--Manakov region:} For $-8<\xi<-4$, the phase function has two real stationary points, and the critical line $\mathcal{L}$ intersects the unit circle $\mathcal{C}$ at two nontrivial points, as depicted in Fig. \ref{proptheta}(c).

\item \textbf{Solitonic region:} For $-4<\xi\leq-\frac{7}{4}$, the phase function has no real stationary points. In the interior of this interval, $\mathcal{L}$ intersects $\mathcal{C}$ at four nontrivial points, as depicted in Fig. \ref{proptheta}(e). At $\xi=-\frac{7}{4}$, these four intersection points merge pairwise into the two points $z=\frac{1\pm i\sqrt{15}}{4}$, as depicted in Fig. \ref{proptheta}(f).

\item \textbf{Transition region:} At $\xi=-8$ and $\xi=-4$, a pair of saddle points coalesces at $z=-1$ and $z=1$, respectively. Moreover, for generic initial data, we have $\lim_{z\to\pm1}|r(z)|=1$, and consequently $\left(1-|r(z)|^2\right)^{-1}\to\infty$, as $z\to\pm1$. This indicates the emergence of new asymptotic phenomena in the transition regions $\xi\approx-8$ and $\xi\approx-4$, as illustrated in Fig. \ref{proptheta} (b) and Fig. \ref{proptheta} (d).

\end{itemize}

Note that at $\xi=-\frac{7}{4}$, the nontrivial intersections of $\mathcal{L}$ and $\mathcal{C}$ merge pairwise at $z=\frac{1\pm i\sqrt{15}}{4}$, away from the real jump contour. Since the real jump contour remains free of stationary points as $\xi$ passes through this value, $\xi=-\frac{7}{4}$ does not generate an additional transition region associated with the continuous spectrum.

\section{Painlev\'e Asymptotics in the Transition Region $\mathcal{P}_{-4}$}\label{sec-P-4}

In this section, we consider the asymptotics in the region $\mathcal{P}_{<-4}$, as the discussion for the other half region is similar. The region $\mathcal{P}_{<-4}$ is given by
\begin{align}
\mathcal{P}_{<-4}:=\left\{
(x,t)\in\mathbb{R}\times\mathbb{R}^{+}:
-C<\left(\frac{x}{t}+4\right)t^{2/3}<0
\right\},
\end{align}
where $C>0$ is a constant, which corresponds to the case in Fig. \ref{proptheta} (c). In this region, two real stationary points $z_1$ and $z_2$ defined by \eqref{eq:positive-stationary-points} approach $1$ at least at the speed of $t^{-1/3}$ as $t\to+\infty$. We first make some modifications to the basic RH problem.

\subsection{Regularization of the basic RH problem}\label{sec3.1}

We first apply a diagonal conjugation to the basic RH problem \ref{RH1}, retaining its poles, and then remove the singularity at $z=0$.

{\bf Step 1: Diagonal transformation.}
The discrete spectrum $\eta_j \in \mathcal{Z}^+$ and $\overline{\eta}_j \in \mathcal{Z}^-$ are finite and located on the unit circle. Since $\eta_j^{-1}=\overline{\eta}_j$, we have
\begin{align}
 {\rm Re}( i\theta(\eta_j;\xi))&= \operatorname{Im}\eta_j(\xi-v_j),\label{opew}
\end{align}
where $v_j$ is the soliton velocity associated with $\eta_j$, given by
\begin{align}
v_j=-4(\operatorname{Re}\eta_j )^2+2\operatorname{Re}\eta_j-2. \label{323}
\end{align}
For $\xi=-4 $,  it follows from (\ref{opew}) and (\ref{323}) that two critical velocities  are given by
$${\rm Re}  \eta_j =-1/2, \ \  {\rm Re} \eta_j =1. $$
We  define
\begin{align}\label{eq:Delta-minus4}
\Delta=\{\eta_j : {\rm Re}  \eta_j<-1/2 \}, \ \ \  \Lambda=\{\eta_j, \bar \eta_j: |v_j+4|\leq Ct^{-2/3}\}.
\end{align}

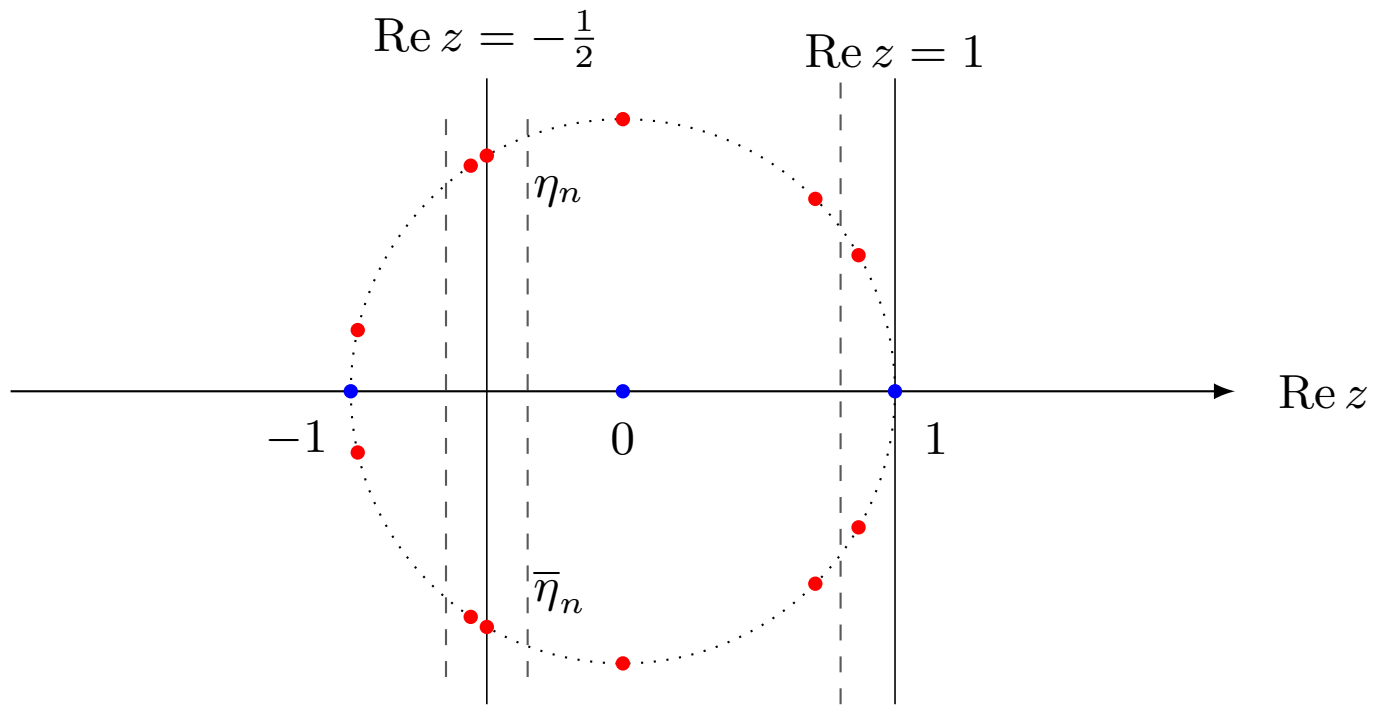
\begin{figure}[htbp]
\centering
\begin{tikzpicture}[scale=0.9]

\draw[dotted] (0,0) circle[radius=2];

\def\bandDelta{0.65}
\pgfmathsetmacro{\bandLeft}{(1-sqrt(9+4*\bandDelta))/2}
\pgfmathsetmacro{\bandRight}{(1-sqrt(9-4*\bandDelta))/2}

\draw[dashed,black!65]  (-1.3,-2.1)--(-1.3,2.1);
\draw[dashed,black!65] (-0.7,-2.1)--(-0.7,2.1);
\draw[line width=0.3pt] (-1,-2.3)--(-1,2.3)
    node[above,inner sep=2pt] {\footnotesize $\operatorname{Re}z=-\tfrac12$};

\draw[dashed,black!65] (1.6, -2.3)--(1.6,2.3);
\draw[line width=0.3pt] (2,-2.3)--(2,2.3) node[above,inner sep=2pt] {\footnotesize $\operatorname{Re}z=1$};
\draw[-latex] (-4.5,0)--(4.5,0);
\node at (5.15,0) {\footnotesize $\operatorname{Re} z$};

\node[anchor=west,inner sep=1pt] at ({\bandRight+0.06},1.47) {\footnotesize $\eta_n$};
\node[anchor=west,inner sep=1pt] at ({\bandRight+0.06},-1.47) {\footnotesize $\overline{\eta}_n$};

\node at (-2.4,-0.35) {\footnotesize $-1$};
\node at (0,-0.35) {\footnotesize $0$};
\node at (2.3,-0.35) {\footnotesize $1$};

\coordinate (A) at (30:2);
\coordinate (C) at (45:2);
\coordinate (E) at (90:2);
\coordinate (G) at (120:2);
\coordinate (I) at (167:2);

\coordinate (B) at (-30:2);
\coordinate (D) at (-45:2);
\coordinate (F) at (-90:2);
\coordinate (H) at (-120:2);
\coordinate (J) at (-167:2);

\coordinate (K) at (0,0);
\coordinate (L) at (2,0);
\coordinate (M) at (-2,0);

\foreach \point in {A,B,C,D,E,F,G,H,I,J}
    \fill[red] (\point) circle (1.5pt);

\foreach \angle in {124,-124}
    \fill[red] (\angle:2) circle (1.5pt);

\foreach \point in {K,L,M}
    \fill[blue] (\point) circle (1.5pt);

\end{tikzpicture}
\caption{The distribution of the discrete spectrum (red dots) on the unit circle $\{z:|z|=1\}$ in the $z$-plane. The dashed lines delimit the strip near $\operatorname{Re}z=-1/2$ and
$\operatorname{Re}z=1$  corresponding to $\mathcal{P}_{-4}$.}
\label{dis-spectrum}
\end{figure}

We define
\begin{align}\label{T-def}
T(z) = \prod_{j \in \Delta} \frac{z - \eta_j}{z \eta_j - 1} \exp \left\{ -i \int_{\mathbb{R}} \nu(\zeta) \left( \frac{1}{\zeta - z} - \frac{1}{2\zeta} \right) d\zeta \right\}.
\end{align}
Then the following lemma holds.

\begin{lemma}
	The function $T(z)$ possesses the following properties.
\begin{itemize}
\item $T(z)$ is meromorphic in $\mathbb{C} \setminus \mathbb{R}$, with a simple zero at each $\eta_j \in \Delta$ and a simple pole at each $\overline{\eta}_j$.
\item $\overline{T(\overline{z})} = T(z)^{-1} = T(z^{-1})$.
\item $T(z)$ satisfies the jump condition
\begin{align}
T_{-}(z) = T_{+}(z)(1 - |r(z)|^2), \quad z\in\mathbb{R}.
\end{align}
\item Let
\begin{align}
T(\infty) := \lim_{z \to \infty} T(z) =\left( \prod_{j \in \Delta } \overline{\eta}_j \right) \exp \left\{ \frac{i}{2} \int_{\mathbb{R}} \frac{\nu(\zeta)}{\zeta} \, d\zeta \right\}.
\end{align}
Then, $|T(\infty)|^2 = 1$ and the asymptotic expansion at infinity is
\begin{align}\label{T-infty}
T(z) = T(\infty) \left[ 1 - \frac{i}{z} \left( 2 \sum_{j \in \Delta } \operatorname{Im} \eta_j - \int_{\mathbb{R}} \nu(\zeta) \, d\zeta \right) + \mathcal{O}(z^{-2}) \right].
\end{align}
\item The ratio $a(z)/T(z)$ is holomorphic and bounded in $\mathbb{C}^+$. Moreover, $a(z)/T(z)$ extends continuously to $\mathbb{R}$, and its boundary values satisfy $\left|\frac{a(z)}{T_+(z)}\right|=1, z\in\mathbb{R}$, where $T_+$ denotes the boundary value of $T$ taken from $\mathbb{C}^+$.
\end{itemize}
\end{lemma}


	In the transition region $\mathcal{P}_{<-4}$, the two stationary points $z_1$ and $z_2$ approach $z = 1$ at the rate $\mathcal{O}(t^{-1/3})$ as $t \to +\infty$. To establish a connection between the RH problem   and the Painlev\'e II model problem in \ref{appx} through a sequence of transformations, we retain the jump on the interval $(z_2, z_1)$ and do not open this part of the contour.
Choose $\rho>0$ such that
\begin{align}\label{definerho}
0<\rho<\frac12\min\{
\min_{\substack{\eta_j,\eta_l\in\mathcal{Z}^{+}\\j\ne l}}|\eta_j-\eta_l|,
\min_{\eta_j\in\mathcal{Z}^{+}}|\operatorname{Im}\eta_j|
\}.
\end{align}
For $\eta_j\in\mathcal{Z}^{+}$ and $\overline{\eta}_j \in\mathcal{Z}^{-}$, we take small circles $C_j$ and $\overline{C}_j$ centered at $\eta_j$ and $\overline{\eta}_j$, respectively, with a radius of $\rho$. The corresponding disks $D_j$ and $\overline{D}_j$ are pairwise disjoint and avoid $\mathbb{R}$. Moreover, the disk centered at $1$ with radius $2\rho$ contains no discrete spectrum.

For later use, we introduce the notation $\Gamma = (-\infty,z_2) \cup (z_1,\infty)$. Considering the second factorization of the jump matrix $V(z)$ in \eqref{V_factorization}, denote the following matrices
\begin{align}
V(z)=\left(\begin{array}{cc}
		1 & 0\\
		\frac{ r(z) }{1-|r|^2}  T_-(z)^{2}  e^{ -2it\theta(z)} & 1
	\end{array}\right)\left(\begin{array}{cc}
	1 & -\frac{ \overline{r(z)}}{1-|r|^2}T_+(z)^{-2} e^{2it\theta(z)}\\
	0 & 1
\end{array}\right):=B_-^{-1}B_+,
\end{align}
and make the following transformation
\begin{equation}\label{tran01}
        M^{(1)}(z)=T(\infty)^{-\sigma_3} M(z) T(z)^{\sigma_3},
\end{equation}
Then $M^{(1)}(z)$ satisfies the RH problem as follows.

\begin{prob}\label{RH2}
Find a function $M^{(1)}(z)=M^{(1)}(z;x,t)$ which satisfies
\begin{itemize}
\item $M^{(1)}$ is meromorphic in $\mathbb{C}\setminus\mathbb{R}$, with simple poles in $\mathcal{Z}$.
\item Jump condition:
\begin{align}
M_{+}^{(1)}(z)=M_{-}^{(1)}(z)V^{(1)}(z), \quad z\in \mathbb{R},\nonumber
\end{align}
where
\begin{align}\label{jumpV1}
V^{(1)}(z)=
\begin{cases}
B_-^{-1}B_+,
& z\in \mathbb{R}\setminus (z_1,z_2),
\\[2mm]
T_-(z)^{-\sigma_3}V(z)T_+(z)^{\sigma_3},
& z\in(z_2,z_1).
\end{cases}
\end{align}

\item Residue conditions: $M^{(1)}(z)$ has simple poles at $\eta_j\in\mathcal{Z}^{+}$ and $\overline{\eta}_j\in\mathcal{Z}^{-}$ satisfying
\begin{align}
\operatorname*{Res}_{z=\eta_j}M^{(1)}(z)
&=\lim_{z\to\eta_j}M^{(1)}(z)C_j^T,\\
\operatorname*{Res}_{z=\overline{\eta}_j}M^{(1)}(z)
&=\lim_{z\to\overline{\eta}_j}M^{(1)}(z)\sigma_1\overline{C_j^T}\sigma_1,
\end{align}
where
\begin{align}
C_j^T=
\begin{cases}
\begin{pmatrix}
0 & 0\\
c_jT(\eta_j)^2e^{-2it\theta(\eta_j)} & 0
\end{pmatrix},
& j\notin\Delta,\\[16pt]
\begin{pmatrix}
0 & \dfrac{e^{2it\theta(\eta_j)}}{c_jT'(\eta_j)^2}\\
0 & 0
\end{pmatrix},
& j\in\Delta.
\end{cases}
\end{align}

\item  Asymptotic behaviors:
\begin{align}
M^{(1)}(z)&=I+\mathcal{O}\left(z^{-1}\right),
\quad z\to\infty,
\\
M^{(1)}(z)&=
\frac{\sigma_{2}}{z}
+
\mathcal{O}(1),
\quad z\to0.
\end{align}
\end{itemize}
\end{prob}

{\bf Step 2: Removing singularities.} To remove the singularity at $z=0$, we introduce the following transformation
\begin{align}\label{tran23}
M^{(1)}(z)=\left( I+ \frac{\sigma_2}{z} M^{(2)}(0)^{-1} \right ) M^{(2)}(z),
\end{align}
then $M^{(1)}(z)$ satisfies the  RH problem  \ref{RH2}  if $M^{(2)}(z)$ satisfies the following RH problem.

\begin{prob} \label{RH4}
Find  $M^{(2)}(z)=M^{(2)}(z;x,t)$ with properties
\begin{itemize}
\item  $M^{(2)}(z)$ is meromorphic in $\mathbb{C}\setminus\mathbb{R}$, with simple poles in $\mathcal{Z}$.
\item Jump condition:
            $M^{(2)}_+(z)=M^{(2)}_-(z)V^{(1)}(z),$ where $V^{(1)}(z)$ is given by   \eqref{jumpV1}.	
\item Residue conditions: $M^{(2)}(z)$ satisfies the same residue conditions as $M^{(1)}(z)$ in RH problem \ref{RH2}.
\item Asymptotic behaviors: $  M^{(2)}(z)=I+\mathcal{O}(z^{-1}),	\quad  z \to  \infty.$
\end{itemize}
\end{prob}

\begin{proof}
The proof is analogous to that of RH Problem 3.3 in \cite{WangFan2023} and is therefore omitted.	
\end{proof}

\subsection{Transformation to a hybrid $\bar{\partial}$-RH problem}\label{sec-hybrid}

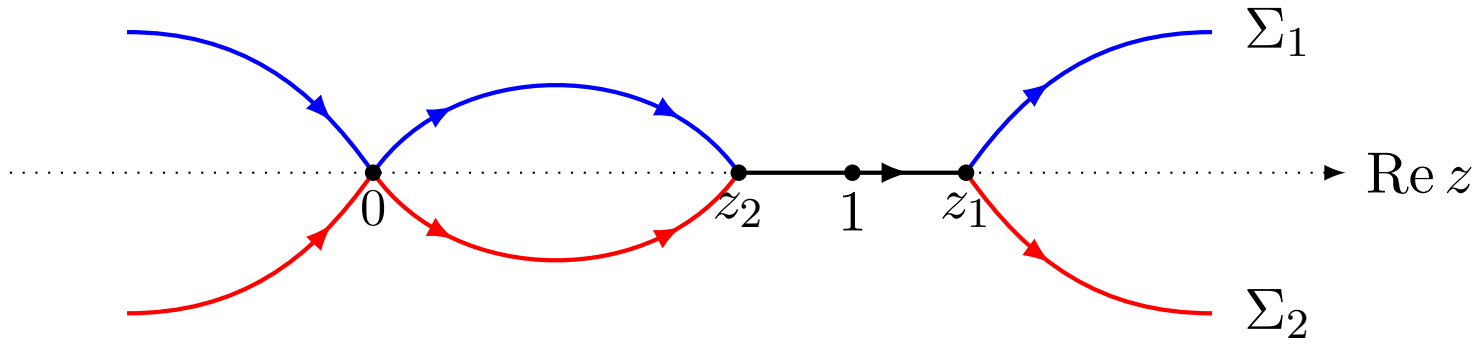
\begin{figure}[htbp]
    \centering
    \begin{tikzpicture}[
        scale=1.55,
        onearrow/.style={
            postaction={decorate},
            decoration={
                markings,
                mark=at position #1 with {\arrow{latex}}
            }
        },
        twoarrows/.style={
            postaction={decorate},
            decoration={
                markings,
                mark=at position 0.22 with {
                    \pgftransformxshift{2.5pt}
                    \arrow{Latex[length=5pt,width=4pt]}
                },
                mark=at position 0.78 with {
                    \pgftransformxshift{2.5pt}
                    \arrow{Latex[length=5pt,width=4pt]}
                }
            }
        }
    ]

        \coordinate (O)  at (0.8,0);
        \coordinate (Z2) at (2.36,0);
        \coordinate (E)  at (2.845,0);
        \coordinate (Z1) at (3.33,0);

        \draw[-latex,dotted]
            (-0.75,0)--(4.95,0)
            node[right] {$\operatorname{Re}z$};

        \draw[blue,thick,onearrow=0.78]
            (-0.25,0.6)
            to[out=0,in=125] (O);

        \draw[blue,thick,twoarrows]
            (O)
            to[out=55,in=125] (Z2);

        \draw[blue,thick,onearrow=0.42]
            (Z1)
            to[out=55,in=180] (4.38,0.6);

        \draw[red,thick,onearrow=0.78]
            (-0.25,-0.6)
            to[out=0,in=-125] (O);

        \draw[red,thick,twoarrows]
            (O)
            to[out=-55,in=-125] (Z2);

        \draw[red,thick,onearrow=0.42]
            (Z1)
            to[out=-55,in=180] (4.38,-0.6);

        \draw[thick,onearrow=0.75]
            (Z2)--(Z1);

        \node[right] at (4.43,0.6) {$\Sigma_1$};
        \node[right] at (4.43,-0.6) {$\Sigma_2$};

        \fill (Z2) circle (1pt)
            node[below] {$z_2$};

        \fill (E) circle (1pt)
            node[below] {$1$};

        \fill (Z1) circle (1pt)
            node[below] {$z_1$};

        \fill (O) circle (1pt)
            node[below,scale=0.9] {$0$};

    \end{tikzpicture}
    \caption{Opening the jump contour $\mathbb{R}\setminus[z_2,z_1]$ with a sufficiently small angle $\phi$. The opened contours $\Sigma_1$ ($\textcolor{blue}{\bullet}$) and $\Sigma_2$ ($\textcolor{red}{\bullet}$) lie in the green and white decay regions shown in Figure \ref{proptheta}(c), respectively.}
    \label{open-jump}
\end{figure}

In this subsection, we construct suitable continuous $\bar\partial$-extensions of the triangular factors of the jump matrix $V^{(1)}(z)$ on $\Gamma=\mathbb{R}\setminus[z_2,z_1]$, thereby transforming RH Problem \ref{RH4} into a hybrid $\bar\partial$-RH problem. To this end, we open lenses along $\Gamma$ by introducing the contours $\Sigma_1\subset\mathbb{C}^+$ and $\Sigma_2=\overline{\Sigma}_1\subset\mathbb{C}^-$, as illustrated in Fig. \ref{open-jump}.  For $j=1,2$, let $\Omega_j$ denote the union of the open regions between $\Sigma_j$ and the corresponding intervals of the real axis. In particular, $\Omega_2=\overline{\Omega}_1$. Choose a sufficiently small angle $\phi$ and the lens contours satisfying the following conditions:
\begin{enumerate}
    \item
    The regions $\Omega_1$ and $\Omega_2$ do not intersect the critical line $\mathcal{L}$ and satisfy
    \begin{align*}
        \operatorname{Re}(2i\theta(z))
        &<0, \quad z\in\Omega_1, \\
        \operatorname{Re}(2i\theta(z))
        &>0, \quad z\in\Omega_2.
    \end{align*}

    \item The closures of $\Omega_1$ and $\Omega_2$ stay away from the disks $D_n$ and $\overline{D}_n$, $n=1,\ldots,N$.

    \item
    $0<\phi<\pi/12<\pi/4$.
\end{enumerate}
Near the attachment points $0$, $z_1$, and $z_2$, the lens contours lie within sectors of opening angle $\phi$ measured from the adjacent real-axis intervals. We further require the lens regions to satisfy
\begin{align}\label{eq:lens-sector-bound}
    |\operatorname{Im}z|\leq\tan\phi\,\operatorname{dist}\bigl(\operatorname{Re}z,\{0,z_1,z_2\}\bigr), \quad z\in\Omega_1\cup\Omega_2.
\end{align}

We introduce the notation
\begin{align}
    \Omega=\Omega_1\cup\Omega_2, \quad
    \Sigma^{(3)} =\Sigma_1\cup\Sigma_2\cup[z_2,z_1].\nonumber
\end{align}
The jump on $[z_2,z_1]$ is retained. We first establish estimates for $\operatorname{Re}(2i\theta(z))$ in the lens regions, with particular attention to the origin and the stationary points $z_j$, $j=1,2$.

\begin{proposition}\label{prop-Re}
Let $(x,t)\in\mathcal{P}_{<-4}$ and fix
$0<\phi<\pi/12$.
Suppose that the lens regions satisfy
\eqref{eq:lens-sector-bound}.
Then, for all sufficiently large $t$, the following
estimates hold on the corresponding lens portions.

\begin{itemize}
    \item {\rm Case I.} $($near $z=0$$)$
    \begin{align}
        \operatorname{Re}(2i\theta(z))
        &\lesssim -|\operatorname{Im}z|,
        \quad z\in\Omega_1,\\
        \operatorname{Re}(2i\theta(z))
        &\gtrsim |\operatorname{Im}z|,
        \quad z\in\Omega_2.
    \end{align}

    \item {\rm Case II.} $($near $z=z_j$, $j=1,2$$)$
    \begin{align}
        \operatorname{Re}(2i\theta(z))
        &\lesssim
        -|\operatorname{Im}z|
        |\operatorname{Re}z-z_j|^2,
        \quad z\in\Omega_1,\\
        \operatorname{Re}(2i\theta(z))
        &\gtrsim
        |\operatorname{Im}z|
        |\operatorname{Re}z-z_j|^2,
        \quad z\in\Omega_2.
    \end{align}
\end{itemize}
\end{proposition}

\begin{proof}
By the symmetry $\theta(\overline{z})=\overline{\theta(z)}$, it suffices to prove the estimates for $z\in\Omega_1$. Since $\xi\to-4$ and $z_2\to1$ as $t\to+\infty$, we may take $t$ sufficiently large so that
\begin{align}
    -\frac54\leq\xi+3<-1, \quad \frac45<z_2<1<z_1.\nonumber
\end{align}
The phase function can be written as
\begin{align}
    \theta(z)=\frac12\left[z^3-z^2+(\xi+3)z-(\xi+3)z^{-1}+z^{-2}-z^{-3}\right].\nonumber
\end{align}

We first consider $z=s+iv\in\Omega_1$ with $v>0$ and $s<z_2/2$. For $z=\rho e^{i\psi}$, a direct calculation gives
\begin{align}
    \operatorname{Re}(2i\theta(z))
    &=-(\rho^3+\rho^{-3})\sin3\psi+(\rho^2+\rho^{-2})\sin2\psi \notag\\
    &\quad-(\xi+3)(\rho+\rho^{-1})\sin\psi.\nonumber
\end{align}
Suppose first that $0<s<z_2/2$. By \eqref{eq:lens-sector-bound},
\begin{align}
    0<\psi\leq\phi<\frac{\pi}{12}, \quad 0<\rho\leq\frac{z_2}{2}\sec\phi<\frac{1}{\sqrt3}.\nonumber
\end{align}
Using $\sin3\psi=(3-4\sin^2\psi)\sin\psi$, $\sin2\psi=2\sin\psi\cos\psi$, and $3-4\sin^2\psi\geq2$, we obtain
\begin{align}
    \operatorname{Re}(2i\theta(z))
    &\leq-\rho^{-3}\sin\psi\left[2(1+\rho^6)-2(\rho+\rho^5)-\frac54(\rho^2+\rho^4)\right].\nonumber
\end{align}
For $0<\rho\leq1/\sqrt3$, the expression in brackets satisfies
\begin{align}
    &2(1+\rho^6)-2(\rho+\rho^5)-\frac54(\rho^2+\rho^4) \notag\\
    &\quad\geq2-2(\rho+\rho^5)-\frac54(\rho^2+\rho^4) \notag\\
    &\quad\geq\frac{13}{9}-\frac{20}{9\sqrt3}>0.\nonumber
\end{align}
Since $v=\rho\sin\psi$ and $\rho<1$, it follows that
\begin{align}
    \operatorname{Re}(2i\theta(z))\lesssim-\rho^{-3}\sin\psi\lesssim-v.\nonumber
\end{align}

For $s<0$, write $z=\rho e^{i(\pi-\psi)}$, where $0<\psi\leq\phi$. Using the same trigonometric identities and $\rho^3+\rho^{-3}\geq\rho+\rho^{-1}$, we obtain
\begin{align}
    \operatorname{Re}(2i\theta(z))
    &\leq-\sin\psi\left[2(\rho^3+\rho^{-3})-\frac54(\rho+\rho^{-1})\right] \notag\\
    &\leq-\frac34(\rho^3+\rho^{-3})\sin\psi \notag    \leq-\frac34v.\nonumber
\end{align}
This proves the upper-half-plane estimate near the origin.

We next consider $z=s+iv\in\Omega_1$ with $v>0$, where $s>z_1$ for $j=1$ and $z_2/2\leq s<z_2$ for $j=2$. Set $u=s-z_j$. By \eqref{eq:lens-sector-bound},
\begin{align}
    0<v\leq|u|\tan\phi, \quad s\geq\frac{z_2}{2}>\frac25.\nonumber
\end{align}
Directly from the phase function,
\begin{align}
    &\operatorname{Re}(2i\theta(z))=-vF(s,v;\xi),\nonumber\\
    &F(s,v;\xi)
    =3s^2-v^2-2s+(\xi+3)+\frac{\xi+3}{s^2+v^2}
    \quad-\frac{2s}{(s^2+v^2)^2}+\frac{3s^2-v^2}{(s^2+v^2)^3}.\nonumber
\end{align}
Recall that $\omega_1>\omega_2>\omega_3$ are the roots of
\begin{align}
    P(\omega;\xi)=3\omega^3-2\omega^2-(6-\xi)\omega+4,\nonumber
\end{align}
and that
\begin{align}
    \omega_1=z_1+z_1^{-1}=z_2+z_2^{-1}, \quad z_1z_2=1.\nonumber
\end{align}
At $\xi=-4$, the other two roots are $(-2+\sqrt{10})/3$ and $(-2-\sqrt{10})/3$. Thus, for all sufficiently large $t$,
\begin{align}
    \omega_2<\frac12, \quad \omega_3<-\frac32.\nonumber
\end{align}
A direct factorization yields
\begin{align}
    F(s,0;\xi)=\frac{3}{s^2}(s-z_1)(s-z_2)\left(s+\frac1s-\omega_2\right)\left(s+\frac1s-\omega_3\right).\nonumber
\end{align}
Define
\begin{align}
    A(s)
    &:=\frac{3}{s^2}\left(s+\frac1s-\frac12\right)\left(s+\frac1s+\frac32\right) \notag\\
    &=3+\frac3s+\frac{15}{4s^2}+\frac3{s^3}+\frac3{s^4}\geq3.\nonumber
\end{align}
In both portions under consideration,
\begin{align}
    (s-z_1)(s-z_2)=u^2+(z_1-z_2)|u|\geq u^2.\nonumber
\end{align}
Consequently,
\begin{align}
    F(s,0;\xi)\geq A(s)u^2.\nonumber
\end{align}

To estimate the $v$-dependent terms, set $q=v^2/s^2$. Since $\xi+3<0$ and $s>0$, we have
\begin{align}
    F(s,v;\xi)-F(s,0;\xi)
    &\geq-v^2+\frac{3s^2-v^2}{(s^2+v^2)^3}-\frac3{s^4} \notag\\
    &=-v^2-\frac{v^2}{s^6}\frac{10+9q+3q^2}{(1+q)^3} \notag\\
    &\geq-\left(1+\frac{10}{s^6}\right)v^2.\nonumber
\end{align}
Moreover, for $s\geq2/5$,
\begin{align}
    1+\frac{10}{s^6}\leq13A(s).\nonumber
\end{align}
Indeed, multiplying this inequality by $4s^6$ reduces it to
\begin{align}
    152s^6+156s^5+195s^4+156s^3+156s^2-40\geq0,\nonumber
\end{align}
whose left-hand side is increasing for $s>0$ and positive at $s=2/5$. Combining the preceding estimates gives
\begin{align}
    F(s,v;\xi)
    \geq A(s)(u^2-13v^2) \geq A(s)(1-13\tan^2\phi)u^2.\nonumber
\end{align}
Since $0<\phi<\pi/12$,
\begin{align}
    13\tan^2\phi<13\tan^2\frac{\pi}{12}=13(7-4\sqrt3)<1.\nonumber
\end{align}
Hence,
\begin{align}
    F(s,v;\xi)\geq3(1-13\tan^2\phi)u^2\gtrsim u^2.\nonumber
\end{align}
Therefore,
\begin{align}
    \operatorname{Re}(2i\theta(z))\lesssim-vu^2=-|\operatorname{Im}z|\,|\operatorname{Re}z-z_j|^2.\nonumber
\end{align}
The corresponding estimates in $\Omega_2$ follow by complex conjugation.
\end{proof}

We next open the contour $\Gamma$ by introducing suitable continuous extensions of the factors in the jump matrix $V^{(1)}(z)$.

\begin{proposition}\label{def-Rj}
Let $q_0\in\tanh(x)+H^{4,4}(\mathbb{R})$. For all sufficiently large $t$ with $(x,t)\in\mathcal{P}_{<-4}$, the lens contours satisfying \eqref{eq:lens-sector-bound} can be chosen so that there exist uniformly bounded functions $R_j:\Omega_j\cup\partial\Omega_j\to\mathbb{C}$, $j=1,2$, that are continuous on $\Omega_j\cup\partial\Omega_j$ and have continuous first-order partial derivatives in $\Omega_j$, with the following boundary values:
\begin{align}
R_1(z)
&=
\begin{cases}
\dfrac{\overline{r(z)}}{1-|r(z)|^2}T_+(z)^{-2},
& z\in\Gamma, \\[0.8em]
g(\operatorname{Re}z),
& z\in\Sigma_1,
\end{cases}\\
R_2(z)
&=
\begin{cases}
\dfrac{r(z)}{1-|r(z)|^2}T_-(z)^2,
& z\in\Gamma, \\[0.8em]
\overline{g(\operatorname{Re}z)},
& z\in\Sigma_2.
\end{cases}
\end{align}
Here $R_2(z)=\overline{R_1(\bar{z})}$, and
\begin{align}
g(s)
&=
\begin{cases}
0, & s\leq 0, \\[0.3em]
[1-\eta(s/z_2)]\gamma(z_2), & 0<s\leq z_2, \\[0.3em]
\gamma(z_1), & s\geq z_1,
\end{cases}
\end{align}
where $\eta\in C^\infty(\mathbb{R};[0,1])$ is a fixed cut-off function satisfying $\eta(s)=1$ for $s\leq 1/3$ and $\eta(s)=0$ for $s\geq 2/3$. The function $\gamma(z)$ is defined by
\begin{equation}\label{eq:gamma}
\gamma(z)=
\begin{cases}
\dfrac{\overline{D_2(z)}}{D_1(z)}
\left(\dfrac{a(z)}{T_+(z)}\right)^2,
& \text{in the generic case}, \\[8pt]
\dfrac{\overline{r(z)}}{1-|r(z)|^2}T_+(z)^{-2},
& \text{in the non-generic case}.
\end{cases}
\end{equation}
where
\begin{equation}\label{D1D2}
D_1(z)=\det\bigl(\Phi_{+,1}(z),\Phi_{-,2}(z)\bigr),
\quad
D_2(z)=\det\bigl(\Phi_{-,1}(z),\Phi_{+,1}(z)\bigr).
\end{equation}
Let $\varphi\in C_0^\infty(\mathbb{R};[0,1])$ be a fixed cut-off function equal to one near $-1$ and with small support there. Then, with $s=\operatorname{Re}z$, we have, for $j=1,2$,
\begin{align}
|\bar{\partial}R_j(z)|
&\lesssim
\begin{cases}
\varphi(s)+|r'(s)|+|z|^{-1/2},
& s<z_2/2, \\[0.3em]
|r'(s)|+|z-z_2|^{-1/2},
& z_2/2\leq s<z_2, \\[0.3em]
|r'(s)|+|z-z_1|^{-1/2},
& s>z_1,
\end{cases}
\quad z\in\Omega_j.
\end{align}	
\end{proposition}
\begin{proof}
Choose fixed smooth contours $\Sigma_1$ and $\Sigma_2$ as in \eqref{eq:lens-sector-bound}. Denote by $\ell(s)>0$ the smooth height of $\Sigma_1$ above $s\in\Gamma\setminus\{0\}$, with uniformly bounded derivative. Set $s=\operatorname{Re}z$, $v=\operatorname{Im}z$, and $h(s)=\frac{\overline{r(s)}}{1-|r(s)|^2}T_+(s)^{-2}$, with the values at $s=0,\pm1$ understood by continuity. Let $\chi\in C^\infty([0,1];[0,1])$ equal one near $0$ and zero near $1$. For $z\in\Omega_1$ with $s>z_1$, define	
\begin{align}
R_1(z)&=h(s)\chi\left(\frac{v}{\ell(s)}\right)+\gamma(z_1)\left[1-\chi\left(\frac{v}{\ell(s)}\right)\right].
\end{align}
For $z\in\Omega_1$ with $0<s<z_2$, define
\begin{align}
R_1(z)&=h(s)\chi\left(\frac{v}{\ell(s)}\right)+[1-\eta(s/z_2)]\gamma(z_2)\left[1-\chi\left(\frac{v}{\ell(s)}\right)\right].\nonumber
\end{align}

For the part above $(-\infty,0)$, in the non-generic case, define
\begin{align}
R_1(z)&=\frac{\overline{r(s)}}{1-|r(s)|^2}T(z)^{-2}\chi\left(\frac{v}{\ell(s)}\right).\nonumber
\end{align}
In the generic case, choose $\chi_{-1}\in C_0^\infty(\mathbb{R};[0,1])$ equal to one near $-1$ and supported where $\varphi=1$ and $D_1\neq0$. Define $R_1(z)=R_{11}(z)+R_{12}(z)$ by
\begin{align}
R_{11}(z)&=[1-\chi_{-1}(s)]\frac{\overline{r(s)}}{1-|r(s)|^2}T(z)^{-2}\chi\left(\frac{v}{\ell(s)}\right),\nonumber\\
R_{12}(z)&=f_3(s)G(z)\chi\left(\frac{v}{\ell(s)}\right),\nonumber
\end{align}
where
\begin{align}
F(s)&=\frac{\overline{D_2(s)}}{D_1(s)},\quad G(z)=\left(\frac{a(z)}{T(z)}\right)^2,\quad f_3(s)=\chi_{-1}(s)F(s).\nonumber
\end{align}
Finally, define $R_2(z)=\overline{R_1(\bar{z})}$ and extend both functions to their boundaries by continuity.
\end{proof}

Furthermore, using the extensions constructed in Proposition \ref{def-Rj}, we define a matrix-valued function
\begin{align}
R^{(2)}(z)=
\begin{cases}
\begin{pmatrix}
1 & R_1(z)e^{2it\theta(z)} \\
0 & 1
\end{pmatrix},
& z\in\Omega_1, \\[4mm]
\begin{pmatrix}
1 & 0 \\
R_2(z)e^{-2it\theta(z)} & 1
\end{pmatrix},
& z\in\Omega_2, \\[4mm]
I, & \text{elsewhere}.
\end{cases}
\end{align}
Then, the new matrix function defined by
\begin{align}\label{tran34}
M^{(3)}(z)=M^{(2)}(z)R^{(2)}(z)
\end{align}
satisfies the following hybrid $\bar{\partial}$-RH problem as a consequence of RH problem \ref{RH4} and Proposition \ref{def-Rj}.

\begin{prob2}\label{RH5}
Find a function $M^{(3)}(z)=M^{(3)}(z;x,t)$ with the following properties:
\begin{itemize}
\item $M^{(3)}(z)$ is continuous in $\mathbb{C}\setminus(\Sigma^{(3)}\cup\mathcal{Z})$ and admits continuous boundary values $M^{(3)}_+(z)$ and $M^{(3)}_-(z)$ on $\Sigma^{(3)}$ from the left and right, respectively.

\item
$M^{(3)}(z)$ satisfies the jump condition
\begin{align}
M^{(3)}_+(z)
=
M^{(3)}_-(z)V^{(3)}(z),
\quad z\in\Sigma^{(3)},\nonumber
\end{align}
where
\begin{align}\label{V4}
V^{(3)}(z)=
\begin{cases}
\begin{pmatrix}
1 & -R_1(z)e^{2it\theta(z)} \\
0 & 1
\end{pmatrix},
& z\in\Sigma_1, \\[1em]
\begin{pmatrix}
1 & 0 \\
R_2(z)e^{-2it\theta(z)} & 1
\end{pmatrix},
& z\in\Sigma_2, \\[1em]
T_-(z)^{-\sigma_3}V(z)T_+(z)^{\sigma_3},
& z\in[z_2,z_1].
\end{cases}
\end{align}

\item $M^{(3)}(z)$ has simple poles at $\eta_j\in\mathcal{Z}^{+}$ and $\overline{\eta}_j\in\mathcal{Z}^{-}$ satisfying
\begin{align}
\operatorname*{Res}_{z=\eta_j}M^{(3)}(z)
&=\lim_{z\to\eta_j}M^{(3)}(z)C_j^T,\\
\operatorname*{Res}_{z=\overline{\eta}_j}M^{(3)}(z)
&=\lim_{z\to\overline{\eta}_j}M^{(3)}(z)\sigma_1\overline{C_j^T}\sigma_1,
\end{align}
where
\begin{align}
C_j^T=
\begin{cases}
\begin{pmatrix}
0 & 0\\
c_jT(\eta_j)^2e^{-2it\theta(\eta_j)} & 0
\end{pmatrix},
& j\notin\Delta_{-4},\\[8pt]
\begin{pmatrix}
0 & \dfrac{e^{2it\theta(\eta_j)}}{c_jT'(\eta_j)^2}\\
0 & 0
\end{pmatrix},
& j\in\Delta_{-4}.
\end{cases}
\end{align}

\item $M^{(3)}(z)=I+\mathcal{O}(z^{-1}), \quad z\to\infty$,

\item
For $z\in \mathbb{C}\setminus(\Sigma^{(3)}\cup\mathcal{Z})$, we have
\begin{align}
\bar{\partial}M^{(3)}(z)
=
M^{(3)}(z)\bar{\partial}R^{(2)}(z),\nonumber
\end{align}
where
\begin{align}\label{Dbar-R3}
\bar{\partial}R^{(2)}(z)
&=
\begin{cases}
\begin{pmatrix}
0 & \bar{\partial}R_1(z)e^{2it\theta(z)} \\
0 & 0
\end{pmatrix},
& z\in\Omega_1, \\[1em]
\begin{pmatrix}
0 & 0 \\
\bar{\partial}R_2(z)e^{-2it\theta(z)} & 0
\end{pmatrix},
& z\in\Omega_2, \\[1em]
0, & \text{elsewhere}.
\end{cases}
\end{align}

\end{itemize}
\end{prob2}

So far, we have obtained the hybrid $\bar{\partial}$-RH problem \ref{RH5} for $M^{(3)}(z)$  starting from
the original RH problem for $M(z)$. We   next construct the solution $M^{(3)}(z)$ as follows:
\begin{align}
    M^{(3)}(z)=M^{(4)}(z)M^{rhp}(z)=
    \begin{cases}
        \bar{\partial}R^{(2)}(z)=0 \longrightarrow M^{rhp}(z),\\[6pt]
        \bar{\partial}R^{(2)}(z)\neq0 \longrightarrow M^{(4)}(z)=M^{(3)}(z)\left(M^{rhp}(z)\right)^{-1},\nonumber
    \end{cases}
\end{align}
where $M^{rhp}$ solves the pure RH problem obtained by omitting the $\bar{\partial}$-component of $M^{(3)}$, while $M^{(4)}$ solves the remaining pure $\bar{\partial}$-problem. These two problems are analyzed in Sections \ref{sec-pureRH} and \ref{sec-Dbar}, respectively.

\subsection{Contribution from a pure RH problem}\label{sec-pureRH}

In this subsection, we first consider the pure RH problem obtained by omitting the $\bar{\partial}$ component of $M^{(3)}(z)$. The resulting function $M^{rhp}(z)$ is characterized by the following RH problem.

\begin{prob}\label{RH-rhp}
    Find  $M^{rhp}(z)=M^{rhp}(z;x,t)$ which satisfies
	  \begin{itemize}
	  	\item  $M^{rhp}(z)$ is analytic in $\mathbb{C}\setminus(\Sigma^{(3)}\cup\mathcal{Z})$.
	  	\item $M^{rhp}(z)$  satisfies the jump condition
\begin{equation*}
	  		M^{rhp}_+(z)=M^{rhp}_-(z)V^{(3)}(z),\quad z\in\Sigma^{(3)},
	  	\end{equation*}
	  	where $V^{(3)}(z)$ is given by \eqref{V4}.
	  	\item $M^{rhp}(z)$ has the same poles and residue conditions as $M^{(3)}(z)$.
	  	\item   $M^{rhp}(z)$  has the same asymptotics as  $M^{(3)}(z)$.
	  \end{itemize}
\end{prob}

Let  $t$  be  large enough so that $2\left(\frac{9}{4}\right)^{\frac{1}{3}}\frac{\sqrt{C}}{3}\left(\frac{9}{4}t\right)^{-\frac{1}{3}+\tau}<\rho$ where $\tau$ is a constant with $0<\tau<\frac{1}{15}$ and $\rho$ has been defined in \eqref{definerho}. For a fixed constant  $\varepsilon \leq 2\left(\frac{9}{4}\right)^{\frac{1}{3}}\frac{\sqrt{C}}{3}$, define the open disk
\begin{align}
\mathcal{U}_{z}(1)= \{z \in \mathbb{C}: |z-1|< \left(\frac{9}{4}t\right)^{-\frac{1}{3}+\tau}\varepsilon\}.	
\end{align}
We solve RH problem \ref{RH-rhp} for $M^{rhp}(z)$ by treating the disk $\mathcal{U}_z(1)$ and its complement separately. For sufficiently large $t$, the disk contains both stationary points $z_1$ and $z_2$. In Section \ref{sec-local}, we use the Painlev\'e II model to construct a local parametrix $M^{loc}(z)$ for the jumps in this disk. Outside the disk, the main contribution comes from the discrete spectrum. The corresponding outer model $M^{out}(z)$, constructed in Sec. \ref{sec-Mout}, has the same poles and residue conditions as $M^{rhp}(z)$. Moreover, we denote the error matrix by $E(z)$ and analyze it in Sec. \ref{sec-E}. Thus we obtain the following decomposition of $M^{rhp}(z)$:
\begin{align}\label{Ez}
    M^{rhp}(z) &=
    \begin{cases}
        E(z)M^{out}(z), & z\in\mathbb{C}\setminus\mathcal{U}_z(1),\\[4pt]
        E(z)M^{out}(z)M^{loc}(z), & z\in\mathcal{U}_z(1).
    \end{cases}
\end{align}

\subsubsection{Soliton solutions for $M^{out}(z)$}\label{sec-Mout}

Here we consider the model $M^{out}(z)$, which satisfies the same residue conditions as $M^{rhp}(z)$ but has no jumps in the complex plane. More precisely, $M^{out}(z)$ is characterized by the following RH problem:
\begin{prob}\label{RHP-out}
Find $M^{out}(z)=M^{out}(z;x,t)$ with the following properties:
\begin{itemize}
    \item $M^{out}(z)$ is analytic in $\mathbb{C}\setminus\{\eta_j,\overline{\eta}_j:1\le j\le N\}$.
    \item \textit{Residue conditions:} $M^{out}(z)$ has simple poles at $\eta_j$ and $\overline{\eta}_j$, $j=1,\ldots,N$, satisfying
    \begin{align}
        \operatorname*{Res}_{z=\eta_j}M^{out}(z)
        &=\lim_{z\to\eta_j}M^{out}(z)C_j^T,\\
        \operatorname*{Res}_{z=\overline{\eta}_j}M^{out}(z)
        &=\lim_{z\to\overline{\eta}_j}M^{out}(z)\sigma_1\overline{C_j^T}\sigma_1,
    \end{align}
    where
    \begin{align}
        C_j^T=
        \begin{cases}
            \begin{pmatrix}
                0 & 0\\
                c_jT(\eta_j)^2e^{-2it\theta(\eta_j)} & 0
            \end{pmatrix},
            & j\notin\Delta,\\
            \begin{pmatrix}
                0 & \dfrac{e^{2it\theta(\eta_j)}}{c_jT'(\eta_j)^2}\\
                0 & 0
            \end{pmatrix},
            & j\in\Delta.
        \end{cases}
    \end{align}
    \item \textit{Asymptotic behavior:} $M^{out}(z)=I+\mathcal{O}(z^{-1})$ as $z\to\infty$.
\end{itemize}	
\end{prob}

The reduced model retains only the poles with velocities in $\Lambda$.
\begin{prob}\label{RHP-out-I}
Find $M_\Lambda^{out}(z)=M_\Lambda^{out}(z;x,t)$ with the following properties:
\begin{itemize}
    \item $M_\Lambda^{out}(z)$ is analytic in $\mathbb{C}\setminus\{\eta_j \in\Lambda\}$.
    \item \textit{Residue conditions:} $M_\Lambda^{out}(z)$ has simple poles at $\eta_j  \in \Lambda$, satisfying
    \begin{align}
        \operatorname*{Res}_{z=\eta_j}M_\Lambda^{out}(z)
        &=\lim_{z\to\eta_j}M_\Lambda^{out}(z)C_j^T,\nonumber\\
        \operatorname*{Res}_{z=\overline{\eta}_j}M_\Lambda^{out}(z)
        &=\lim_{z\to\overline{\eta}_j}M_\Lambda^{out}(z)\sigma_1\overline{C_j^T}\sigma_1,\nonumber
    \end{align}
    where $C_j^T$ is as in RH problem \ref{RHP-out}.
    \item \textit{Asymptotic behavior:} $M_\Lambda^{out}(z)=I+\mathcal{O}(z^{-1})$ as $z\to\infty$.
\end{itemize}
\end{prob}

Further, we can show that
\begin{proposition}
Given the discrete scattering data $\{(\eta_j,c_j)\}_{j=1}^N$. Then, as $t\to\infty$, uniformly for $(x,t)\in\mathcal{P}_{<-4}$, we have
\begin{align}\label{eq:out-reduction}
    M^{out}(z)=\left(I+\mathcal{O}(e^{-ct})\right)M_\Lambda^{out}(z),\quad c>0.
\end{align}
 \end{proposition}
By using reconstruction formula (\ref{recon-sol}), we  have
\begin{corollary}
Let $q^{sol}(x,t)$ denote the $N$-soliton solution of the defocusing Hirota equation associated with RH problem \ref{RHP-out}. Then, as $t\to\infty$, uniformly for $(x,t)\in\mathcal{P}_{<-4}$, we have
\begin{align}\label{soliton-reduction}
    q^{sol}(x,t)=q^{sol}_\Lambda(x,t)+\mathcal{O}(e^{-ct}),
\end{align}
where $q^{sol}_\Lambda(x,t)$ denotes the $N(\Lambda)$-soliton solution of the  Hirota equation corresponding to discrete spectrum   $\eta_j \in \Lambda$.
\end{corollary}

\subsubsection{Local parametrix}\label{sec-local}

\begin{figure}[htbp]
    \centering
    \begin{tikzpicture}[scale=0.8]
        \coordinate (C)  at (0,0);
        \coordinate (Z2) at (-0.75,0);
        \coordinate (Z1) at (0.75,0);

        \draw[dotted,-latex] (-4.2,0)--(4.3,0)
            node[black,right] {$\operatorname{Re}z$};

        \draw[blue,thick] (-2.05,1.25)--(Z2);
        \draw[blue,thick,-latex] (-1.53,0.75)--(-1.244,0.475);

        \draw[blue,thick] (Z1)--(2.05,1.25);
        \draw[blue,thick,-latex] (1.27,0.50)--(1.556,0.775);

        \draw[red,thick] (-2.05,-1.25)--(Z2);
        \draw[red,thick,-latex] (-1.53,-0.75)--(-1.244,-0.475);

        \draw[red,thick] (Z1)--(2.05,-1.25);
        \draw[red,thick,-latex] (1.27,-0.50)--(1.556,-0.775);

        \draw[black,line width=0.8pt] (Z2)--(Z1);
        \draw[black,-latex] (-0.55,0)--(-0.25,0);

        \draw[
            teal!75!black,
            line width=1pt,
            dash pattern=on 0pt off 3pt,
            line cap=round
        ] (C) circle[radius=2.4];

        \node at (0,2.72) {$\partial\mathcal{U}_{z}(1)$};

        \fill (Z2) circle (2pt);
        \fill (C) circle (2pt);
        \fill (Z1) circle (2pt);

        \node at (-0.75,-0.31) {$z_2$};
        \node at (0,-0.31) {$1$};
        \node at (0.75,-0.31) {$z_1$};

        \node at (1.35,1.02) {$\Sigma_1$};
        \node at (1.35,-1.02) {$\Sigma_2$};
    \end{tikzpicture}
    \caption{The jump contour $\Sigma^{loc}$ for $M^{loc}(z)$.}
    \label{jump-zloc}
\end{figure}
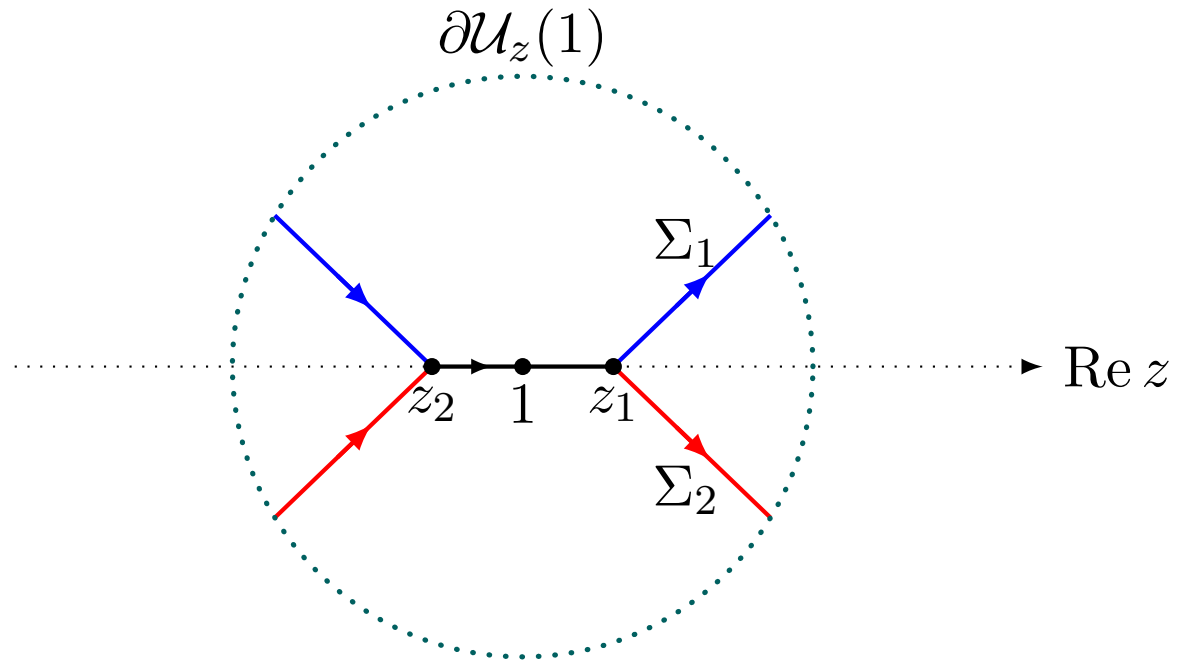

Denote the local jump contour
\begin{equation*}
  \Sigma^{loc}:= \Sigma^{(3)} \cap \mathcal{U}_{z}(1),
\end{equation*}
as depicted in Fig. \ref{jump-zloc}. The local model  $M^{loc}(z)$  satisfies the following RH problem.
\begin{prob}\label{mloc}
    Find  $M^{loc}(z)=M^{loc}(z;x,t)$ which satisfies
	  \begin{itemize}
	  	\item  $M^{loc}(z)$ is analytic in $\mathbb{C}\setminus \Sigma^{loc}$.
	  	\item $M^{loc}(z)$  satisfies the jump condition
\begin{equation*}
	  		M^{loc}_+(z)=M^{loc}_-(z)V^{loc}(z), \ \ z \in \Sigma^{loc},
	  	\end{equation*}
	  	where $V^{loc}(z) := V^{(3)}(z)|_{ z \in \Sigma^{loc}}$.
	  	\item   $M^{loc}(z)$  has the same asymptotics as  $M^{rhp}(z)$.

	  \end{itemize}
\end{prob}

In the region $ -C < (\xi+4) t^{2/3}< 0$ with $C>0$,  we notice that $ \xi \to -4^-$ as $t \to  \infty$, which leads to the coalescence of saddle points: $z_1$ and $z_2$  merge to $z=1$. We analyze the local properties of the phase function $t \theta(z)$ near $z=1$. As $z \to 1$, we find that
\begin{align}\label{t-theta}
t\theta(z)=3t(z-1)^3+(x+4t)(z-1)+A(z),	
\end{align}
where
\begin{align}
A(z)=\frac{x+4t}{2}
\sum_{n=2}^{\infty}(-1)^{n+1}(z-1)^n+\frac{t}{4}
\sum_{n=4}^{\infty}(-1)^{n+1}(n-1)(n+2)(z-1)^n.\nonumber
\end{align}
Define $s$ as the space-time parameter and  $k$  as the scaled parameter by
\begin{align}\label{scaled}
	s=\frac{4}{9}\left(\frac{9}{4}t\right)^{\frac{2}{3}}(\xi+4),\quad k=\left(\frac{9}{4}t\right)^{\frac{1}{3}}(z-1),
\end{align}
then it can be proved that $A(z)$ converges and $A(z)=\mathcal{O}(t^{-1/3} k^4)$. Therefore, the Taylor expansion \eqref{t-theta} becomes
\begin{align}\label{equ-ttheta}
t\theta(z)=\frac{4}{3}k^{3}+sk+\mathcal{O}(t^{-1/3}k^{4}).
\end{align}

Next, we define the open disk centered at the origin in the scaled $k$-plane by
\begin{align}
\mathcal{U}_{k}(0)= \{k \in \mathbb{C}: |k|< \left(\frac{9}{4}t\right)^{\tau}\varepsilon\}, \nonumber
\end{align}
whose boundary is oriented counterclockwise. Then the transformation given by \eqref{scaled} defines a map $z \mapsto k$, which maps  $ \mathcal{U}_{z}(1)$  onto  $\mathcal{U}_{k}(0)$  in the $k$-plane.

\begin{figure}[htbp]
    \centering
    \begin{tikzpicture}[scale=0.8]
        \coordinate (C) at (0,0);
        \coordinate (K2) at (-0.95,0);
        \coordinate (K1) at (0.95,0);

        \draw[dotted,-latex] (-4.2,0)--(4.3,0)
            node[black,right] {$\operatorname{Re}k$};

        \draw[dotted,-latex] (0,-3.2)--(0,3.8)
            node[black,above] {$\operatorname{Im}k$};

        \draw[blue,thick] (-2.10,1.20)--(K2);
        \draw[blue,thick,-latex] (-1.64,0.72)--(-1.387,0.456);

        \draw[blue,thick] (K1)--(2.10,1.20);
        \draw[blue,thick,-latex] (1.41,0.48)--(1.663,0.744);

        \draw[red,thick] (-2.10,-1.20)--(K2);
        \draw[red,thick,-latex] (-1.64,-0.72)--(-1.387,-0.456);

        \draw[red,thick] (K1)--(2.10,-1.20);
        \draw[red,thick,-latex] (1.41,-0.48)--(1.663,-0.744);

        \draw[black,line width=0.8pt] (K2)--(K1);
        \draw[black,-latex] (-0.70,0)--(-0.42,0);

        \draw[
            teal!75!black,
            line width=1pt,
            dash pattern=on 0pt off 3pt,
            line cap=round
        ] (C) circle[radius=2.4];

        \node[fill=white,inner sep=1pt] at (0,2.8)
            {$\partial\mathcal{U}_{k}(0)$};

        \fill (K2) circle (2pt);
        \fill (C) circle (2pt);
        \fill (K1) circle (2pt);

        \node at (-0.95,-0.33) {$k_2$};
        \node at (0,-0.33) {$0$};
        \node at (0.95,-0.33) {$k_1$};

        \node at (1.40,1.02) {$\widehat{\Sigma}_1$};
        \node at (1.40,-1.02) {$\widehat{\Sigma}_2$};
    \end{tikzpicture}
    \caption{The jump contour $\widehat{\Sigma}^{loc}$ for $M^{loc}(k)$.}
    \label{jump-kloc}
\end{figure}
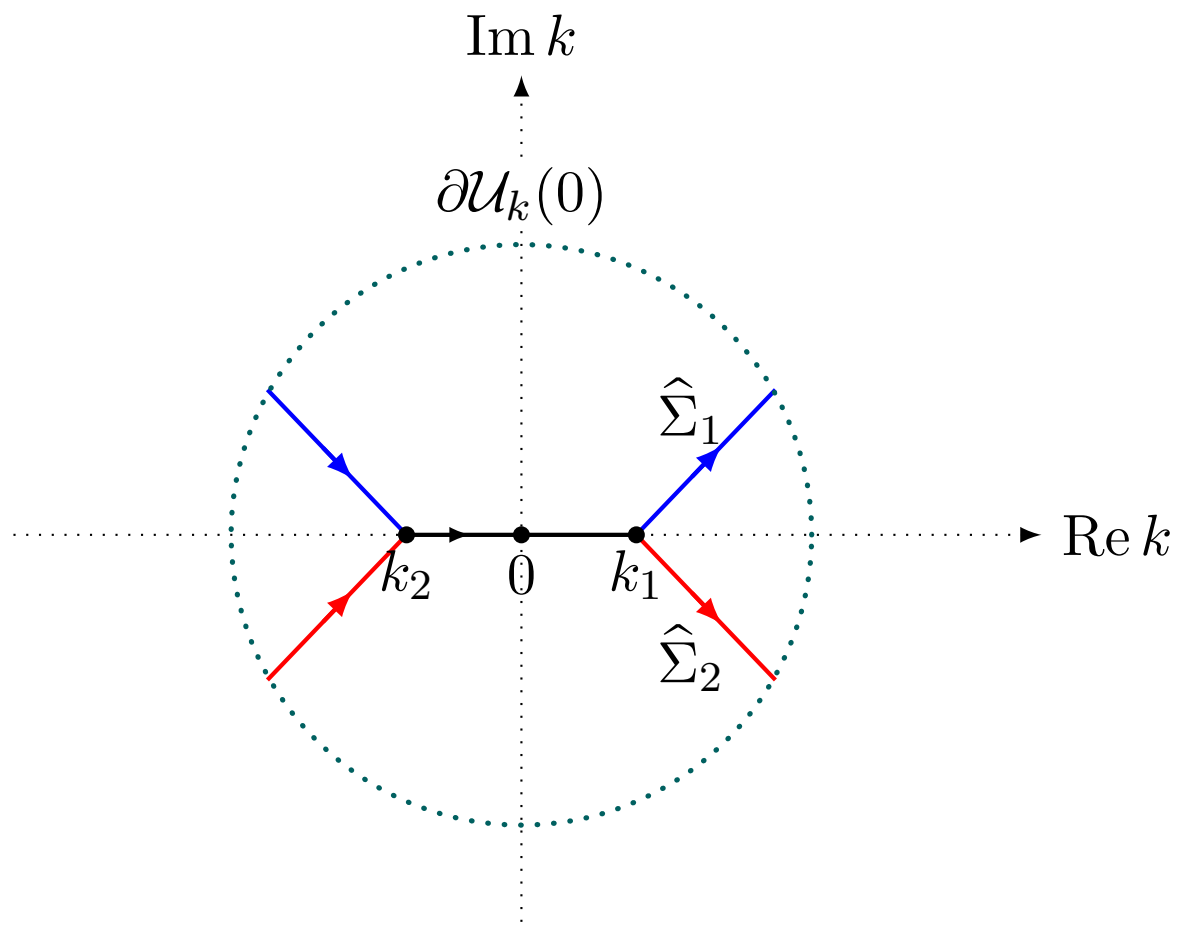

First, we construct the local parametrix $M^{loc}(k)$. Define the contour $\widehat{\Sigma}^{loc}$ in the $k$-plane
\begin{align*}
    \widehat{\Sigma}^{loc}
    := \widehat{\Sigma}_1 \cup \widehat{\Sigma}_2 \cup [k_2,k_1],
\end{align*}
which corresponds to $\Sigma^{loc}$ after scaling $z$ to the scaled parameter $k$. The corresponding regions in the $k$-plane can be  seen in Fig. \ref{jump-kloc}. Correspondingly, the saddle points $z_1$ and $z_2$ in the $z$-plane are rescaled to $k_1$ and $k_2$ respectively. The jump matrix $V^{loc}(z)$  transforms to the following $V^{loc}(k)$ in the $k$-plane.

\begin{align}
V^{loc}(k)
=
\begin{cases}
e^{it\theta\left((\frac{9}{4}t)^{-\frac{1}{3}}k+1\right)\widehat\sigma_3  }\begin{pmatrix}
1 & -\gamma(z_j)\\
0 & 1
\end{pmatrix},
& k\in\widehat{\Sigma}_1,\\[4mm]
e^{it\theta\left((\frac{9}{4}t)^{-\frac{1}{3}}k+1\right)\widehat\sigma_3  }\begin{pmatrix}
1 & 0\\
\overline{\gamma(z_j)} & 1
\end{pmatrix},
& k\in \widehat{\Sigma}_2,\\[4mm]
T_-((\frac{9}{4}t)^{-\frac{1}{3}}k+1)^{-\sigma_3}V((\frac{9}{4}t)^{-\frac{1}{3}}k+1)T_+((\frac{9}{4}t)^{-\frac{1}{3}}k+1)^{\sigma_3},
& k\in [k_2,k_1].
\end{cases}
\end{align}
In the first two cases, $j=1$ when $\operatorname{Re}k>k_1$ and $j=2$ when $\operatorname{Re}k<k_2$.

In the generic case, $z_j \to 1$ and $|r(z_j)| \to 1$ for $j=1,2$ as $t \to \infty$, which causes the appearance of the singularity of $\frac{r(z_j)}{1-|r(z_j)|^2}$. However, this singularity can be balanced by the factor $T(z)^{-2}$. Define a cutoff function $\chi(z)\in C_0^\infty(\mathbb{R},[0,1])$ satisfying
\begin{equation*}
\chi (z)=1, \ z\in  \mathbb{R}\cap \mathcal{U}_z(1),
\end{equation*}
and  a new  reflection coefficient  $\widetilde{r}(z)$  satisfying
\begin{equation}\label{rr-g}
\widetilde{r}(z)=(1-\chi (z)) \frac{\overline{r(z)} }{1-|r(z)|^2}  T_+( z)^{-2}+\chi (z)  f(z)  h(z)^2,
\end{equation}
where
\begin{equation*}
f(z) := \frac{ \overline{D_2 (z)} }{D_1 (z)},  \ \ \  h(z):=\frac{ a(z) }{T_+(z)},	
\end{equation*}
and $D_1(z)$ and  $D_2(z)$ are defined in \eqref{D1D2}, while in the non-generic case,
\begin{equation}\label{rr-non}
\widetilde{r}(z)= \frac{\overline{r(z)} }{1-|r(z)|^2}  T_+ (z)^{-2}.
\end{equation}
Moreover, we have $\widetilde{r}(z_j)=\gamma(z_j)$ as $z_j \to 1, \ j=1,2$.

Next, we will show that RH problem for $M^{loc}(k)$ can be explicitly approximated by the following model RH problem for  $\widehat{M}^{loc}(k)$ in $\mathcal{U}_k(0)$, and then prove the solution $\widehat{M}^{loc}(k)$ is associated to the Painlev\'e II equation.

\begin{prob}
    Find  $ \widehat{M}^{loc}(k)=\widehat{M}^{loc}(k;x,t)$ with properties
	  \begin{itemize}
	  	\item  $\widehat{M}^{loc}(k)$ is analytic in $\mathbb{C} \setminus  \widehat{\Sigma}^{loc}$.
	  	\item  $\widehat{M}^{loc}(k)$  satisfies the jump condition
\begin{equation*}
	  		\widehat{M}^{loc}_+(k)=\widehat{M}^{loc}_-(k) \widehat{V}^{loc}(k), \ k\in  \widehat{\Sigma}^{loc},
\end{equation*}
where
\begin{align*}
        \widehat{V}^{loc}(k)= \begin{cases}
       e^{i \left( \frac{4}{3} k^3 +  s k \right)\widehat\sigma_3  }   \left( \begin{array}{cc}
       		1& -\widetilde{r}(1)\\
       		 0   & 1
       	\end{array}\right):=b_+^{-1},  & k\in \widehat{\Sigma}_{1}, \\
  e^{i \left( \frac{4}{3} k^3 +  s k \right)\widehat\sigma_3  }  \left(	\begin{array}{cc}
       			1& 0\\
       			\overline{\widetilde{r}(1)} & 1
       		\end{array}\right):=b_- , & k \in \widehat{\Sigma}_2,\\
b_-b_+^{-1},     & k\in [k_2,  k_1].
       	\end{cases}
       \end{align*}

         \item   $\widehat{M}^{loc}(k)\to I, \ k\to \infty$.
	  \end{itemize}
\end{prob}

Define $N(k):=M^{loc}(k) (\widehat{M}^{loc}(k))^{-1}$ which satisfies the following RH problem.
\begin{prob}
 Find  $ N(k)=N(k;x,t)$  which satisfies
 \begin{itemize}
   \item $N(k)$ is analytic in $\mathbb{C} \setminus  \widehat{\Sigma}^{loc}$.
   \item  $N(k)$ satisfies the following jump condition
   \begin{equation*}
     N_+(k) = N_-(k)V^N(k),
   \end{equation*}
   where
   \begin{equation*}
   V^N(k) = \widehat{M}_-^{loc}(k) V^{loc}(k) \widehat{V}^{loc}(k)^{-1} \widehat{M}_-^{loc}(k)^{-1}.
   \end{equation*}
   \item   $N(k)\to I, \ k\to \infty$.
 \end{itemize}
\end{prob}

\begin{proposition}\label{estimate-Nk}
As $t \to \infty$, $N(k)$ exists and satisfies
\begin{equation}\label{Nk-est}
 N(k) = I + \mathcal{O}(t^{-\frac{1}{3}+4\tau}),
\end{equation}
where $\tau$ is a constant with $0<\tau<\frac{1}{15}$.
\end{proposition}

\begin{proof}
Suppose that $\widehat{M}^{loc}(k)$ and its inverse are uniformly bounded, which will be shown below through the deformation to the Painlev\'{e} II model. We only need to estimate the error between $V^{loc}(k)$ and $\widehat{V}^{loc}(k)$.

For $k \in [k_2,k_1]$, we have $\left|e^{2it\theta(z)}\right|=\left|e^{i\left(\frac{8}{3}k^3+2sk\right)}\right|=1$. Direct calculations give
\begin{align*}
&\left|
\widetilde{r}(z)e^{2it\theta(z)}
-\widetilde{r}(1)e^{i\left(\frac{8}{3}k^3+2sk\right)}
\right| \\
&\quad\leq
\left|\widetilde{r}(z)-\widetilde{r}(1)\right|
+\left|\widetilde{r}(1)\right|
\left|
e^{2it\theta(z)}
-e^{i\left(\frac{8}{3}k^3+2sk\right)}
\right| \\
&\quad\leq
|h(z)|^2|f(z)-f(1)|
+|f(1)|\left|h^2(z)-h^2(1)\right|
+\left|\widetilde{r}(1)\right|
\left|
e^{2it\theta(z)}
-e^{i\left(\frac{8}{3}k^3+2sk\right)}
\right|.
\end{align*}
Here, the last inequality applies to the generic case, while in the non-generic case the same estimate follows directly from \eqref{rr-non}. By the regularity of the scattering data near $z=1$, $f$ is continuously differentiable there. Moreover, it follows from the trace formulas \eqref{equ-trace} and \eqref{T-def} that
\begin{align*}
h(z)=C_0\prod_{j\notin\Delta_{-4}}\frac{z-\eta_j}{z-\overline{\eta_j}},
\end{align*}
where $C_0$ is independent of $z$. Since the discrete spectrum is separated from $z=1$, both $h$ and its derivative are uniformly bounded near $z=1$. Therefore,
\begin{align*}
|f(z)-f(1)|+|h(z)-h(1)|\lesssim |z-1|\lesssim t^{-\frac{1}{3}+\tau}.
\end{align*}
By \eqref{equ-ttheta} and $|k|\lesssim t^\tau$ in $\mathcal U_k(0)$, we have
\begin{align*}
\left|e^{2it\theta(z)}-e^{i(\frac83k^3+2sk)}\right|
\lesssim t^{-\frac13}|k|^4
\lesssim t^{-\frac13+4\tau}.
\end{align*}
Consequently,
\begin{align*}
\left|\widetilde{r}(z)e^{2it\theta(z)}-\widetilde{r}(1)e^{i\left(\frac{8}{3}k^3+2sk\right)}\right|\lesssim t^{-\frac{1}{3}+4\tau},\quad k\in[k_2,k_1].
\end{align*}

We next consider the nonreal parts of the jump contour. For $k\in\widehat{\Sigma}_1$ with $\operatorname{Re}k>k_1$, write $k=k_1+\widehat{u}+i\widehat{v}$ and $z=z_1+u+iv$. A direct calculation gives
\begin{align*}
\operatorname{Re}i\left(\frac{8}{3}k^3+2sk\right)&=\operatorname{Re}i\left(6t(z-1)^3+2(x+4t)(z-1)\right)\\
&=-18tu^2v-36t(z_1-1)uv+6tv^3-2tvw(z_1),
\end{align*}
where
\begin{align*}
w(z_1)=9(z_1-1)^2+\xi+4.
\end{align*}
Since $z_1$ is a stationary point, \eqref{equ-xi} implies
\begin{align*}
w(z_1)=\frac{(z_1-1)^3(6z_1^3+2z_1^2+7z_1+3)}{z_1^2(z_1^2+1)}\geq 0.
\end{align*}
Moreover, the choice of the opening angle gives $0\leq v\leq u$. Combining this with $z_1\geq 1$ and $w(z_1)\geq 0$, we obtain
\begin{align*}
\operatorname{Re}i\left(\frac{8}{3}k^3+2sk\right)\leq-\frac{16}{3}\hat{u}^2\hat{v},
\end{align*}
which yields
\begin{align*}
\left|e^{i\left(\frac{8}{3}k^3+2sk\right)}\right|\leq e^{-\frac{16}{3}\hat{u}^2\hat{v}},\quad k\in\widehat{\Sigma}_1\ \text{and}\ \operatorname{Re}k>k_1.
\end{align*}
Similarly to the case on the real axis, we obtain
\begin{align*}
\left|\widetilde{r}(z_1)e^{2it\theta(z)}-\widetilde{r}(1)e^{i\left(\frac{8}{3}k^3+2sk\right)}\right|\lesssim t^{-\frac{1}{3}+4\tau},\quad k\in\widehat{\Sigma}_1\ \text{and}\ \operatorname{Re}k>k_1.
\end{align*}
The estimate on other jump contours can be given in a similar way.

The preceding estimates imply
\begin{equation*}
\left\|V^N-I\right\|_{L^1(\widehat{\Sigma}^{loc})\cap L^2(\widehat{\Sigma}^{loc})\cap L^\infty(\widehat{\Sigma}^{loc})}
\lesssim t^{-\frac{1}{3}+4\tau}.
\end{equation*}
Therefore, the existence and uniqueness of $N(k)$ can be proven by the theorem of the small-norm RH problem, which also yields \eqref{Nk-est}.

\end{proof}

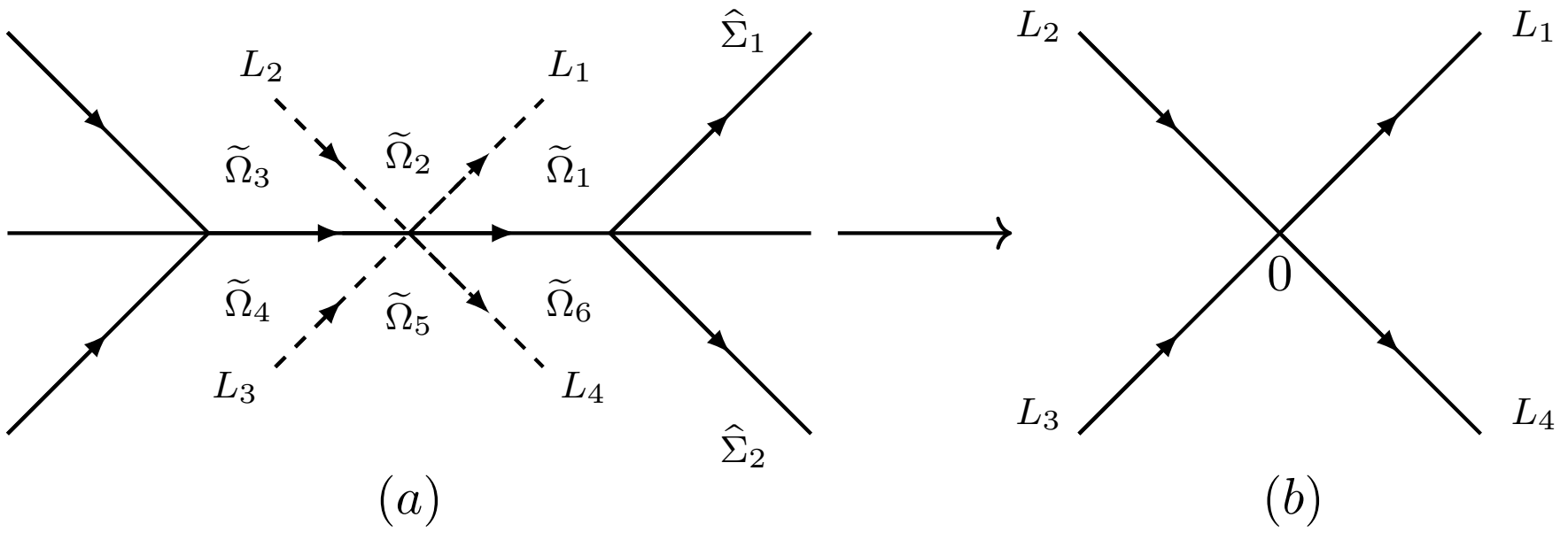
\begin{figure}[htbp]
    \centering
    \begin{tikzpicture}
        \draw[thick] (-6.5,0)--(-0.5,0);
        \draw[thick,-latex] (-5,0)--(-4,0);
        \draw[thick,-latex] (-4,0)--(-2.7,0);

        \draw[thick] (-6.5,1.5)--(-5,0);
        \draw[thick,-latex] (-6.5,1.5)--(-5.75,0.75);

        \draw[thick] (-2,0)--(-0.5,1.5);
        \draw[thick,-latex] (-2,0)--(-1.1,0.9);

        \draw[thick] (-6.5,-1.5)--(-5,0);
        \draw[thick,-latex] (-6.5,-1.5)--(-5.75,-0.75);

        \draw[thick] (-2,0)--(-0.5,-1.5);
        \draw[thick,-latex] (-2,0)--(-1.1,-0.9);

        \draw[thick,dashed] (-4.5,-1)--(-2.5,1);
        \draw[thick,dashed,-latex] (-4.5,-1)--(-4,-0.5);
        \draw[thick,dashed,-latex] (-3.5,0)--(-2.9,0.6);

        \draw[thick,dashed] (-4.5,1)--(-2.5,-1);
        \draw[thick,dashed,-latex] (-4.5,1)--(-4,0.5);
        \draw[thick,dashed,-latex] (-3.5,0)--(-2.9,-0.6);

        \node at (-2.3,0.5) {\scriptsize $\widetilde{\Omega}_1$};
        \node at (-3.5,0.6) {\scriptsize $\widetilde{\Omega}_2$};
        \node at (-4.7,0.5) {\scriptsize $\widetilde{\Omega}_3$};
        \node at (-4.7,-0.5) {\scriptsize $\widetilde{\Omega}_4$};
        \node at (-3.5,-0.6) {\scriptsize $\widetilde{\Omega}_5$};
        \node at (-2.3,-0.5) {\scriptsize $\widetilde{\Omega}_6$};

        \node[below] at (-1,1.8) {\scriptsize $\widehat{\Sigma}_1$};
        \node[below] at (-1,-1.3) {\scriptsize $\widehat{\Sigma}_2$};

        \node[below] at (-2.3,1.5) {\scriptsize $L_1$};
        \node[below] at (-4.6,1.5) {\scriptsize $L_2$};
        \node[below] at (-4.8,-0.9) {\scriptsize $L_3$};
        \node[below] at (-2.2,-0.9) {\scriptsize $L_4$};

        \draw[thick,->] (-0.3,0)--(1,0);

        \draw[thick] (1.5,1.5)--(3,0);
        \draw[thick,-latex] (1.5,1.5)--(2.25,0.75);
        \draw[thick] (3,0)--(4.5,1.5);
        \draw[thick,-latex] (3,0)--(3.9,0.9);

        \draw[thick] (1.5,-1.5)--(3,0);
        \draw[thick,-latex] (1.5,-1.5)--(2.25,-0.75);
        \draw[thick] (3,0)--(4.5,-1.5);
        \draw[thick,-latex] (3,0)--(3.9,-0.9);

        \node[below] at (4.9,1.8) {\scriptsize $L_1$};
        \node[below] at (1.2,1.8) {\scriptsize $L_2$};
        \node[below] at (1.2,-1.1) {\scriptsize $L_3$};
        \node[below] at (4.9,-1.1) {\scriptsize $L_4$};
        \node at (3,-0.3) {$0$};

        \node at (-3.5,-2) {$(a)$};
        \node at (3.1,-2) {$(b)$};
    \end{tikzpicture}
    \caption{In panel (a), four dashed auxiliary rays are added to the jump contour of $\widehat{M}^{loc}(k)$, which is then deformed into the jump contour of the Painlev\'e II model RH problem for $M^P(k)$, as shown in panel (b).}
    \label{dashjump}
\end{figure}

Therefore, the solution $\widehat{M}^{loc}(k)$ is crucial to our analysis. Next, we relate it to the Painlev\'e II equation via an equivalent transformation. For this purpose, we introduce four auxiliary rays $L_j$, $j=1,2,3,4$, emanating from $k=0$ and making an angle $\pi/3$ with the real axis. Together with the real axis and the original contour $\widehat{\Sigma}^{loc}$, these rays divide the complex plane into ten regions, six of which are denoted by $\widetilde{\Omega}_j$, $j=1,\ldots,6$, as shown in Fig. \ref{dashjump}.

We further define
\begin{align}
&P(k) =\left\{\begin{matrix}
b_+,\  \ & k\in \widetilde{\Omega}_1\cup\widetilde{\Omega}_3,\cr
b_-,\  \ & k\in \widetilde{\Omega}_4\cup\widetilde{\Omega}_6,\cr
I,\  \ & \text{elsewhere},
\end{matrix}\right. \nonumber
\end{align}
and  make a transformation
\begin{equation}
\widetilde M^{loc} (k)=\widehat M^{loc} (k)P(k),\label{pre1}
\end{equation}
then we obtain the following RH problem.

\begin{prob}
    Find $\widetilde M^{loc} (k)=\widetilde M^{loc} (k;s )$ with properties
    \begin{itemize}
        \item $\widetilde M^{loc} (k)$ is analytic in $\mathbb{C} \setminus \widetilde \Sigma^{P}$, where $\widetilde \Sigma^{P}=\cup_{j=1}^4 L_{j}$.
        \item $\widetilde M^{loc}(k)$ satisfies the jump condition
        \begin{equation*}
            \widetilde M^{loc}_+(k)
            =
            \widetilde M^{loc}_-(k) \widetilde V^{P}(k),
            \quad
            k\in \widetilde \Sigma^{P},
        \end{equation*}
        where
        \begin{align}
        \widetilde{V}^{P}(k)=
\begin{cases}
b_{+}^{-1}, & k\in L_{1}\cup L_{2},\\
b_{-}, & k\in L_{3}\cup L_{4}.
\end{cases}
        \end{align}

        \item $\widetilde{M}^{loc}(k)\to I$, as $k\to\infty$.
    \end{itemize}
\end{prob}

Let $$\varphi_0=\arg \widetilde{r}(1) .$$  Then $$\widetilde{r}(1)= |\widetilde{r}(1)| e^{i\varphi_0}.$$ The solution  $\widetilde M^{loc}(k)$ can be expressed by
\begin{equation}\label{mrt}
   \widetilde M^{loc}(k) = \sigma_1 e^{- i  \left( \frac{\pi}{4}+\frac{\varphi_0}{2} \right)\widehat\sigma_3} M^P(k)\sigma_1,
\end{equation}
where $M^P(k)$  satisfies a standard Painlev\'{e}  II  model RH problem given in \ref{appx}  with the parameters $p=i|\widetilde{r}(1)|$ and $r=0$. By Proposition \ref{estimate-Nk},  we have
\begin{align}\label{Mloc-es}
     M^{loc}_1(s) = \widetilde M^{loc}_1(s)  + \mathcal{O}(t^{-\frac{1}{3}+4\tau}) =   \sigma_1 e^{- i  \left( \frac{\pi}{4}+\frac{\varphi_0}{2} \right) \widehat\sigma_3} M^P_1(s) \sigma_1 + \mathcal{O}(t^{-\frac{1}{3}+4\tau}) ,
\end{align}
where  the subscript ``$1$'' represents the coefficient  of the  $k^{-1}$ term in the asymptotic expansion of the corresponding solution as $k \to \infty$, and $M^P_1(s)$ is given by \eqref{posee}. Finally, as $k\to\infty$, $M^{loc}(z)$ can be described by the following equation
\begin{align}\label{Mloc-boundary}
M^{loc}(z) = I + \frac{ M^{loc}_1(s)}{(\frac{9}{4}t)^{\frac{1}{3}}(z-1)} + \mathcal{O}(t^{-2\tau}).
\end{align}

\subsubsection{A small-norm RH problem}\label{sec-E}

We consider the error function  $E(z)$ defined by \eqref{Ez}, which admits the following RH problem. For the error RH problem, we orient $\partial\mathcal{U}_{z}(1)$ clockwise.

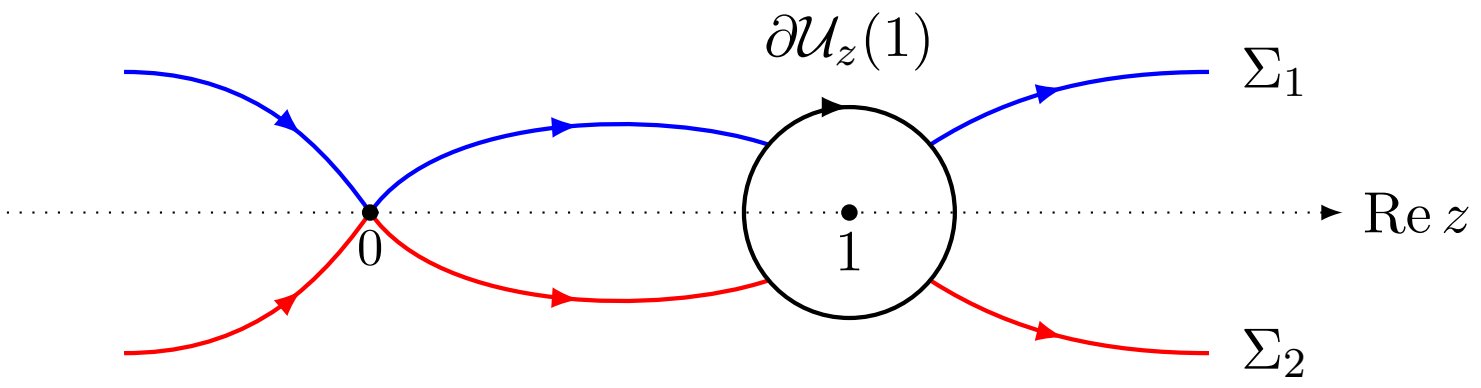
\begin{figure}[htbp]
    \centering
    \begin{tikzpicture}[
        scale=1.55,
        onearrow/.style={
            postaction={decorate},
            decoration={
                markings,
                mark=at position #1 with {\arrow{latex}}
            }
        }
    ]
        \def\CircleRadius{0.45}

        \coordinate (O) at (0.8,0);
        \coordinate (E) at (2.845,0);

        \path (E) ++(140:\CircleRadius) coordinate (UL);
        \path (E) ++(220:\CircleRadius) coordinate (LL);
        \path (E) ++(40:\CircleRadius) coordinate (UR);
        \path (E) ++(-40:\CircleRadius) coordinate (LR);

        \draw[-latex,dotted]
            (-0.75,0)--(4.95,0)
            node[right] {$\operatorname{Re}z$};

        \draw[blue,thick,onearrow=0.65]
            (-0.25,0.6)
            to[out=0,in=125] (O);

        \draw[blue,thick,onearrow=0.55]
            (O)
            .. controls (1.10,0.43) and (2.05,0.44) ..
            (UL);

        \draw[blue,thick,onearrow=0.50]
            (UR)
            .. controls (3.55,0.52) and (3.94,0.60) ..
            (4.38,0.60);

        \draw[red,thick,onearrow=0.65]
            (-0.25,-0.6)
            to[out=0,in=-125] (O);

        \draw[red,thick,onearrow=0.55]
            (O)
            .. controls (1.10,-0.43) and (2.05,-0.44) ..
            (LL);

        \draw[red,thick,onearrow=0.50]
            (LR)
            .. controls (3.55,-0.52) and (3.94,-0.60) ..
            (4.38,-0.60);

        \begin{scope}[shift={(E)}]
            \draw[black,thick,onearrow=0.25]
                (-\CircleRadius,0)
                arc[start angle=180,end angle=-180,radius=\CircleRadius];
        \end{scope}

        \node at (2.845,{\CircleRadius+0.28})
            {$\partial\mathcal{U}_z(1)$};

        \node[right] at (4.43,0.6) {$\Sigma_1$};
        \node[right] at (4.43,-0.6) {$\Sigma_2$};

        \fill (O) circle (1pt)
            node[below,scale=0.9] {$0$};

        \fill (E) circle (1pt)
            node[below] {$1$};
    \end{tikzpicture}
    \caption{The jump contour $\Sigma^E$ for $E(z)$.}
    \label{fig-jump-VE}
\end{figure}

\begin{prob}\label{RHP-E}
  Find $E(z)$ with the properties
          \begin{itemize}
          	\item   $E(z)$ is analytic in $\mathbb{C}\setminus  \Sigma^{E}$,
          where $\Sigma^E = \partial \mathcal{U}_z(1)  \cup \left(\Sigma^{(3)} \setminus \mathcal{U}_z(1) \right)$. See Fig. \ref{fig-jump-VE}.
          	\item  $E(z)$  satisfies the jump condition
          	\begin{align*}
          		E_{+}(z)=E_-(z)V^{E}(z), \quad z\in \Sigma^{E},
          	\end{align*}
          	where the jump matrix $V^{E}(z)$ is given by
\begin{align}\label{jump-VE}
    V^E(z)=
    \begin{cases}
        M^{out}(z)V^{(3)}(z)M^{out}(z)^{-1},
        & z\in\Sigma^{(3)}\setminus\mathcal{U}_z(1),\\
        M^{out}(z)M^{loc}(z)M^{out}(z)^{-1},
        & z\in\partial\mathcal{U}_z(1).
    \end{cases}
\end{align}

          	\item       $E(z)=I+\mathcal{O}(z^{-1}),	\quad  z\to  \infty.$	

          \end{itemize}
\end{prob}

To obtain the existence of the solution $E(z)$,  we estimate the jump matrix $V^{E}(z)-I$.

\begin{proposition}\label{VE-I}
Let $2 \le p \le \infty$ and $0<\tau<\frac{1}{15}$. It follows that
\begin{align}\label{vee}
    &\|V^E-I\|_{L^p(\Sigma^E\setminus \partial \mathcal U_z(1))}
    =\mathcal{O}(e^{-ct^{3\tau}}),\\
    &\|V^E-I\|_{L^p(\partial\mathcal U_z(1))}
    =\mathcal{O}(t^{-\kappa_p}),\label{vee-2}
\end{align}
where $c>0$ is a constant and $\kappa_p = \tau + \frac{1/3-\tau}{p}, 2 \leq p \leq \infty$.

\end{proposition}
\begin{proof}
First, we consider the estimate when $p=\infty$.
For $z\in \Sigma^{E} \setminus \partial \mathcal{U}_{z}(1)$, by \eqref{jump-VE}, Proposition \ref{prop-Re} and the uniform bounds for the outer model and its inverse,
\begin{equation*}
|V^{E}(z)-I| \lesssim |V^{(3)}(z)-I| \lesssim   e^{-c t^{3\tau}},
\end{equation*}
where $c = c(\phi,\xi)$.
For $z \in \partial\mathcal{U}_{z}(1) $, by \eqref{jump-VE},
\begin{equation*}
|V^{E}(z)-I| \lesssim |M^{loc}(z)-I| \lesssim  t^{-\tau}.
\end{equation*}
Other cases when $2 \le p <\infty$ can be proven similarly.
\end{proof}

This proposition establishes RH problem \ref{RHP-E} as a small-norm RH problem, for which there exists a well-known existence and uniqueness theorem \cite{PX3}. Define the integral operator $ C_{w^{E}}: L^2(\Sigma^E) \to L^2(\Sigma^E)$ by
\begin{align*}
C_{w^{E}}f =C_-\left( f \left( V^{E}(z)-I \right) \right),
\end{align*}
where $w^{E} =V^{E}(z)-I$ and  $C_-$ is the  Cauchy projection operator on $\Sigma^{E}$. By \eqref{vee}-\eqref{vee-2}, a simple calculation shows that
       $$||C_{w^{E}}||_{L^2_{op}(\Sigma^{E})}\lesssim  ||C_-||_{L^2_{op}(\Sigma^{E})}||V^{E}(z)-I||_{L^\infty(\Sigma^{E})} \lesssim t^{-\tau}.$$
According to the theorem of Beals-Coifman \cite{BC84}, the solution of RH problem \ref{RHP-E} can be expressed in terms of
\begin{equation*}
E(z)=I +\frac{1}{2\pi i} \int_{\Sigma^{E}} \frac{\mu_E(\zeta) \left( V^{E}(\zeta)-I\right)}{\zeta-z}\, \mathrm{d}\zeta,
\end{equation*}
where $\mu_E -I \in L^2\left( \Sigma^{E}\right)$   satisfies $\left(1-C_{w^{E}}\right)(\mu_E-I)=C_{w^{E}}I$. Furthermore, from \eqref{vee}-\eqref{vee-2}, we have the estimates
       \begin{equation}\label{owej}
        ||V^{E}(z)-I||_{L^2(\Sigma^{E})}\lesssim t^{-\frac{1}{6}-\frac{\tau}{2}},\quad
          ||\mu_E-I||_{L^2(\Sigma^{E})} \lesssim  t^{-\frac{1}{6}-\frac{\tau}{2}},
       \end{equation}
which imply that  RH problem \ref{RHP-E} has a unique solution.

To recover the potential $q(x, t)$ of \eqref{recon-sol}, the behavior of $E(z)$ at $z=0$ and $z\to\infty$ must be characterized. We  first derive the expansion of  $E(z)$ as $z\to\infty$
\begin{equation}\label{E-expa}
E(z)=I + z^{-1}E_1 +\mathcal{O}\left(z^{-2}\right),
\end{equation}
where
\begin{equation}
 E_1=-\frac{1}{2\pi i} \int_{\Sigma^{E}} \mu_E(\zeta)\left( V^{E}(\zeta)-I \right)\, \mathrm{d}\zeta.	
\end{equation}

\begin{proposition}
$E_1$ and $E(0)$ can be estimated as follows:
\begin{align}\label{E1-es}
    &E_1=\left(\frac{9}{4}t\right)^{-1/3}M^{out}(1) \widetilde M^{loc}_1(s) M^{out}(1)^{-1}+\mathcal{O}(t^{-1/3-\tau}),\\
    &E(0)=I-\left(\frac{9}{4}t\right)^{-1/3}M^{out}(1) \widetilde M^{loc}_1(s) M^{out}(1)^{-1}+\mathcal{O}(t^{-1/3-\tau}),\label{E0-es}
\end{align}
where
\begin{align}\label{Mloc-1-explicit}
    \widetilde M^{loc}_1(s)=\frac{1}{2}
    \begin{pmatrix}
        i\int_s^\infty u(\zeta)^2\,d\zeta & iu(s)e^{i\varphi_0}\\
        -iu(s)e^{-i\varphi_0} & -i\int_s^\infty u(\zeta)^2\,d\zeta
    \end{pmatrix}.
\end{align}
\end{proposition}

\begin{proof}
By \eqref{owej}, we obtain
\begin{align*}
    &E_1 =-\frac{1}{2\pi i}\oint_{\partial\mathcal{U}_z(1)}\bigl(V^E(\zeta)-I\bigr)\,d\zeta+\mathcal{O}(t^{-1/3-\tau}),\\
    &E(0) =I+\frac{1}{2\pi i}\oint_{\partial\mathcal{U}_z(1)}\frac{V^E(\zeta)-I}{\zeta}\,d\zeta+\mathcal{O}(t^{-1/3-\tau}).
\end{align*}
The outer model and its inverse are analytic and uniformly bounded near $1$. Substituting \eqref{Mloc-boundary} into \eqref{jump-VE} and evaluating the residues on the clockwise circle $\partial\mathcal{U}_z(1)$ give \eqref{E1-es}--\eqref{E0-es} by \eqref{Mloc-es} and $0<\tau<1/15$.
\end{proof}

\subsection{Contribution from a pure $\bar{\partial}$-problem}\label{sec-Dbar}

In this subsection, we consider the long-time asymptotic behavior of the pure $\bar{\partial}$-problem. Define the function
\begin{equation}\label{tran45}
M^{(4)}(z)=M^{(3)}(z)\left(M^{rhp}(z)\right)^{-1},
\end{equation}
which  satisfies the following $\bar{\partial}$-problem.

\begin{prob3}\label{pure-Dbar}
 Find  $M^{(4)}(z):=M^{(4)}(z;x,t)$ which satisfies
            \begin{itemize}
                \item   $M^{(4)}(z)$ is continuous  in $\mathbb{C}$ and analytic in $\mathbb{C}\setminus \overline{\Omega}$.
                \item    $M^{(4)}(z)=I+\mathcal{O}(z^{-1}), \ z\to  \infty$.	

                \item For $z\in \mathbb{C}$,  $M^{(4)}(z)$ satisfies the $ \bar{\partial}$-equation
                \begin{equation*}
                    \bar{\partial} M^{(4)}(z) = M^{(4)}(z) W^{(4)}(z),
                \end{equation*}
                where
	\begin{align}
	W^{(4)}(z):=M^{rhp}(z)  \bar{\partial} R^{(2)}(z) \left(M^{rhp}(z)\right)^{-1},
	\end{align}
 and  $ \bar{\partial}  R^{(2)}(z)$ has been given in \eqref{Dbar-R3}.
            \end{itemize}

\end{prob3}

The solution of $ \bar{\partial}$-problem   \ref{pure-Dbar} can be given by
\begin{equation} \label{M5}
	M^{(4)}(z)=I-\frac{1}{\pi}  \iint_\mathbb{C} \frac{ M^{(4)}(\zeta) W^{(4)}(\zeta)}{\zeta-z} \, \mathrm{d} A(\zeta),
\end{equation}
where $\mathrm{d} A(\zeta)$ is Lebesgue measure on the plane. Eq. \eqref{M5} can be written as an operator equation
\begin{equation}\label{operator-S}
	(I-S) M^{(4)}(z)=I,
\end{equation}
where $S$ is the solid Cauchy operator
\begin{equation}\label{Sf}
	Sf(z)=-\frac{1}{\pi} \iint_\mathbb{C} \frac{f(\zeta)W^{(4)}(\zeta)}{\zeta-z} \, \mathrm{d} A(\zeta).
\end{equation}

\begin{proposition}\label{prop-S}
The operator $S$ defined by \eqref{Sf}  satisfies the estimate
\begin{equation}
  \parallel S\parallel_{L^\infty(\mathbb{C})\to L^\infty(\mathbb{C})}\lesssim  t^{-1/6},
\end{equation}
	which implies the existence of $(I-S)^{-1}$ for sufficiently large $t$.
\end{proposition}	
\begin{proof}
By \eqref{Dbar-R3}, the support of $\bar{\partial}R^{(2)}$ lies in the closed lenses and stays a positive distance from the discrete poles. Since $s$ in \eqref{scaled} stays bounded, the Painlev\'e II estimates and \eqref{Nk-est} give uniform bounds for $M^{loc}$ and its inverse in $\mathcal{U}_z(1)$. Applying these bounds and those for $M^{out}$, $E$, and their inverses in the factorization \eqref{Ez} gives
\begin{align*}
    \bigl|M^{rhp}(z)\bigr|+\bigl|(M^{rhp}(z))^{-1}\bigr| &\lesssim 1,\quad z\in\operatorname{supp}\bigl(\bar{\partial}R^{(2)}\bigr)\setminus\Sigma^{(3)}.
\end{align*}
Hence $\lvert W^{(4)}(z)\rvert\lesssim\lvert\bar{\partial}R^{(2)}(z)\rvert$.

We restrict the integration in $S$ to $\Omega_1\cap\{\operatorname{Re}\zeta>z_1\}$ and retain the notation $S$. The other cases can be treated similarly. Let $y=\operatorname{Im}z$ and write
\begin{align*}
\zeta=z_1+u+iv,\quad v\geq 0,\quad u\geq \kappa v,
\end{align*}
where $\kappa=\cot\phi>0$. By Propositions \ref{prop-Re} and \ref{def-Rj},
\begin{align*}
\left|W^{(4)}(\zeta)\right|
\lesssim
\left(
\left|r'(z_1+u)\right|
+\left|u+iv\right|^{-\frac12}
\right)e^{-ctu^2v}.
\end{align*}
It follows that
\begin{align*}
|Sf(z)|
\lesssim
\|f\|_{L^\infty(\mathbb{C})}(I_1+I_2),
\end{align*}
where
\begin{align*}
I_1&=\int_0^\infty\int_{\kappa v}^\infty\frac{\left|r'(z_1+u)\right|e^{-ctu^2v}}{\left|z_1+u+iv-z\right|}\mathrm{d}u\mathrm{d}v, \\
I_2&=\int_0^\infty\int_{\kappa v}^\infty\frac{|u+iv|^{-\frac12}e^{-ctu^2v}}{\left|z_1+u+iv-z\right|}\mathrm{d}u\mathrm{d}v.
\end{align*}	
Since $u\geq\kappa v$, the Cauchy--Schwarz inequality in $u$ gives
\begin{align}
I_1 \lesssim \|r'\|_{L^2(\mathbb{R})}\int_0^\infty e^{-ctv^3}|v-y|^{-\frac12}\mathrm{d}v \lesssim t^{-\frac16}.\nonumber
\end{align}
For $I_2$, choose $p>2$ and let $p^{-1}+q^{-1}=1$. The elementary estimates
\begin{align}
\left\||u+iv|^{-\frac12}\right\|_{L^p(\kappa v,\infty)} \lesssim v^{\frac1p-\frac12}, \quad \left\||z_1+u+iv-z|^{-1}\right\|_{L^q(\kappa v,\infty)} \lesssim |v-y|^{\frac1q-1}, \nonumber
\end{align}
and H\"older's inequality yield
\begin{align}
I_2 \lesssim \int_0^\infty v^{\frac1p-\frac12}|v-y|^{-\frac1p}e^{-ctv^3}\mathrm{d}v \lesssim t^{-\frac16}.\nonumber
\end{align}
Both estimates are uniform in $y\in\mathbb{R}$.

\end{proof}

Proposition \ref{prop-S}  implies that the operator  equation \eqref{operator-S}  has a unique solution, which can be expanded in the form
\begin{equation}\label{M5-expa}
	M^{(4)}(z)=I + z^{-1} M^{(4)}_1(x,t) +\mathcal{O}(z^{-2}), \quad z\to \infty,
\end{equation}
where
\begin{align}
M^{(4)}_1(x,t)=\frac{1}{\pi} \iint_{\mathbb{C}} M^{(4)}(\zeta)W^{(4)}(\zeta)\, \mathrm{d} A(\zeta).
\end{align}
Take $z=0$ in \eqref{M5}, then
\begin{equation}\label{M50}
	M^{(4)}(0)=I-\frac{1}{\pi}  \iint_\mathbb{C} \frac{ M^{(4)}(\zeta) W^{(4)}(\zeta)}{\zeta}  \mathrm{d} A(\zeta).
\end{equation}

\begin{proposition}\label{prop-M5-est}
We have the following estimates
\begin{equation}
		|M^{(4)}_1(x,t)| \lesssim t^{-1/2}, \quad |M^{(4)}(0)-I| \lesssim t^{-1/2}.
\end{equation}
\end{proposition}
\begin{proof}
Similarly to the proof of Proposition \ref{prop-S}, we take $z \in \Omega_1\cap\{\operatorname{Re}z>z_1\}$ as an example. By the uniform boundedness of $M^{(4)}$, $M^{rhp}$, and $(M^{rhp})^{-1}$, it suffices to estimate the corresponding area integrals of $\bar{\partial}R^{(2)}$ with their exponential factors. Thus, we have	
\begin{align}\label{m51esti}
	|M^{(4)}_1(x,t)| \lesssim I_1+I_2,
\end{align}
where
\begin{align*}
I_1&= \int_0^\infty\int_{\kappa v}^\infty |r'(z_1+u)|e^{-ctu^2v}\mathrm{d}u\mathrm{d}v,\\
I_2&= \int_0^\infty\int_{\kappa v}^\infty |u+iv|^{-\frac12}e^{-ctu^2v}\mathrm{d}u\mathrm{d}v.	
\end{align*}
The Cauchy--Schwarz inequality in $u$ gives
\begin{align*}
I_1 \lesssim t^{-\frac14}\|r'\|_{L^2(\mathbb{R})}\int_0^\infty v^{-\frac14}e^{-ctv^3}\mathrm{d}v \lesssim t^{-\frac12}.
\end{align*}
The change of variables $(u,v)=t^{-\frac13}(U,V)$ yields
\begin{align*}
I_2 = t^{-\frac12}\int_0^\infty\int_{\kappa V}^\infty |U+iV|^{-\frac12}e^{-cU^2V}\mathrm{d}U\mathrm{d}V \lesssim t^{-\frac12}.
\end{align*}

The corresponding estimates in Propositions \ref{prop-Re} and \ref{def-Rj} also give
\begin{align}
    \iint_{\Omega}\frac{|W^{(4)}(\zeta)|}{|\zeta|}\,dA(\zeta)\lesssim t^{-1/2}.\nonumber
\end{align}
Together with the uniform boundedness of $M^{(4)}$ and its integral representation at $z=0$, this yields $|M^{(4)}(0)-I|\lesssim t^{-1/2}$.

\end{proof}

To recover the potential via  the reconstruction formula \eqref{recon-sol}, we require an estimate of  $M^{(2)}(0)$.

\begin{proposition} As $t \to +\infty$, $M^{(2)}(0)$ satisfies the   estimate
	\begin{align}\label{m30}
		M^{(2)}(0)=E(0)M^{out}(0)+\mathcal{O}(t^{-1/2}),
	\end{align}
where $E(0)$ is given by \eqref{E0-es}.
\end{proposition}
\begin{proof}
Since $0\notin\mathcal{U}_z(1)$ and $R^{(2)}(0)=I$, \eqref{tran34}, \eqref{Ez}, and \eqref{tran45} give
\begin{align}
    M^{(2)}(0)=M^{(4)}(0)E(0)M^{out}(0).\nonumber
\end{align}
Applying Proposition \ref{prop-M5-est} and the uniform bounds for $E(0)$ and $M^{out}(0)$ gives \eqref{m30}.
\end{proof}

\subsection{Proof of Theorem \ref{Theo-1}--Case I}\label{sec-proof}

We now prove the long-time asymptotics of the solution to the defocusing Hirota equation \eqref{Hirota}-\eqref{Hirota-bc} in $\mathcal{P}_{<-4}$.

\begin{proof}
Inverting the sequence of transformations \eqref{tran01}, \eqref{tran23}, \eqref{tran34}, \eqref{Ez}, and \eqref{tran45}, and taking $z\to\infty$ vertically so that $R^{(2)}(z)=I$, we obtain
\begin{align}
M(z)=T(\infty)^{\sigma_3}\left(I+\frac{\sigma_2}{z}M^{(2)}(0)^{-1}\right)M^{(4)}(z)E(z)M^{out}(z)T(z)^{-\sigma_3}.\label{sec35-M-factorization}
\end{align}
Substituting the asymptotic expansions \eqref{T-infty},  \eqref{E-expa}, and \eqref{M5-expa} into the above formula and using Proposition \ref{prop-M5-est}, the reconstruction formula \eqref{recon-sol} yields
\begin{align}
q(x,t)=-iT(\infty)^2\left(\sigma_2M^{(2)}(0)^{-1}+M_1^{out}+E_1\right)_{12}+\mathcal{O}(t^{-1/3-\tau}).\label{sec35-q-reconstruction}
\end{align}
Here, the $z^{-1}$-coefficient arising from $T(z)^{-\sigma_3}$ is diagonal and hence does not contribute to the $(1,2)$-entry. Moreover, the $\mathcal{O}(t^{-1/2})$ contribution from $M_1^{(4)}(x,t)$ is absorbed into $\mathcal{O}(t^{-1/3-\tau})$ since $0<\tau<1/15$.

Finally, \eqref{m30} gives $M^{(2)}(0)^{-1}=M^{out}(0)^{-1}E(0)^{-1}+\mathcal{O}(t^{-1/2})$. Substituting \eqref{E1-es}-\eqref{E0-es} into \eqref{sec35-q-reconstruction}, and using \eqref{soliton-reduction}, we obtain
\begin{align}\label{sec35-q-final}
q(x,t) = q^{sol}_\Lambda(x,t)-iT(\infty)^2\left(\frac{9t}{4}\right)^{-1/3} f_{12}+\mathcal{O}\left(t^{-1/3-\tau}\right),
\end{align}
where $f_{12}=\left[\left(I+\sigma_2M^{out}(0)^{-1}\right)M^{out}(1)\widetilde{M}^{loc}_1(s)M^{out}(1)^{-1}\right]_{12}$.  Further   denoting
\begin{align}
    M^{out}(j)
    =\begin{pmatrix}
        \alpha_j & \beta_j\\
        \overline{\beta}_j & \overline{\alpha}_j
    \end{pmatrix}, \ \ j=0,1, \label{mout}
    \end{align}
and  using  \eqref{Mloc-1-explicit},  we obtain
\begin{align}
    f_{12}=\frac{i}{2}\Big\{&
    \alpha_1\bigl[(1+i\overline{\beta}_0)\alpha_1
        -i\alpha_0\overline{\beta}_1\bigr]u(s)e^{i\varphi_0}\nonumber\\
    &+\beta_1\bigl[(1+i\overline{\beta}_0)\beta_1
        -i\alpha_0\overline{\alpha}_1\bigr]u(s)e^{-i\varphi_0}\nonumber\\
    &+\bigl[i\alpha_0\bigl(|\alpha_1|^2+|\beta_1|^2\bigr)
        -2(1+i\overline{\beta}_0)\alpha_1\beta_1\bigr]
        \int_s^\infty u(\zeta)^2\,\mathrm{d}\zeta\Big\},
\end{align}
where   $u(s)$ is a solution of the Painlev\'e II equation, together with its Airy asymptotics, follows from \ref{appx} with the parameters $p=i|\widetilde r(1)|$ and $r=0$. This proves Case I of Theorem \ref{Theo-1}.
\end{proof}

\section{Painlev\'e Asymptotics in the Transition Region $\mathcal{P}_{-8}$}\label{sec-P-8}

We now consider the asymptotics in the region $\mathcal{P}_{<-8}$ given by
\begin{align}
\mathcal{P}_{<-8}:=\left\{
(x,t)\in\mathbb{R}\times\mathbb{R}^{+}:
-C<\left(\frac{x}{t}+8\right)t^{2/3}<0
\right\},
\end{align}
where $C>0$ is a constant, which corresponds to the case in Fig. \ref{proptheta} (a). In this region, the phase function $\theta(z)$ has four real stationary points
$z_j$, $j=1,2,3,4$, satisfying
\begin{align}
z_4<-1<z_3<0<z_2<1<z_1,
\quad z_1z_2=z_3z_4=1.
\end{align}
As $t\to+\infty$ with $(x,t)\in\mathcal P_{<-8}$, the two negative stationary points $z_3$ and $z_4$ coalesce at $z=-1$ at least at the rate $t^{-1/3}$. The two positive stationary points satisfy
\begin{align}
z_j&=z_j^*+\mathcal{O}(t^{-2/3}),\quad j=1,2,\\
z_{1,2}^*&=\frac{4+\sqrt{10}\pm\sqrt{8\sqrt{10}-10}}{6},
\end{align}
where $0<z_2^*<1<z_1^*$. Thus, $z_1$ and $z_2$ remain simple and uniformly separated from each other and from the coalescing pair. As the analysis is similar to that given in Section \ref{sec-P-4}, we only give a sketch of the procedure in this case.

In this section, we continue to use the notation established previously (e.g., $T$, $M^{(1)}$, $G$, \dots). Note that these symbols may take on different meanings depending on the context, but we believe this will not cause confusion.

\subsection{Regularization of the basic RH problem}

In this subsection, we make two successive transformations to the basic RH problem \ref{RH1} to obtain a regular RH problem without poles and singularities.

{\bf Step 1: Removing poles.}
Opening lenses along $J=\mathbb{R}\setminus[z_2,z_1]$ requires the second factorization in \eqref{V_factorization}. To this end, we introduce the scalar function
\begin{align}
T(z) = \exp\left\{
-i\int_J \nu(\zeta)
\left(\frac{1}{\zeta-z}-\frac{1}{2\zeta}\right)\,d\zeta
\right\}, \quad
\nu(\zeta)=-\frac{1}{2\pi}\log(1-|r(\zeta)|^2),
\end{align}
which has the following properties.

\begin{lemma}
The function $T(z)$ has the following properties:
\begin{itemize}
\item $T(z)$ is analytic in $\mathbb{C}\setminus \overline{J}$.
\item $\overline{T(\overline{z})} = T(z)^{-1} = T(z^{-1})$.	
\item $T(z)$ satisfies the jump condition
    \begin{align}
    T_-(z)= T_+(z)(1-|r(z)|^2), \quad  z\in J
    \end{align}
\item Let
    \begin{align}
    T(\infty):=\lim_{z\to\infty}T(z)
    =\exp\left\{
    \frac{i}{2}\int_J\frac{\nu(\zeta)}{\zeta}\,d\zeta
    \right\}.
    \end{align}
    Then $|T(\infty)|^2=1$, $T(0)=T(\infty)^{-1}$, and the
    asymptotic expansion at infinity is
    \begin{align}
    T(z)=T(\infty)\left[
    1+\frac{i}{z}\int_J\nu(\zeta)\,d\zeta+\mathcal{O}(z^{-2})
    \right].
    \end{align}
\item The ratio $a(z)/T(z)$ is holomorphic in $\mathbb{C}^+$
and extends holomorphically to a neighborhood of $-1$,
where it is bounded and nonzero. Its boundary values satisfy
\begin{align}
\left|\frac{a(z)}{T_+(z)}\right|=1,
\quad z\in J,
\end{align}
where $T_+$ denotes the boundary value from $\mathbb{C}^+$
and the quotient at $z=-1$ is understood by continuous extension.
\end{itemize}
\end{lemma}

\begin{figure}[htbp]
\begin{center}
\begin{tikzpicture}[scale=1.0]

\draw[dotted] (0,0) circle [radius=2];

\draw[-latex] (-4.5,0)--(4.5,0);

\node at (5.1,0) {\footnotesize $\operatorname{Re} z$};
\node at (-2.55,-0.35) {\footnotesize $-1$};
\node at (0,-0.35) {\footnotesize $0$};
\node at (2.35,-0.35) {\footnotesize $1$};

\node at (-1.30,1.25) {\footnotesize $\eta_j$};
\node at (-1.30,-1.25) {\footnotesize $\overline{\eta}_j$};

\coordinate (K) at (0,0);
\coordinate (L) at (2,0);
\coordinate (M) at (-2,0);

\fill[blue] (K) circle (1.5pt);
\fill[blue] (L) circle (1.5pt);
\fill[blue] (M) circle (1.5pt);

\coordinate (A) at (-1.94,0.48);
\coordinate (B) at (-1.10,1.67);
\coordinate (C) at (0,2.00);
\coordinate (D) at (1.38,1.45);
\coordinate (E) at (1.82,0.82);

\coordinate (F) at (-1.94,-0.48);
\coordinate (G) at (-1.10,-1.67);
\coordinate (H) at (0,-2.00);
\coordinate (I) at (1.38,-1.45);
\coordinate (J) at (1.82,-0.82);

\fill[red] (A) circle (1.5pt);
\fill[red] (B) circle (1.5pt);
\fill[red] (C) circle (1.5pt);
\fill[red] (D) circle (1.5pt);
\fill[red] (E) circle (1.5pt);

\fill[red] (F) circle (1.5pt);
\fill[red] (G) circle (1.5pt);
\fill[red] (H) circle (1.5pt);
\fill[red] (I) circle (1.5pt);
\fill[red] (J) circle (1.5pt);

\draw (A) circle [radius=0.25];
\draw (B) circle [radius=0.25];
\draw (C) circle [radius=0.25];
\draw (D) circle [radius=0.25];
\draw (E) circle [radius=0.25];

\draw (F) circle [radius=0.25];
\draw (G) circle [radius=0.25];
\draw (H) circle [radius=0.25];
\draw (I) circle [radius=0.25];
\draw (J) circle [radius=0.25];

\draw[->] (-1.69,0.48)
    to[out=90,in=0] (-1.94,0.73);

\draw[->] (-0.85,1.67)
    to[out=90,in=0] (-1.10,1.92);

\draw[->] (0.25,2.00)
    to[out=90,in=0] (0,2.25);

\draw[->] (1.63,1.45)
    to[out=90,in=0] (1.38,1.70);

\draw[->] (2.07,0.82)
    to[out=90,in=0] (1.82,1.07);

\draw[->] (-1.94,-0.73)
    to[out=180,in=270] (-2.19,-0.48);

\draw[->] (-1.10,-1.92)
    to[out=180,in=270] (-1.35,-1.67);

\draw[->] (0,-2.25)
    to[out=180,in=270] (-0.25,-2.00);

\draw[->] (1.38,-1.70)
    to[out=180,in=270] (1.13,-1.45);

\draw[->] (1.82,-1.07)
    to[out=180,in=270] (1.57,-0.82);

\end{tikzpicture}
\end{center}
\caption{\footnotesize The jump contour $\Sigma^{(1)}$.}
\label{circle-jump}
\end{figure}
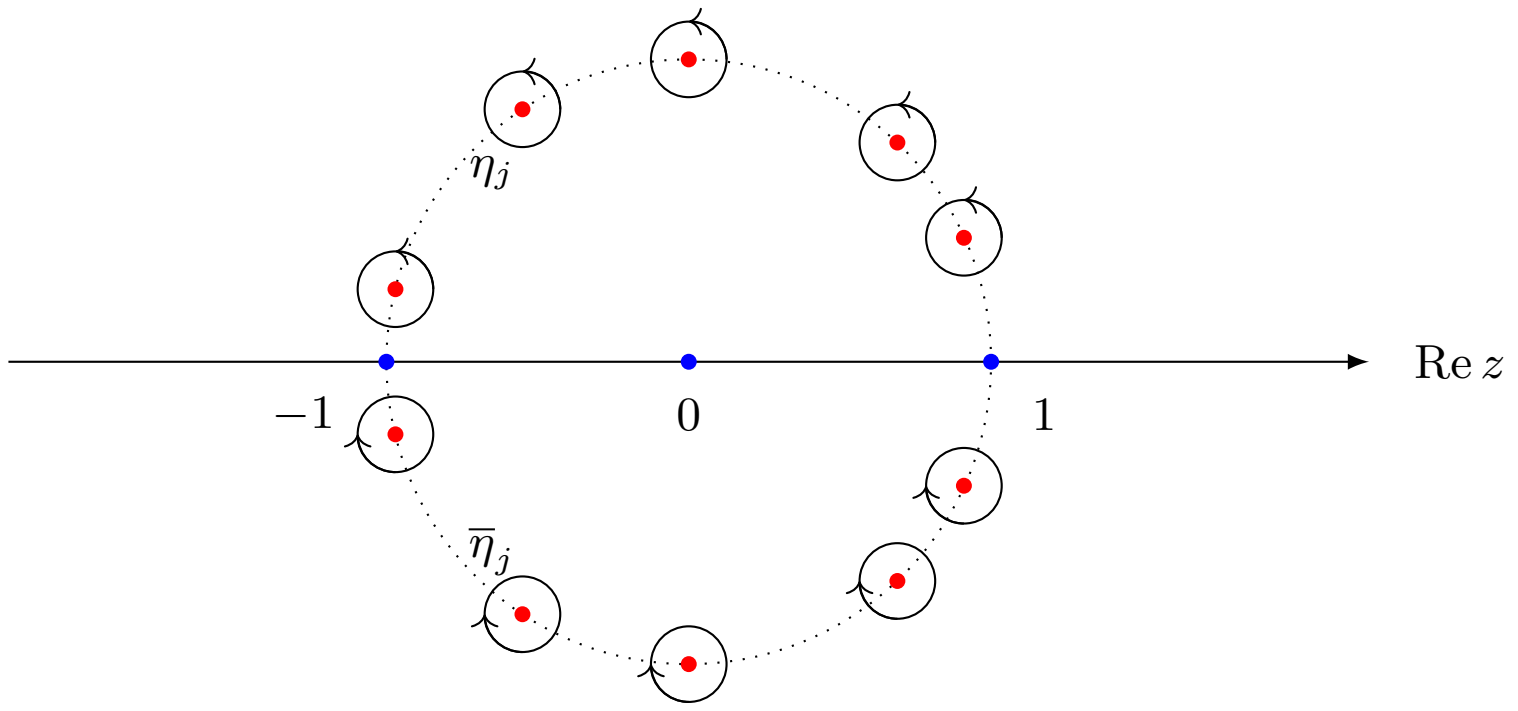

Define
\begin{align}
    0<\rho<\frac{1}{2}\min\{
    \min_{\substack{\eta_j,\eta_l\in\mathcal{Z}^+\\ j\neq l}}
    |\eta_j-\eta_l|,
    \min_{\eta_j\in\mathcal{Z}^+}|\operatorname{Im}\eta_j|,
    \min_{\substack{\eta_j\in\mathcal{Z}^+, z\in\mathcal{L}}}
    |\eta_j-z|\}.
\end{align}
For $\eta_j \in \mathcal{Z}^+$ and $\overline{\eta}_j \in \mathcal{Z}^-$, we make small circles  $C_j$ and  $\overline{C}_j$ centered at  $\eta_j $ and  $\overline{\eta}_j$ respectively,  with a radius of $\rho$. The circles are oriented counterclockwise in $\mathbb{C}^+$ and clockwise in $\mathbb{C}^-$. Define the contour
\begin{equation}
	\Sigma^{(1)}=\mathbb{R} \cup \left( \bigcup_{j=1}^{N} \left(  C_j \cup \overline{C}_j \right) \right),
\end{equation}
where $\mathbb{R}$ is oriented left-to-right and the boundaries of the corresponding disks $D_j$ and  $\overline{D}_j$ are oriented counterclockwise in $\mathbb{C}^+$ and clockwise in $\mathbb{C}^-$, as depicted in Fig. \ref{circle-jump}.

To remove poles on the unit circle $|z|=1$ by converting their residues into jumps, we introduce the interpolation function
\begin{align}\label{G-P-8}
G(z)=
\begin{cases}
\begin{pmatrix}
1 & 0\\[2mm]
-\dfrac{c_j e^{-2it\theta(\eta_j)}}{z-\eta_j} & 1
\end{pmatrix},
& z\in D_j,\\[5mm]
\begin{pmatrix}
1 & -\dfrac{\bar{c}_j e^{2it\theta(\bar{\eta}_j)}}{z-\bar{\eta}_j}\\[2mm]
0 & 1
\end{pmatrix},
& z\in\overline{D_j},\\[5mm]
I,
& \text{elsewhere}.
\end{cases}
\end{align}

Considering two factorizations of the jump matrix $V(z)$ in \eqref{V_factorization}, denote the following matrices
\begin{align}
&\left(\begin{array}{cc}
1 & -\overline{r(z)}T(z)^{-2}e^{2it\theta(z)}\\
0 & 1
\end{array}\right)
\left(\begin{array}{cc}
1 & 0\\
r(z)T(z)^2e^{-2it\theta(z)} & 1
\end{array}\right)
:=b_-^{-1}b_+,\\
&\left(\begin{array}{cc}
		1 & 0\\
		\frac{ r(z) }{1-|r|^2}  T_-(z)^{2}  e^{ -2it\theta(z)} & 1
	\end{array}\right)\left(\begin{array}{cc}
	1 & -\frac{ \overline{r(z)}}{1-|r|^2}T_+(z)^{-2} e^{2it\theta(z)}\\
	0 & 1
\end{array}\right):=B_-^{-1}B_+.
\end{align}
We make the following transformation
\begin{align}
M^{(1)}(z)=T(\infty)^{-\sigma_3}M(z)G(z)T(z)^{\sigma_3}.
\end{align}
Then $M^{(1)}$ has no poles, and its jumps on the small circles differ from $I$ by $\mathcal{O}(e^{-ct})$ for some $c>0$. The jumps on the small circles can be removed by introducing $M^{(2)}$, related to $M^{(1)}$ by
\begin{align}
M^{(1)}(z)=M^{(2)}(z)\left(I+\mathcal{O}\left(e^{-ct}\right)\right).
\end{align}
Set $\Gamma = (-\infty,z_4)\cup(z_3,z_2)\cup(z_1,\infty)$. The resulting function $M^{(2)}$ satisfies the following RH problem.

\begin{prob}
Find a function $M^{(2)}(z)=M^{(2)}(z;x,t)$ which satisfies
\begin{itemize}
\item  $M^{(2)}(z)$ is analytic in $\mathbb{C}\setminus \mathbb{R}$.
\item Jump condition:
\begin{align}
M_{+}^{(2)}(z)=M_{-}^{(2)}(z)V^{(2)}(z), \quad z\in\mathbb{R},\nonumber
\end{align}
where
\begin{align}\label{jump-V2-P-8}
V^{(2)}(z)=
\begin{cases}
B_-^{-1}B_+,
& z\in\Gamma,
\\[2mm]
b_-^{-1}b_+,
& z\in(z_2,z_1),
\\[2mm]
T_-(z)^{-\sigma_3}V(z)T_+(z)^{\sigma_3},
& z\in(z_4,z_3).
\end{cases}
\end{align}
\item Asymptotic behaviors:
\begin{align}
M^{(2)}(z)&=I+\mathcal{O}\left(z^{-1}\right),
\quad z\to\infty,
\\
M^{(2)}(z)&=
\frac{\sigma_{2}}{z}+\mathcal{O}(1),
\quad z\to0.
\end{align}	
\end{itemize}
\end{prob}

{\bf Step 2: Removing singularities.}
To remove the singularity at $z=0$, introduce
\begin{align}
M^{(2)}(z)=\left( I+ \frac{\sigma_2}{z} M^{(3)}(0)^{-1} \right ) M^{(3)}(z),
\end{align}
where $M^{(3)}(z)$ is the solution of the following RH problem.
\begin{prob}
Find  $M^{(3)}(z)=M^{(3)}(z;x,t)$ with properties
\begin{itemize}
\item  $M^{(3)}(z)$ is  analytic  in $\mathbb{C}\setminus \mathbb{R}$.
\item Jump condition:
            $M^{(3)}_+(z)=M^{(3)}_-(z)V^{(2)}(z),$ where $V^{(2)}(z)$ is given by   \eqref{jump-V2-P-8}.	
\item Asymptotic behaviors: $  M^{(3)}(z)=I+\mathcal{O}(z^{-1}),	\quad  z \to  \infty.$
\end{itemize}
\end{prob}

\subsection{Transformation to a hybrid $\bar{\partial}$-RH problem}

In this part, we transform the RH problem for $M^{(3)}(z)$ into a hybrid $\bar{\partial}$-RH problem for $M^{(4)}(z)$ by constructing continuous extensions of the triangular factors of the jump matrix $V^{(2)}(z)$.

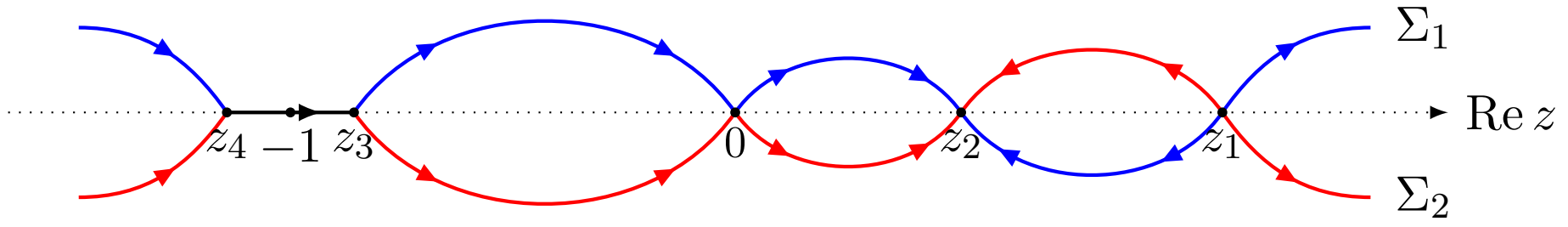
\begin{figure}[htbp]
    \centering
    \begin{tikzpicture}[
        scale=1.10,
        onearrow/.style={
            postaction={decorate},
            decoration={
                markings,
                mark=at position #1 with {\arrow{latex}}
            }
        },
        pics/centerarrow/.style={
            code={
                \fill (3pt,0)--(-1.5pt,2pt)--(-1.5pt,-2pt)--cycle;
            }
        }
    ]

        \coordinate (Z4) at (-3.60,0);
        \coordinate (M)  at (-3.15,0);
        \coordinate (Z3) at (-2.70,0);
        \coordinate (O)  at (0,0);
        \coordinate (Z2) at (1.60,0);
        \coordinate (Z1) at (3.45,0);

        \draw[-latex,dotted]
            (-5.15,0)--(5.05,0)
            node[right] {$\operatorname{Re}z$};

        \draw[blue,thick]
            (-4.65,0.6)
            to[out=0,in=125]
            pic[pos=0.50,sloped,allow upside down] {centerarrow}
            (Z4);

        \draw[blue,thick]
            (Z3)
            to[out=55,in=125]
            pic[pos=0.22,sloped,allow upside down] {centerarrow}
            pic[pos=0.78,sloped,allow upside down] {centerarrow}
            (O);

        \draw[blue,thick]
            (O)
            to[out=55,in=125]
            pic[pos=0.22,sloped,allow upside down] {centerarrow}
            pic[pos=0.78,sloped,allow upside down] {centerarrow}
            (Z2);

        \draw[blue,thick]
            (Z1)
            to[out=-125,in=-55]
            pic[pos=0.22,sloped,allow upside down] {centerarrow}
            pic[pos=0.78,sloped,allow upside down] {centerarrow}
            (Z2);

        \draw[blue,thick]
            (Z1)
            to[out=55,in=180]
            pic[pos=0.50,sloped,allow upside down] {centerarrow}
            (4.50,0.6);

        \draw[red,thick]
            (-4.65,-0.6)
            to[out=0,in=-125]
            pic[pos=0.50,sloped,allow upside down] {centerarrow}
            (Z4);

        \draw[red,thick]
            (Z3)
            to[out=-55,in=-125]
            pic[pos=0.22,sloped,allow upside down] {centerarrow}
            pic[pos=0.78,sloped,allow upside down] {centerarrow}
            (O);

        \draw[red,thick]
            (O)
            to[out=-55,in=-125]
            pic[pos=0.22,sloped,allow upside down] {centerarrow}
            pic[pos=0.78,sloped,allow upside down] {centerarrow}
            (Z2);

        \draw[red,thick]
            (Z1)
            to[out=125,in=55]
            pic[pos=0.22,sloped,allow upside down] {centerarrow}
            pic[pos=0.78,sloped,allow upside down] {centerarrow}
            (Z2);

        \draw[red,thick]
            (Z1)
            to[out=-55,in=180]
            pic[pos=0.50,sloped,allow upside down] {centerarrow}
            (4.50,-0.6);

        \draw[thick,onearrow=0.75]
            (Z4)--(Z3);

        \node[right] at (4.55,0.6) {$\Sigma_1$};
        \node[right] at (4.55,-0.6) {$\Sigma_2$};

        \fill (Z4) circle (1pt)
            node[below] {$z_4$};

        \fill (M) circle (1pt)
            node[below] {$-1$};

        \fill (Z3) circle (1pt)
            node[below] {$z_3$};

        \fill (O) circle (1pt)
            node[below,scale=0.9] {$0$};

        \fill (Z2) circle (1pt)
            node[below] {$z_2$};

        \fill (Z1) circle (1pt)
            node[below] {$z_1$};

    \end{tikzpicture}
    \caption{Opening the jump contour $\mathbb{R}\setminus[z_4,z_3]$ with a sufficiently small angle $\phi$. The opened contours $\Sigma_1$ ($\textcolor{blue}{\bullet}$) and $\Sigma_2$ ($\textcolor{red}{\bullet}$) lie in the green and white decay regions shown in Fig. \ref{proptheta} (a), respectively.}
    \label{open-jump-P-8}
\end{figure}

We now open the jump contour $\mathbb{R}\setminus[z_4,z_3]$, using the factorization $B_-^{-1}B_+$ on $\Gamma$ and $b_-^{-1}b_+$ on $(z_2,z_1)$. To this end, we introduce the contours $\Sigma_1$ and $\Sigma_2=\overline{\Sigma}_1$, as illustrated in Fig. \ref{open-jump-P-8}. The contour $\Sigma_1$ consists of four smooth curves lying above the intervals $(-\infty,z_4)$, $(z_3,0)$, $(0,z_2)$, and $(z_1,+\infty)$, respectively, and one smooth curve lying below the interval $(z_2,z_1)$. For $j=1,2$, let $\Omega_j$ denote the union of the open regions between $\Sigma_j$ and the corresponding intervals of the real axis. In particular, $\Omega_2=\overline{\Omega}_1$. The contours $\Sigma_1$ and $\Sigma_2$ are oriented so that $\Omega_1$ and $\Omega_2$ lie to their right and left, respectively. The interval $(z_4,z_3)$ is oriented from left to right. Choose a sufficiently small angle $\phi$ and the lens contours satisfying the following conditions:
\begin{enumerate}
    \item The regions $\Omega_1$ and $\Omega_2$ do not intersect the critical line $\mathcal{L}$ and satisfy
    \begin{align}
        \operatorname{Re}(2i\theta(z))&<0,\quad z\in\Omega_1,\nonumber\\
        \operatorname{Re}(2i\theta(z))&>0,\quad z\in\Omega_2.\nonumber
    \end{align}
    \item The closures of $\Omega_1$ and $\Omega_2$ stay away from the disks $D_n$ and $\overline{D}_n$, $n=1,\ldots,N$.
    \item $0<\phi<\pi/12<\pi/4$.
\end{enumerate}
Near the attachment points $0$ and $z_j$, $j=1,2,3,4$, the lens contours lie within sectors of opening angle $\phi$ measured from the adjacent real-axis intervals. We further require the lens regions to satisfy
\begin{align}
    |\operatorname{Im}z|\leq\tan\phi\,\operatorname{dist}\bigl(\operatorname{Re}z,\{0,z_1,z_2,z_3,z_4\}\bigr),\quad z\in\Omega_1\cup\Omega_2.
    \label{eq:lens-sector-bound-P-8}
\end{align}
We introduce the notation
\begin{align}
    \Omega=\Omega_1\cup\Omega_2,\quad \Sigma^{(4)}=\Sigma_1\cup\Sigma_2\cup[z_4,z_3].
    \label{eq:lens-contour-P-8}
\end{align}
The jump on $[z_4,z_3]$ is retained.

For the positive stationary points, define
\begin{align}
    \chi_1(\zeta) = \mathbf{1}_{[z_1,z_1+1]}(\zeta),\quad
     \chi_2(\zeta) = \mathbf{1}_{[z_2-1,z_2]}(\zeta),
\end{align}
where $\mathbf{1}_I$ denotes the characteristic function of $I$, and set
\begin{align}
    T_0(z_k)
    := \exp\left\{
        \frac{i}{2}\int_J \frac{\nu(\zeta)}{\zeta}\,d\zeta
        - i\int_J
        \frac{\nu(\zeta)-\nu(z_k)\chi_k(\zeta)}{\zeta-z_k}\,d\zeta
    \right\},
    \quad k=1,2.
\end{align}
Here $J=\mathbb{R}\setminus[z_2,z_1]$ and
$\nu(\zeta)=-(2\pi)^{-1}\log(1-|r(\zeta)|^2)$, as in the definition of $T$. These constants satisfy
\begin{align}
    |T_0(z_k)| = 1, \quad
    \overline{T_0(z_k)} = T_0(z_k)^{-1},
    \quad k=1,2.
\end{align}
Using the principal logarithm, we obtain the following expansions as $z$ approaches the positive stationary points along the corresponding portions of $\Sigma_1\cup\Sigma_2$:
\begin{align}
    &T(z)= T_0(z_1)(z_1-z)^{i\nu(z_1)}
    \left(1+\mathcal{O}\left(|z-z_1|^{1/2}\right)\right),
    \quad z \to z_1, \\
    &T(z)= T_0(z_2)(z-z_2)^{-i\nu(z_2)}
    \left(1+\mathcal{O}\left(|z-z_2|^{1/2}\right)\right),
    \quad z \to z_2.
\end{align}

\begin{proposition}
Let $q_0\in\tanh(x)+H^{4,4}(\mathbb{R})$. For all sufficiently large $t$ with $(x,t)\in \mathcal{P}_{<-8}$, the lens contours satisfying \eqref{eq:lens-sector-bound-P-8} can be chosen so that there exist uniformly bounded functions $R_j:(\Omega_j\cup\partial\Omega_j)\setminus\{z_1,z_2\}\to\mathbb{C}$, $j=1,2$, that are continuous on $(\Omega_j\cup\partial\Omega_j)\setminus\{z_1,z_2\}$ and have continuous first-order partial derivatives in $\Omega_j$, with the following boundary values:
\begin{align}
R_1(z)
&=
\begin{cases}
\dfrac{\overline{r(z)}}{1-|r(z)|^2}T_+(z)^{-2},
& z\in\Gamma, \\[6pt]
-\overline{r(z)}\,T(z)^{-2},
& z\in(z_2,z_1), \\[6pt]
g(z),
& z\in\Sigma_1,
\end{cases}
\\
R_2(z)
&=
\begin{cases}
\dfrac{r(z)}{1-|r(z)|^2}T_-(z)^2,
& z\in\Gamma, \\[6pt]
-r(z)T(z)^2,
& z\in(z_2,z_1), \\[6pt]
\overline{g(\bar z)},
& z\in\Sigma_2.
\end{cases}
\end{align}
Here $R_2(z)=\overline{R_1(\bar z)}$. Set
\begin{align}
g_1(z)
&=\overline{r(z_1)}\,T_0(z_1)^{-2}(z_1-z)^{-2i\nu(z_1)},\\
g_2(z)
&=\overline{r(z_2)}\,T_0(z_2)^{-2}(z-z_2)^{2i\nu(z_2)}.
\end{align}
Then
\begin{align}
g(z)
=
\begin{cases}
\gamma(z_4),
& s<z_4, \\[6pt]
[1-\eta(s/z_3)]\gamma(z_3),
& z_3<s<0, \\[8pt]
[1-\eta(s/z_2)]\dfrac{g_2(z)}{1-|r(z_2)|^2},
& 0<s<z_2, \\[10pt]
-\eta\!\left(\dfrac{s-z_2}{z_1-z_2}\right)g_2(z)-\left[1-\eta\!\left(\dfrac{s-z_2}{z_1-z_2}\right)\right]g_1(z),
& z_2<s<z_1, \\[10pt]
\dfrac{g_1(z)}{1-|r(z_1)|^2},
& s>z_1,
\end{cases}
\end{align}
where $s=\operatorname{Re}z$ and $\eta\in C^\infty(\mathbb{R};[0,1])$ is a fixed cut-off function satisfying $\eta(s)=1$ for $s\leq1/3$ and $\eta(s)=0$ for $s\geq2/3$. The values of $g$ at $z=0,z_3,z_4$ are defined by continuity. The function $\gamma(z)$ is defined near $-1$ by
\begin{align}
\gamma(z)
=
\begin{cases}
\dfrac{\overline{D_2(z)}}{D_1(z)}
\left(\dfrac{a(z)}{T_+(z)}\right)^2,
& \text{in the generic case}, \\[8pt]
\dfrac{\overline{r(z)}}{1-|r(z)|^2}T_+(z)^{-2},
& \text{in the non-generic case},
\end{cases}
\end{align}
where $D_1,D_2$ are given by \eqref{D1D2}. Then, with $s=\operatorname{Re}z$, we have, for $j=1,2$,
\begin{align}
&|\bar{\partial}R_j(z)|
\lesssim
|r'(s)|+|z-z_k|^{-1/2},
\ \  z\in\Omega_j,  \ \  k=0,1,2,3,4,
\end{align}
where we denote $z_0=0$.
\end{proposition}

Define a matrix-valued function
\begin{align}
R^{(3)}(z)
:=
\begin{cases}
\left(\begin{array}{cc}
1 & R_1(z)e^{2it\theta(z)} \\
0 & 1
\end{array}\right),
& z\in\Omega_1, \\[8pt]
\left(\begin{array}{cc}
1 & 0 \\
R_2(z)e^{-2it\theta(z)} & 1
\end{array}\right),
& z\in\Omega_2, \\[8pt]
I,
& \text{elsewhere}.
\end{cases}
\end{align}
The contour $\Sigma^{(4)}=\Sigma_1\cup\Sigma_2\cup[z_4,z_3]$ is depicted in Fig. \ref{open-jump-P-8}. Introduce the transformation
\begin{align}
M^{(4)}(z)=M^{(3)}(z)R^{(3)}(z).
\end{align}
Then $M^{(4)}$ satisfies the following hybrid $\bar{\partial}$--RH problem.
\begin{prob2}
Find a function $M^{(4)}(z)=M^{(4)}(z;x,t)$ with the following properties:
\begin{itemize}
\item $M^{(4)}(z)$ is continuous in $\mathbb{C}\setminus\Sigma^{(4)}$.

\item $M^{(4)}(z)$ satisfies the jump condition
\begin{align}
M^{(4)}_+(z)=M^{(4)}_-(z)V^{(4)}(z),
\quad z\in\Sigma^{(4)},
\end{align}
where
\begin{align}\label{V4-P-8}
V^{(4)}(z)
=
\begin{cases}
\left(\begin{array}{cc}
1 & -R_1(z)e^{2it\theta(z)} \\
0 & 1
\end{array}\right),
& z\in\Sigma_1, \\[8pt]
\left(\begin{array}{cc}
1 & 0 \\
R_2(z)e^{-2it\theta(z)} & 1
\end{array}\right),
& z\in\Sigma_2, \\[8pt]
T_-(z)^{-\sigma_3}V(z)T_+(z)^{\sigma_3},
& z\in(z_4,z_3).
\end{cases}
\end{align}

\item $M^{(4)}(z)=I+\mathcal{O}(z^{-1}), \quad z\to\infty$.

\item For $z\in\mathbb{C}\setminus\Sigma^{(4)}$, we have
\begin{align}
\bar{\partial}M^{(4)}(z)=M^{(4)}(z)\bar{\partial}R^{(3)}(z),
\end{align}
where
\begin{align}\label{Dbar-R3-P-8}
\bar{\partial}R^{(3)}(z)
=
\begin{cases}
\left(\begin{array}{cc}
0 & \bar{\partial}R_1(z)e^{2it\theta(z)} \\
0 & 0
\end{array}\right),
& z\in\Omega_1, \\[8pt]
\left(\begin{array}{cc}
0 & 0 \\
\bar{\partial}R_2(z)e^{-2it\theta(z)} & 0
\end{array}\right),
& z\in\Omega_2, \\[8pt]
0,
& \text{elsewhere}.
\end{cases}
\end{align}
\end{itemize}
\end{prob2}

\subsection{Contribution from a pure RH problem}

We first consider the following pure RH problem for $M^{rhp}(z)$, obtained by omitting the $\bar{\partial}$ component of $M^{(4)}(z)$.
\begin{prob}
    Find  $M^{rhp}(z)=M^{rhp}(z;x,t)$ which satisfies
	  \begin{itemize}
	  	\item  $M^{rhp}(z)$ is analytic in $\mathbb{C}\setminus \Sigma^{(4)}$. See Fig. \ref{open-jump-P-8}.
	  	\item $M^{rhp}(z)$  satisfies the jump condition
\begin{equation*}
	  		M^{rhp}_+(z)=M^{rhp}_-(z)V^{(4)}(z),
	  	\end{equation*}
	  	where $V^{(4)}(z)$ is given by \eqref{V4-P-8}.
	  	\item   $M^{rhp}(z)$  has the same asymptotics as  $M^{(4)}(z)$.
	  \end{itemize}
\end{prob}

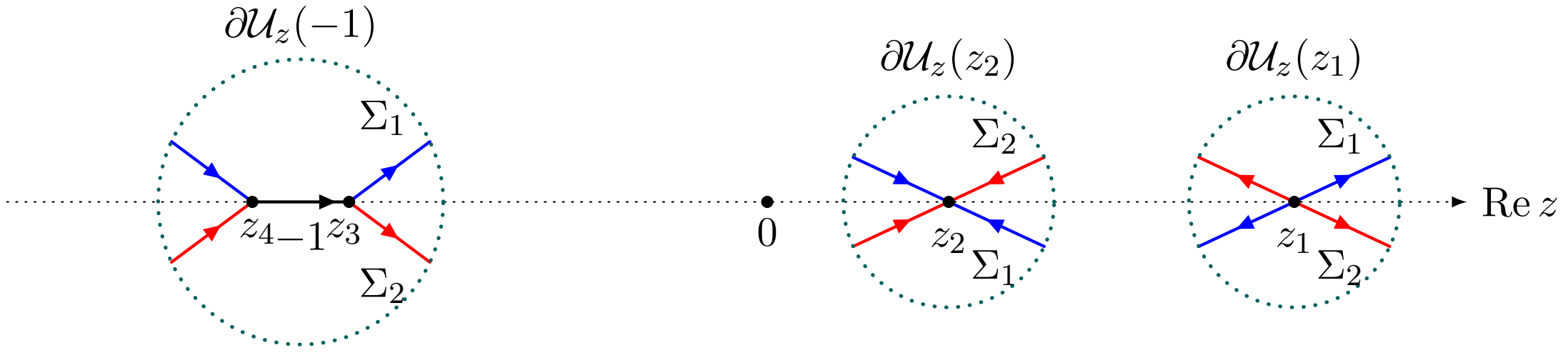
\begin{figure}[htbp]
    \centering
    \begin{tikzpicture}[
        scale=0.8,
        line cap=round,
        line join=round,
        local boundary/.style={
            teal!75!black,
            line width=1pt,
            dash pattern=on 0pt off 3pt,
            line cap=round
        },
        pics/centerarrow/.style={
            code={
                \fill (3pt,0)--(-1.5pt,2pt)--(-1.5pt,-2pt)--cycle;
            }
        }
    ]

        \draw[dotted,-latex] (-8.8,0)--(8.1,0)
            node[black,right] {$\operatorname{Re}z$};

        \begin{scope}[shift={(-5.4,0)}]
            \coordinate (Z4) at (-0.56,0);
            \coordinate (Z3) at (0.56,0);
            \coordinate (LU) at (155:1.65);
            \coordinate (LL) at (205:1.65);
            \coordinate (RU) at (25:1.65);
            \coordinate (RL) at (-25:1.65);

            \draw[blue,thick]
                (LU)--
                pic[pos=0.50,sloped,allow upside down] {centerarrow}
                (Z4);

            \draw[red,thick]
                (LL)--
                pic[pos=0.50,sloped,allow upside down] {centerarrow}
                (Z4);

            \draw[blue,thick]
                (Z3)--
                pic[pos=0.50,sloped,allow upside down] {centerarrow}
                (RU);

            \draw[red,thick]
                (Z3)--
                pic[pos=0.50,sloped,allow upside down] {centerarrow}
                (RL);

            \draw[black,thick]
                (Z4)--
                pic[pos=0.75,sloped,allow upside down] {centerarrow}
                (Z3);

            \draw[local boundary]
                (0,0) circle[radius=1.65];

            \fill (Z4) circle (2pt);
            \fill (Z3) circle (2pt);

            \node at (-0.50,-0.34) {$z_4$};
            \node at (0,-0.42) {$-1$};
            \node at (0.50,-0.34) {$z_3$};

            \node at (0.95,0.97) {$\Sigma_1$};
            \node at (0.95,-0.97) {$\Sigma_2$};
            \node at (0,2.03) {$\partial\mathcal{U}_z(-1)$};
        \end{scope}

        \fill (0,0) circle (2pt);
        \node at (0,-0.35) {$0$};

        \begin{scope}[shift={(2.1,0)}]

            \draw[blue,thick]
                (155:1.22)--
                pic[pos=0.50,sloped,allow upside down] {centerarrow}
                (0,0);

            \draw[blue,thick]
                (-25:1.22)--
                pic[pos=0.50,sloped,allow upside down] {centerarrow}
                (0,0);

            \draw[red,thick]
                (205:1.22)--
                pic[pos=0.50,sloped,allow upside down] {centerarrow}
                (0,0);

            \draw[red,thick]
                (25:1.22)--
                pic[pos=0.50,sloped,allow upside down] {centerarrow}
                (0,0);

            \draw[local boundary]
                (0,0) circle[radius=1.22];

            \fill (0,0) circle (2pt);
            \node at (0,-0.45) {$z_2$};

            \node at (0.53,0.76) {$\Sigma_2$};
            \node at (0.53,-0.76) {$\Sigma_1$};
            \node at (0,1.65) {$\partial\mathcal{U}_z(z_2)$};
        \end{scope}

        \begin{scope}[shift={(6.1,0)}]

            \draw[blue,thick]
                (0,0)--
                pic[pos=0.50,sloped,allow upside down] {centerarrow}
                (205:1.22);

            \draw[blue,thick]
                (0,0)--
                pic[pos=0.50,sloped,allow upside down] {centerarrow}
                (25:1.22);

            \draw[red,thick]
                (0,0)--
                pic[pos=0.50,sloped,allow upside down] {centerarrow}
                (155:1.22);

            \draw[red,thick]
                (0,0)--
                pic[pos=0.50,sloped,allow upside down] {centerarrow}
                (-25:1.22);

            \draw[local boundary]
                (0,0) circle[radius=1.22];

            \fill (0,0) circle (2pt);
            \node at (0,-0.45) {$z_1$};

            \node at (0.53,0.76) {$\Sigma_1$};
            \node at (0.53,-0.76) {$\Sigma_2$};
            \node at (0,1.65) {$\partial\mathcal{U}_z(z_1)$};
        \end{scope}

    \end{tikzpicture}
    \caption{The three local portions of $\Sigma^{loc}$. The model $M^l$ is associated with $\mathcal{U}_z(-1)$, and $M^{r_j}$ with $\mathcal{U}_z(z_j)$, $j=1,2$.}
    \label{jump-three-loc}
\end{figure}

Let $t$ be sufficiently large and fix $0<\tau<1/15$. For fixed constants $\varepsilon>0$ and sufficiently small $\delta>0$, define the open disks
\begin{align}
    &\mathcal{U}_z(-1)
    = \left\{
        z\in\mathbb{C}: |z+1|
        < \left(\frac{15t}{4}\right)^{-1/3+\tau} \varepsilon
    \right\}, \\
    &\mathcal{U}_z(z_j)
    = \left\{
        z\in\mathbb{C}: |z-z_j|<\delta
    \right\}, \quad j=1,2.
\end{align}
Here $\delta$ is chosen sufficiently small so that these disks are pairwise disjoint and avoid both the pole disks and the transition regions of the cut-off functions in $g$. For sufficiently large $t$, both $z_3$ and $z_4$ belong to $\mathcal{U}_z(-1)$. Denote the local jump contour by
\begin{align}
    \Sigma^{loc}
    := \Sigma^{(4)} \cap
    \left(\mathcal{U}_z(-1)\cup \mathcal{U}_z(z_1)\cup \mathcal{U}_z(z_2)\right),
\end{align}
as depicted in Fig. \ref{jump-three-loc}. The local model $M^{loc}(z)$ satisfies the following RH problem.
\begin{prob}
    Find  $M^{loc}(z)=M^{loc}(z;x,t)$ which satisfies
	  \begin{itemize}
	  	\item  $M^{loc}(z)$ is analytic in $\mathbb{C}\setminus \Sigma^{loc}$.
	  	\item $M^{loc}(z)$  satisfies the jump condition
\begin{equation*}
	  		M^{loc}_+(z)=M^{loc}_-(z)V^{loc}(z), \ \ z \in \Sigma^{loc},
	  	\end{equation*}
	  	where $V^{loc}(z) := V^{(4)}(z)|_{ z \in \Sigma^{loc}}$.
	  	\item   $M^{loc}(z)$  has the same asymptotics as  $M^{rhp}(z)$.	
	  \end{itemize}
\end{prob}

Based on the Beals-Coifman theory, the leading asymptotics of $M^{loc}(z)$ are determined by the combined contributions of the separate local models in the neighborhoods of $-1$, $z_1$, and $z_2$ as $t\to\infty$. Denote the corresponding local solutions by $M^l$, $M^{r_1}$, and $M^{r_2}$, respectively.
\begin{prob}
Find $M^j(z)=M^j(z;x,t)$, $j\in\{l,r_1,r_2\}$, with properties
\begin{itemize}
    \item $M^j(z)$ is analytic in $\mathbb{C}\setminus\Sigma^j$, where $\Sigma^l:= \Sigma^{(4)}\cap \mathcal{U}_z(-1), \Sigma^{r_j} := \Sigma^{(4)}\cap \mathcal{U}_z(z_j), j=1,2$.

    \item $M^j(z)$ satisfies the jump condition
    \begin{align}
        M_+^j(z)=M_-^j(z)V^j(z), \quad z\in\Sigma^j,
    \end{align}
    where $V^j(z):=\left.V^{(4)}(z)\right|_{z\in\Sigma^j}$.

    \item As $z\to\infty$, $M^j(z)=I+\mathcal{O}(z^{-1})$.
\end{itemize}
\end{prob}

In the region $-C<(\xi+8)t^{2/3}<0$, the stationary points $z_3,z_4$ coalesce at $-1$, while $z_1,z_2$ remain distinct simple stationary points as $t\to\infty$. Accordingly, $M^l$ is approximated by a Painlev\'{e} II model, whereas $M^{r_1}$ and $M^{r_2}$ are approximated by parabolic cylinder models.

For $z$ close to $-1$,
\begin{align}
    t\theta(z)
    &= 5t(z+1)^3 + (x+8t)(z+1) + A(z), \\
    A(z)
    &= \frac{x+8t}{2}\sum_{n=2}^{\infty}(z+1)^n
    + \frac{t}{4}\sum_{n=4}^{\infty}(n-1)(n+6)(z+1)^n.
\end{align}
Define the space-time parameter $s$ and the scaled variable $k$ by
\begin{align}\label{scaled-P-8}
	s=\frac{4}{15}\left(\frac{15}{4}t\right)^{\frac{2}{3}}(\xi+8),\quad k=\left(\frac{15}{4}t\right)^{\frac{1}{3}}(z+1),
\end{align}
then we have
\begin{align}
t\theta(z)=\frac{4}{3}k^{3}+sk+\mathcal{O}(t^{-1/3}k^{4}).
\end{align}
The transformation in \eqref{scaled-P-8} maps $\mathcal{U}_z(-1)$ onto
\begin{align}
\mathcal{U}_{k}(0)= \{k \in \mathbb{C}: |k|< \left(\frac{15}{4}t\right)^{\tau}\varepsilon\},	
\end{align}
and maps the contour $\Sigma^l$ onto $\widehat{\Sigma}_t^{l}$ in the $k$-plane. Choose $\chi(z)\in C_0^\infty(\mathbb R,[0,1])$, supported near
$-1$ and equal to $1$ on $\mathbb R\cap \mathcal{U}_z(-1)$, and define
\begin{align}\label{eq:rtilde-minus1}
    \widetilde{r}(z)
    =
    \begin{cases}
        \displaystyle
        (1-\chi(z))
        \frac{\overline{r(z)}}{1-|r(z)|^2}T_+(z)^{-2}
        +\chi(z)f(z)h(z)^2,
        & \text{in the generic case}, \\[2mm]
        \displaystyle
        \frac{\overline{r(z)}}{1-|r(z)|^2}T_+(z)^{-2},
        & \text{in the non-generic case},
    \end{cases}
\end{align}
where $f(z)=\overline D_2(z)/D_1(z)$, $h(z)=a(z)/T_+(z)$. Moreover, $\widetilde{r}(z_j)=\gamma(z_j)$ for $j=3,4$, and $|\widetilde{r}(-1)|=|r(-1)|$. As in Sec. \ref{sec-local},  we relate $M^l(z)$ to the Painlev\'e II model through the following steps:
\begin{align}
    M^l(z)
    \rightarrow M^l(k)
    \rightarrow \widehat M^l(k)
    \rightarrow \widetilde M^l(k)
    \rightarrow M^P(k).
\end{align}
Here $\widehat M^l(k)$ is obtained by replacing the phase $t\theta(z)$ by $\frac43k^3+sk$ and the reflection coefficients by their values at $-1$. The subsequent piecewise triangular transformation, contour deformation, and constant conjugation are performed as in Sec. \ref{sec-local}, with $\varphi_0=\arg\widetilde{r}(-1)$. For the Painlev\'e II model $M^P(k;s)$ in \ref{appx}, we take the Stokes parameters $p=i|\widetilde{r}(-1)|$ and $r=0$. It follows that
\begin{align}
    M_1^l(s)=
    \sigma_1
    e^{-i(\frac{\pi}{4}+\frac{\varphi_0}{2})\widehat{\sigma}_3}
    M_1^P(s)\sigma_1
    +\mathcal{O}(t^{-1/3+4\tau}),
\end{align}
where the subscript ``$1$'' denotes the coefficient of $k^{-1}$ in the corresponding expansion at infinity. Consequently, we have, uniformly for $z\in\partial \mathcal{U}_z(-1)$,
\begin{align}
    M^l(z)
    =
    I+\frac{M_1^l(s)}
    {\left(\frac{15t}{4}\right)^{1/3}(z+1)}
    +\mathcal{O}(t^{-2\tau}),
    \quad t\to+\infty.
\end{align}

For the local parametrices $M^{r_j}$ near $z_j$,
$j=1,2$, the local phase expansions are
\begin{align}
    \theta(z)
    =
    \theta(z_j)+\frac12\theta''(z_j)(z-z_j)^2
    +\mathcal{O}((z-z_j)^3),
\end{align}
where $\theta''(z_1)>0>\theta''(z_2)$.
Set
\begin{align}
    \zeta_j
    =
    \sqrt{2t|\theta''(z_j)|}\,(z-z_j),
    \quad
    \nu_j=\nu(z_j),
    \quad j=1,2,
\end{align}
and define
\begin{align}
    \mathfrak r_1
    &=
    r(z_1)T_0(z_1)^2e^{-2it\theta(z_1)}
    \bigl(2t|\theta''(z_1)|\bigr)^{-i\nu_1},\\
    \mathfrak r_2
    &=
    r(z_2)T_0(z_2)^2e^{-2it\theta(z_2)}
    \bigl(2t|\theta''(z_2)|\bigr)^{i\nu_2}.
\end{align}
In particular, $|\mathfrak r_j|=|r(z_j)|\leq r_*<1$ uniformly for sufficiently large $t$. For $r_0\in\mathbb C$ with $|r_0|<1$, let $M^{(PC)}(\zeta;r_0)$ denote the solution of the normalized parabolic cylinder model RH problem corresponding to the $k=2,4$ case of RH Problem 3.11 in \cite{PengChen2026}. Its large-$\zeta$ expansion is
\begin{align}
M^{(PC)}(\zeta;r_0)
=I+\frac{M_1^{(PC)}(r_0)}{\zeta}
+\mathcal{O}(\zeta^{-2}),
\quad \zeta\to\infty.
\end{align}
After deforming and reorienting the rays of the PC model to match the local lens contours, the quadratic approximations are given by
\begin{align}
    \widehat M^{r_1}(z)=
    \sigma_1
    M^{(PC)}(-\zeta_1;\overline{\mathfrak r}_1)
    \sigma_1,\quad
    \widehat M^{r_2}(z)
    =M^{(PC)}(\zeta_2;\mathfrak r_2).
\end{align}
A small-norm RH analysis comparing $M^{r_j}$ with $\widehat M^{r_j}$, together with the above expansion of $M^{(PC)}$, gives, uniformly for $z\in\partial \mathcal{U}_z(z_j)$,
\begin{align}
    M^{r_j}(z)
    =
    I+\frac{B_j(x,t)}
    {\sqrt{2t|\theta''(z_j)|}\,(z-z_j)}
    +\mathcal{O}(t^{-1}),
    \quad j=1,2,
\end{align}
where
\begin{align}
    B_1(x,t)=
    -\sigma_1
    M_1^{(PC)}(\overline{\mathfrak r}_1)
    \sigma_1,\quad
    B_2(x,t)=
    M_1^{(PC)}(\mathfrak r_2).
\end{align}
In particular, uniformly for $z\in\partial \mathcal{U}_z(z_j)$,
\begin{align}
    M^{r_j}(z)=I+\mathcal{O}(t^{-1/2}), \quad j=1,2.
\end{align}

We now construct $M^{rhp}$ in the form
\begin{align}
    M^{rhp}(z)
    =
    \begin{cases}
        E(z),
        & z \in \mathbb{C} \setminus
        \bigl(\mathcal{U}_z(-1) \cup \mathcal{U}_z(z_1) \cup \mathcal{U}_z(z_2)\bigr), \\[2pt]
        E(z)M^l(z),
        & z \in \mathcal{U}_z(-1), \\[2pt]
        E(z)M^{r_j}(z),
        & z \in \mathcal{U}_z(z_j), \quad j=1,2,
    \end{cases}
\end{align}
and it remains to determine $E(z)$.

\begin{figure}[htbp]
    \centering
    \begin{tikzpicture}[
        scale=0.8,
        line cap=round,
        line join=round,
        pics/centerarrow/.style={
            code={
                \fill (3pt,0)--(-1.5pt,2pt)--(-1.5pt,-2pt)--cycle;
            }
        }
    ]

        \coordinate (O)  at (0,0);
        \coordinate (C)  at (-5.10,0);
        \coordinate (Z4) at (-5.62,0);
        \coordinate (Z3) at (-4.58,0);
        \coordinate (Z2) at (2.96,0);
        \coordinate (Z1) at (6.12,0);

        \draw[blue,thick]
            (-8.44,1.31)
            to[out=0,in=125]
            pic[pos=0.50,sloped,allow upside down] {centerarrow}
            (Z4);

        \draw[blue,thick]
            (Z3)
            to[out=55,in=125]
            pic[pos=0.50,sloped,allow upside down] {centerarrow}
            (O);

        \draw[blue,thick]
            (O)
            to[out=55,in=125]
            pic[pos=0.50,sloped,allow upside down] {centerarrow}
            (Z2);

        \draw[blue,thick]
            (Z1)
            to[out=-125,in=-55]
            pic[pos=0.50,sloped,allow upside down] {centerarrow}
            (Z2);

        \draw[blue,thick]
            (Z1)
            to[out=55,in=180]
            pic[pos=0.50,sloped,allow upside down] {centerarrow}
            (8.20,1.05);

        \draw[red,thick]
            (-8.44,-1.31)
            to[out=0,in=-125]
            pic[pos=0.50,sloped,allow upside down] {centerarrow}
            (Z4);

        \draw[red,thick]
            (Z3)
            to[out=-55,in=-125]
            pic[pos=0.50,sloped,allow upside down] {centerarrow}
            (O);

        \draw[red,thick]
            (O)
            to[out=-55,in=-125]
            pic[pos=0.50,sloped,allow upside down] {centerarrow}
            (Z2);

        \draw[red,thick]
            (Z1)
            to[out=125,in=55]
            pic[pos=0.50,sloped,allow upside down] {centerarrow}
            (Z2);

        \draw[red,thick]
            (Z1)
            to[out=-55,in=180]
            pic[pos=0.50,sloped,allow upside down] {centerarrow}
            (8.20,-1.05);

        \fill[white] (C)  circle[radius=1.42];
        \fill[white] (Z2) circle[radius=0.92];
        \fill[white] (Z1) circle[radius=0.92];

        \draw[black,dotted,-latex]
            (-8.70,0)--(8.45,0)
            node[black,right] {$\operatorname{Re}z$};

        \foreach \X/\R in {-5.10/1.42,2.96/0.92,6.12/0.92}{
            \begin{scope}[shift={(\X,0)}]
                \draw[black,thick]
                    (0,0) circle[radius=\R];

                \draw[black,thick,-latex]
                    (110:\R)
                    arc[start angle=110,end angle=92,radius=\R];
            \end{scope}
        }

        \foreach \P in {O,C,Z2,Z1}{
            \fill[black] (\P) circle (2pt);
        }

        \node at (-5.10,-0.32) {$-1$};
        \node at (0,-0.32) {$0$};
        \node at (2.96,-0.32) {$z_2$};
        \node at (6.12,-0.32) {$z_1$};

        \node at (-5.10,1.84) {$\partial\mathcal{U}_z(-1)$};
        \node at (2.96,1.60) {$\partial\mathcal{U}_z(z_2)$};
        \node at (6.12,1.60) {$\partial\mathcal{U}_z(z_1)$};

        \node[black,right] at (8.25,1.05) {$\Sigma_1$};
        \node[black,right] at (8.25,-1.05) {$\Sigma_2$};

    \end{tikzpicture}
    \caption{The jump contour $\Sigma^E$ for $E(z)$.}
    \label{fig-jump-VE-three}
\end{figure}
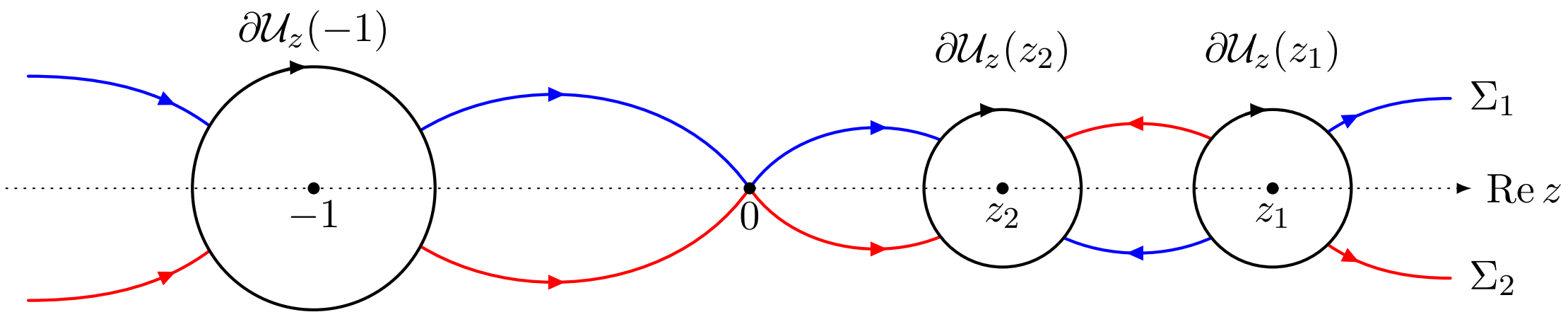

Let $\mathcal{U}=\mathcal{U}_z(-1)\cup \mathcal{U}_z(z_1)\cup \mathcal{U}_z(z_2)$, and orient each component of $\partial\mathcal{U}$ clockwise. The error function $E(z)$ admits the following RH problem.
\begin{prob}\label{E-P-8}
Find $E(z)$ with the properties	
\begin{itemize}
\item $E(z)$ is analytic in $\mathbb{C}\setminus\Sigma^E$, where $\Sigma^E=\partial\mathcal{U}\cup\bigl(\Sigma^{(4)}\setminus\mathcal{U}\bigr)$. See Fig. \ref{fig-jump-VE-three}.
\item $E(z)$  satisfies the jump condition
\begin{align*}
E_{+}(z)=E_-(z)V^{E}(z), \quad z\in \Sigma^{E},
\end{align*}
where the jump matrix $V^{E}(z)$ is given by
\begin{align}
    V^E(z)=
    \begin{cases}
        V^{(4)}(z), & z\in\Sigma^{(4)}\setminus\mathcal{U}, \\
        M^l(z), & z\in\partial \mathcal{U}_z(-1), \\
        M^{r_j}(z), & z\in\partial \mathcal{U}_z(z_j), \quad j=1,2.
    \end{cases}
\end{align}
\item $E(z)=I+\mathcal{O}(z^{-1}),	\quad  z\to  \infty.$	
\end{itemize}
\end{prob}

A result analogous to Proposition \ref{VE-I} can be established, ensuring the existence and uniqueness of the solution to RH Problem \ref{E-P-8}. To recover the potential $q(x,t)$, we need to determine the behavior of $E(z)$ at $z=0$ and as $z\to\infty$. We consider the expansion
\begin{equation}
E(z)=I + z^{-1}E_1 +\mathcal{O}\left(z^{-2}\right),
\end{equation}
where
\begin{equation}
 E_1=-\frac{1}{2\pi i} \int_{\Sigma^{E}} \mu_E(\zeta)\left( V^{E}(\zeta)-I \right)\, \mathrm{d}\zeta.	
\end{equation}

\begin{proposition}
$E_1$ and $E(0)$ can be estimated as follows:
\begin{align}\label{E1-P-8}
    E_1
    &= \left(\frac{15t}{4}\right)^{-1/3}\frac{1}{2}
    \begin{pmatrix}
        i\displaystyle\int_s^\infty u(\zeta)^2\,d\zeta
        & iu(s)e^{i\varphi_0} \\[6pt]
        -iu(s)e^{-i\varphi_0}
        & -i\displaystyle\int_s^\infty u(\zeta)^2\,d\zeta
    \end{pmatrix}
    + \mathcal{O}(t^{-1/3-\tau}), \\
    E(0)
    &= I+\left(\frac{15t}{4}\right)^{-1/3}\frac{1}{2}
    \begin{pmatrix}
        i\displaystyle\int_s^\infty u(\zeta)^2\,d\zeta
        & iu(s)e^{i\varphi_0} \\[6pt]
        -iu(s)e^{-i\varphi_0}
        & -i\displaystyle\int_s^\infty u(\zeta)^2\,d\zeta
    \end{pmatrix}
    + \mathcal{O}(t^{-1/3-\tau}).\label{E0-P-8}
\end{align}
\end{proposition}

\subsection{Contribution from a pure $\bar{\partial}$-problem}

In this part, we consider the long-time asymptotic behavior of the pure $\bar{\partial}$-problem. Define
\begin{equation}
M^{(5)}(z)=M^{(4)}(z)\left(M^{rhp}(z)\right)^{-1},
\end{equation}
which satisfies the following $\bar{\partial}$-problem.
\begin{prob3}
 Find  $M^{(5)}(z):=M^{(5)}(z;x,t)$ which satisfies
            \begin{itemize}
                \item   $M^{(5)}(z)$ is continuous  in $\mathbb{C}$ and analytic in $\mathbb{C}\setminus \overline{\Omega}$.
                \item    $M^{(5)}(z)=I+\mathcal{O}(z^{-1}), \ z\to  \infty$.	

                \item For $z\in \mathbb{C}$,  $M^{(5)}(z)$ satisfies the $ \bar{\partial}$-equation
                \begin{equation}
                    \bar{\partial} M^{(5)}(z) = M^{(5)}(z) W^{(5)}(z), \nonumber
                \end{equation}
                where
	\begin{align}
	W^{(5)}(z):=M^{rhp}(z)  \bar{\partial} R^{(3)}(z) \left(M^{rhp}(z)\right)^{-1},
	\end{align}
 and  $ \bar{\partial}  R^{(3)}(z)$ has been given in \eqref{Dbar-R3-P-8}.
            \end{itemize}

\end{prob3}
The uniform bounds for the Painlev\'e II and parabolic cylinder models, together with the small-norm RH estimates, imply that $M^{rhp}$ and its inverse are uniformly bounded. The solution $M^{(5)}$ of the pure $\bar\partial$-problem satisfies
\begin{align}
    (I-S)M^{(5)}=I,
\end{align}
where the Cauchy operator $S$ is defined by
\begin{align}
    (Sf)(z)
    =
    -\frac{1}{\pi}\iint_{\mathbb C}
    \frac{f(\zeta)W^{(5)}(\zeta)}{\zeta-z}\,dA(\zeta).
\end{align}
\begin{proposition}\label{S-P-8}
The operator $S$ satisfies the estimate
\begin{equation}
  \parallel S\parallel_{L^\infty(\mathbb{C})\to L^\infty(\mathbb{C})}\lesssim  t^{-1/6},
\end{equation}
which implies the existence of $(I-S)^{-1}$ for sufficiently large $t$.
\end{proposition}
Proposition \ref{S-P-8} ensures the unique solvability of the integral equation for sufficiently large $t$. At infinity, the solution admits the expansion
\begin{align}
    M^{(5)}(z)
    =
    I+\frac{M_1^{(5)}(x,t)}{z}+\mathcal{O}(z^{-2}),
    \quad z\to\infty.
\end{align}
The coefficient $M_1^{(5)}$ and the value $M^{(5)}(0)$ are expressed in terms of the integral kernel by
\begin{align}\label{M15-P-8}
    &M_1^{(5)}(x,t)=
    \frac{1}{\pi}\iint_{\mathbb C}
    M^{(5)}(\zeta)W^{(5)}(\zeta)\,dA(\zeta),\\
    &M^{(5)}(0)=
    I-\frac{1}{\pi}\iint_{\mathbb C}
    \frac{M^{(5)}(\zeta)W^{(5)}(\zeta)}{\zeta}\,dA(\zeta).\label{M50-P-8}
\end{align}

\begin{proposition}
We have the following estimates
\begin{equation}\label{M5-bound-P-8}
		|M^{(5)}_1(x,t)| \lesssim t^{-1/2}, \quad |M^{(5)}(0)-I| \lesssim t^{-1/2}.
\end{equation}
\end{proposition}

\begin{proposition}
As $t \to +\infty$, $M^{(3)}(0)$ satisfies the   estimate
	\begin{align}\label{M3-zero-P-8}
		M^{(3)}(0)= E(0) + \mathcal{O}(t^{-1/2}),
	\end{align}
where $E(0)$ is given by \eqref{E0-P-8}.
\end{proposition}

The proofs are analogous to those in Sec. \ref{sec-Dbar} and are omitted. The additional lens regions associated with the nondegenerate stationary points $z_1,z_2$ contribute $\mathcal{O}(t^{-1/4})$ to the operator norm and $\mathcal{O}(t^{-3/4})$ to the area integrals in \eqref{M15-P-8} and \eqref{M50-P-8}. These contributions decay faster and hence do not change the final estimates.

\subsection{Proof of Theorem \ref{Theo-1}--Case II}

We now derive the long-time asymptotics of $q(x,t)$ in $\mathcal{P}_{<-8}$.

\begin{proof}
By \eqref{E1-P-8}-\eqref{E0-P-8} and \eqref{M3-zero-P-8}, we have $M^{(3)}(0)^{-1}=I-E_1+\mathcal{O}(t^{-1/3-\tau})$. Taking $z\to\infty$ vertically as in Sec. \ref{sec-proof}, we use $M^{rhp}(z)=E(z)$ outside the local disks and $M^{(5)}_1(x,t)=\mathcal{O}(t^{-1/2})$ from \eqref{M5-bound-P-8}. The reconstruction formula \eqref{recon-sol} then gives
\begin{align*}
q(x,t)&=-iT(\infty)^2\bigl[\sigma_2+(I-\sigma_2)E_1\bigr]_{12}
+\mathcal{O}(t^{-1/3-\tau})\\
&=-T(\infty)^2+\frac{T(\infty)^2}{2}\left(\frac{15t}{4}\right)^{-1/3}
\left[u(s)e^{i\varphi_0}-i\int_s^\infty u(\zeta)^2\,d\zeta\right]
+\mathcal{O}(t^{-1/3-\tau}).
\end{align*}
The characterization of $u(s)$ follows from \ref{appx} with $p=i|\widetilde r(-1)|$ and $r=0$. This proves Case II of Theorem \ref{Theo-1}.
\end{proof}

\addtocontents{toc}{%
  \protect\setlength{\protect\cftsecnumwidth}{7em}%
}

\renewcommand{\appendixname}{Appendix\space}
\appendix

\section{ Painlev\'{e} \uppercase\expandafter{\romannumeral2}   Model} \label{appx}
The Painlev\'{e} \uppercase\expandafter{\romannumeral2} equation takes the form
\begin{align}\label{up22}
	u_{ss}(s) = 2u^3(s)  +su(s), \quad s \in \mathbb{R},
\end{align}
which can be solved by means of the solution of a  RH problem as follows.
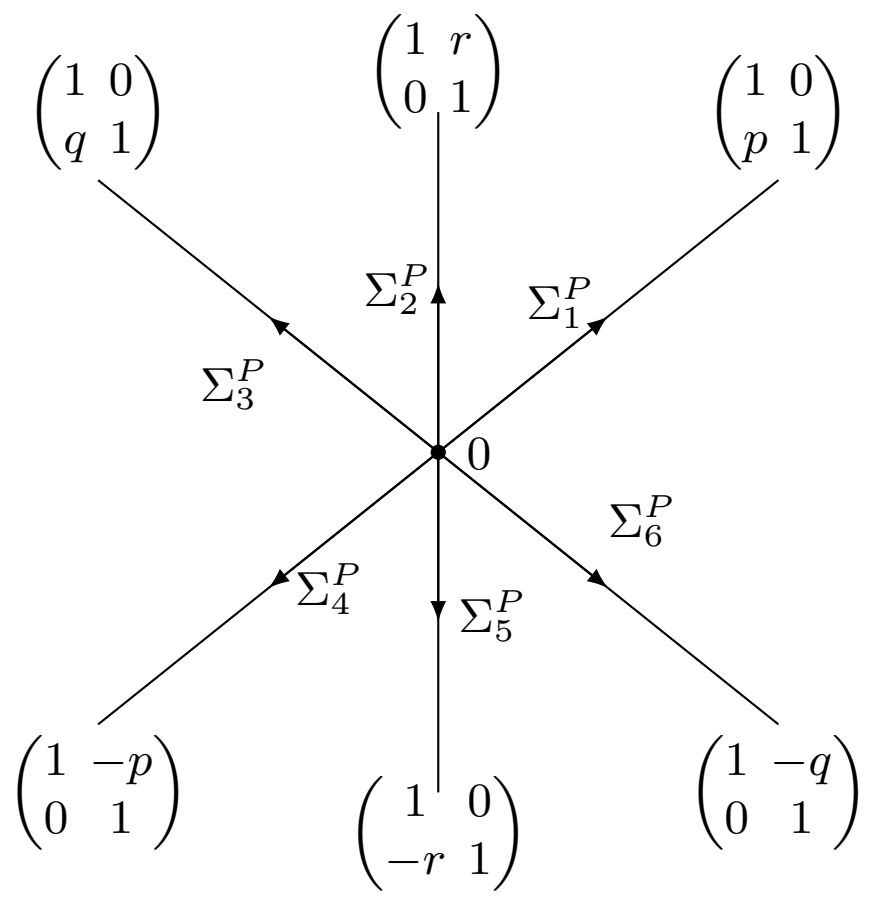
\begin{figure}[H]
	\begin{center}
		\begin{tikzpicture}[scale=0.9]
			\node[shape=circle,fill=black,scale=0.15] at (0,0) {0};
			\node[below] at (0.3,0.25) {\footnotesize $0$};
			\draw [] (0,-2.5 )--(0,2.5);
			\draw [-latex] (0,0)--(0,1.25);
			\draw [-latex] (0,0 )--(0,-1.25);
			\draw [] (0,0 )--(2.5,2);
			\draw [-latex] (0,0)--(1.25,1);
			\draw [] (0,0 )--(2.5,-2);
			\draw [-latex] (0,0)--(1.25,-1);
			\draw [] (0,0 )--(-2.5,2);
			\draw [-latex] (0,0)--(-1.25,1);
			\draw [] (0,0 )--(-2.5,-2);
			\draw [-latex] (0,0)--(-1.25,-1);

			\node at (0.9,1.1) {\footnotesize$\Sigma^P_1$};
			\node at (1.5,-0.5) {\footnotesize$\Sigma^P_6 $};
			\node at (-1.5,0.5) {\footnotesize$\Sigma^P_3$};
			\node at (-0.8,-1) {\footnotesize$\Sigma^P_4$};
			\node at (-0.3,1.2) {\footnotesize$\Sigma^P_2$};
			\node at ( 0.4,-1.2) {\footnotesize$\Sigma^P_5$};
			
			\node  at (2.5,2.5) {\footnotesize $\begin{pmatrix} 1 & 0 \\ p & 1 \end{pmatrix}$};
			\node  at (-2.5,2.5) {\footnotesize $\begin{pmatrix} 1 & 0\\ q& 1 \end{pmatrix}$};
			\node  at (0,2.8) {\footnotesize $\begin{pmatrix} 1 & r\\ 0& 1 \end{pmatrix}$};
			\node  at (0,-2.8) {\footnotesize $\begin{pmatrix} 1 & 0 \\ -r& 1 \end{pmatrix}$};
			\node  at (2.5,-2.5) {\footnotesize $\begin{pmatrix} 1 & -q\\ 0& 1 \end{pmatrix}$};
			\node  at (-2.5,-2.5) {\footnotesize $\begin{pmatrix} 1 & -p \\ 0 & 1 \end{pmatrix}$};
			
		\end{tikzpicture}
		\caption{ \footnotesize { The jump contour $\Sigma^P$ and the corresponding jump matrix.}}
		\label{Sixrays}
	\end{center}
\end{figure}

Denote $\Sigma^P = \bigcup_{j=1}^6\left\{  \Sigma_j^P = e^{i\left(\frac{\pi}{6}+(j-1)\frac{\pi}{3}\right)} \mathbb{R}_+ \right\}$, see Figure \ref{Sixrays}.
The Painlev\'{e} \uppercase\expandafter{\romannumeral2} model RH problem satisfies the following properties:
\begin{prob}\label{mpain}
	Find   $M^{P}(k)=M^{P}(k;s)$ with properties
	\begin{itemize}
		\item Analyticity: $M^{P}(k)$ is analytic in $\mathbb{C}\setminus \Sigma^{P}$.
		\item Jump condition:
		\begin{equation*}
			M^{P}_+( k)=M^{P}_-(k)e^{-i(\frac{4}{3}k^3+sk)\widehat{\sigma}_3}V^{P}(k),
		\end{equation*}
		where
		$V^{P}(k)$ is shown in Figure \ref{Sixrays}. The parameters $p$, $q$, and $r$ in $V^{P}(k)$   satisfy the relation
		\begin{align}\label{rer}
		r=	p+q+pqr .
		\end{align}
		\item Asymptotic behaviors:
		\begin{align*}
			& M^{P}(k) = I +   \mathcal{O} \left(k^{-1}\right), \quad \text{as} \ k \to \infty,\\
			& M^{P}( k) = \mathcal{O}(1),\quad  \text{as} \ k \to 0,
		\end{align*}
and for each $C_1 > 0$,
\begin{align}\label{mPbounded}
	\sup_{k \in \mathbb{C}\setminus \Sigma^P} \sup_{s \geq -C_1} |M^P(k)|  < \infty.
\end{align}
	\end{itemize}
\end{prob}

Then
\begin{equation}\label{up2}
	u(s)=2\left(M_1^P(s)\right)_{12} = 2 \left(M_1^P(s)\right)_{21}
\end{equation}
solves the Painlev\'{e} \uppercase\expandafter{\romannumeral2} equation, where
	\begin{align*}
			& M^{P}(k) = I + k^{-1} M_1^P(s) + \mathcal{O} \left(k^{-2}\right), \quad \text{as} \ k \to \infty.
		\end{align*}
A result due to Hastings and McLeod \cite{p1} presents that, for any $a\in\mathbb{R}$, there exists a unique solution to the homogeneous Painlev\'{e} II equation \eqref{up22} that behaves like
\begin{align}\label{us}
	u(s) = a \mathrm{Ai}(s) +\mathcal{O}\left( s^{-\frac{1}{4}}e^{-\frac{4}{3}s^{3/2}}\right),\quad s\to +\infty,
\end{align}
where  $\mathrm{Ai}(s)$ denotes the Airy function. Particularly, for $q\in i\mathbb{R}$, $|q|<1$, $p=-q$,  and $r=0$, it follows that the solution $u(s)$ has the asymptotics \eqref{us} with $a=\im q$  and the matrix  $M_1^P(s)$ has the form given by
\begin{align}\label{posee}
M_1^P(s) = \frac{1}{2} \begin{pmatrix} -i\int_s^\infty u(\zeta)^2\mathrm{d}\zeta & u(s) \\ u(s) & i\int_s^\infty u(\zeta)^2\mathrm{d}\zeta \end{pmatrix}.
\end{align}
Furthermore, a special argument shows that for the singular case $q\in i \mathbb{R}, |q|=1$, $p=-q$, and $r=0$, \eqref{up2} also leads to a global, real solution of \eqref{up22} with the asymptotics  \eqref{us}. More details can be found in \cite{SegurAblowitz1981,p1,as1,p2}.

\section*{Acknowledgements.}
  This work is supported by the National Natural Science Foundation of China   (Grant No. 12271104 and 12671294). \vspace{2mm}

\noindent\textbf{Data Availability Statements}

The data that support the findings of this study are available within the article.\vspace{2mm}

\noindent{\bf Conflict of Interest}

The authors have no conflicts to disclose.

\end{document}